%% file: main.tex
\documentclass[12pt]{article}
\usepackage{epsfig,amssymb,amsmath,psfrag}
\usepackage{caption}
\usepackage{subcaption}
\usepackage{tikz}
\usepackage{graphicx}

\usetikzlibrary{decorations.pathmorphing} 

\def \bl  {\begin{align*}}
\def \el  {\end{align*}}

\def \be  {\begin{equation}}
\def \ee  {\end{equation}}
\def \ba  {\begin{eqnarray}}
\def \ea  {\end{eqnarray}}
\def \baa {\begin{eqnarray*}}
\def \eaa {\end{eqnarray*}}
\def \bb  {\begin {thebibliography} }
\def \eb  {\end{thebibliography}}
\def \lab #1 {\label{#1}}

\newcommand{\beq}{\begin{equation}}
\newcommand{\eeq}{\end{equation}}
\newcommand{\beqa}{\begin{eqnarray}}
\newcommand{\eeqa}{\end{eqnarray}}

\def \tr {\mathop{\rm tr}\nolimits}

\def\l<{\langle}
\def\r>{\rangle}

\def\XXint#1#2#3{{\setbox0=\hbox{$#1{#2#3}{\int}$}
     \vcenter{\hbox{$#2#3$}}\kern-.5\wd0}}

\renewcommand{\title}[1]{\vbox{\center\LARGE{#1}}\vspace{5mm}}
\renewcommand{\author}[1]{\vbox{\center#1}\vspace{5mm}}

\input{diagramDrawer}
\numberwithin{equation}{section}
\begin{document}

\thispagestyle{empty}

\begin{flushright}

\end{flushright}

\vskip2.2truecm
\begin{center}
\vskip 0.2truecm {\Large\bf
{\Large Light-like Wilson loop correlators}
}\\
\vskip 1truecm
{\bf J.M. Drummond\footnote{J.M.Drummond@soton.ac.uk}, {\"O}. G\"urdo\u gan\footnote{O.C.Gurdogan@soton.ac.uk}, M. Rochford\footnote{M.J.Rochford@soton.ac.uk}, R. Wright\footnote{Rowan.Wright@soton.ac.uk}
}
\vskip 0.4truecm
{\emph{School of Physics and Astronomy, University of Southampton, Southampton, SO17 1BJ, UK
}}

\vskip 0.4truecm

\begingroup\bf\large

\endgroup
\vspace{0mm}

\begingroup
\textit{
 }\\
\par
\endgroup

\end{center}

\vskip 0.4truecm 

\vspace{0mm}

\centerline{\bf Abstract} 
It is well-known that the expectation values of null polygonal Wilson loops computed in planar \(\mathcal{N}=4\) super Yang-Mills theory are dual to MHV amplitudes in that theory, and moreover that the duality can be extended to higher helicity sectors through the introduction of super Wilson loops. In this first of a series of papers, we investigate the natural generalisation posed by correlation functions of \emph{multiple} light-like loop operators, both in the bosonic case and in the case of super Wilson loops. Explicit calculations are presented in several cases and we verify that, in the Abelian theory, these objects obey a natural generalisation of the \(\bar{Q}\)-equation which relates different loop orders, kinematic configurations and Grassmann sectors.

 \noindent

\newpage
\tableofcontents
\newpage
\setcounter{page}{1}\setcounter{footnote}{0}

\section{Introduction}

Light-like Wilson loops have been studied in great detail in planar $\mathcal{N}=4$ super Yang-Mills theory. These investigations were primarily driven by the duality of such objects with maximally-helicity-violating scattering amplitudes \cite{Alday:2007hr,Drummond:2007aua,Brandhuber:2007yx,Drummond:2007cf,Bern:2008ap,Drummond:2008aq}. These objects provide a perfect testing ground to investigate fundamental aspects of quantum field theory, many of which are much more widely applicable. Examples include the analytic structure of perturbative loop amplitudes and loop integrals \cite{Goncharov:2010jf,Arkani-Hamed:2010pyv,Drummond:2010cz,Golden:2013xva,Caron-Huot:2016owq,Drummond:2017ssj,Dixon:2011nj,Caron-Huot:2012awx,Bourjaily:2017bsb,Caron-Huot:2018dsv,Spiering:2024sea,He:2024fij}, the interplay between amplitudes, Wilson loops and correlation functions \cite{Alday:2010zy,Eden:2010zz}, aspects related to field theory at strong coupling via the AdS/CFT correspondence \cite{Alday:2007hr,Alday:2009ga,Alday:2009yn,Alday:2009dv,Alday:2010vh} and the integrable systems which arise in the planar limit of $\mathcal{N}=4$ super Yang-Mills theory \cite{Basso:2013vsa,Basso:2013aha,Basso:2014koa}. The study of Wilson loops and amplitudes has also inspired fascinating connections between quantum field theory and positive geometry \cite{Arkani-Hamed:2012zlh}, leading to a geometric formulation of loop integrands in terms of the Amplituhedron \cite{Arkani-Hamed:2013jha,Arkani-Hamed:2013kca}, whose introduction has inspired many further connections to geometry and combinatorics \cite{Damgaard:2019ztj,Ferro:2022abq,Lukowski:2023nnf,Ferro:2023qdp,Even-Zohar:2023del,Even-Zohar:2024nvw,Galashin:2024ttp}.

Based on the duality, a wealth of techniques have been developed to study Wilson loops and amplitudes, some most naturally arising in the amplitude setting, and some more obviously based on the Wilson loop formulation. There are several examples of direct relevance here. First, we have the extension of Wilson loops to super Wilson loops \cite{Mason:2010yk,Caron-Huot:2010ryg}, necessary to extend the duality to amplitudes beyond the MHV sector and the twistorial formulation of the theory \cite{Boels:2006ir} to develop techniques to calculate both amplitudes and Wilson loops \cite{Bullimore:2010pj,Adamo:2011pv,Bullimore:2013jma}. Second, there are the BCFW recursion relations \cite{Britto:2005fq}, used to compute tree-level amplitudes and loop integrands \cite{Arkani-Hamed:2010zjl}. Third, we have the $\bar{Q}$ equation which applies to super Wilson loops and relates these quantities at different perturbative loop orders \cite{Caron-Huot:2011dec,Bullimore:2011kg}.

Given the depth and breadth of the connections revealed in the above studies, it has been of significant interest to generalise the objects involved. One generalisation that has been well-studied involves replacing the Wilson loop with a correlator of a light-like loop operator and a local operator, specifically the (chiral) Lagrangian of the theory \cite{Alday:2011ga,Alday:2012hy,Alday:2013ip,Chicherin:2022bov,Chicherin:2022zxo,Chicherin:2024hes,Carrolo:2025pue}. Such an observable is related to the original Wilson loop - it may be thought of as the object obtained by performing all but one of the loop integrations needed to obtain the $L$-loop Wilson loop/amplitude from the integrand.

Here we aim to generalise the setting to include correlation functions of multiple light-like loop operators. These objects are not immediately dual to scattering amplitudes in the same way as a single light-like loop. However, they do provide a natural setting to study many of the same connections and questions described above. Here we will first focus on the general structure expected of these objects, including their colour structure, the nature of their ultraviolet divergences, as well as their anomlaous conformal symmetry. We will then investigate perturbative aspects of their computation. We will employ the same super twistor formalism mentioned above to generalise these objects and construct a supersymmetric completion. Then we will use these techniques to make some explicit perturbative computations. We also propose a natural extension of the $\bar{Q}$ equation to the correlators of multiple loops and we will show that this has a non-trivial consequence even in the simple case of Abelian loops which we can test. In forthcoming work we will discuss the generalisation of the BCFW recursion relations to such objects and we will also explore some deeper consequences of the $\bar{Q}$ equation.

\section{Multiple Wilson loop correlators}
\label{Sec-MWLcorrs}

We would like to consider four-dimensional $\mathcal{N}=4$ super Yang-Mills theory with gauge group $G=SU(N)$ or $G=U(N)$. We refer the reader to Appendix \ref{app-YMandWL} for our conventions. The objects we would like to consider here are loop operators
\be
\mathcal{L}(C) = \frac{1}{N} \tr \mathcal{P}\,{\rm exp} \,i  \oint_C  dx^{\mu}A_\mu\,,
\label{loopOp}
\ee
where we take a closed $n$-sided piecewise light-like contour $C$, familiar from the study of the duality between amplitudes and Wilson loops in planar $\mathcal{N}=4$ super Yang-Mills theory \cite{Alday:2007hr,Drummond:2007aua,Brandhuber:2007yx,Drummond:2007cf,Bern:2008ap,Drummond:2008aq,Drummond:2007bm}. Note that here we will take the trace in the fundamental representation of $G$. 

We recall that in the $\mathcal{N}=4$ theory, the expectation values of such Wilson loop operators take the form \cite{Drummond:2007cf,Drummond:2007bm,DV1,KK1,Bassetto:1993xd,Drummond:2007au}
\be
W_n = \langle \mathcal{L}(C) \rangle = \Bigl[\prod_{i=1}^n D_i \Bigr] F_n R_n\,.
\label{Wnfactors}
\ee
Here $D_i$ is a UV divergent factor (see eq. (\ref{Di})), associated to the corner at $x_i$. The factor $F_n$ (see eqs. (\ref{Fn}), (\ref{Fn1loop}) and (\ref{Iij})) is a particular choice of finite part which is constrained by the anomalous conformal Ward identity \cite{Drummond:2007cf,Drummond:2007au} given in (\ref{CWI}). Finally, $R_n$ is a finite and conformally invariant piece (whose logarithm in the planar limit is the `remainder function', much studied in the literature in relation to scattering amplitudes \cite{Bern:2008ap,Drummond:2007bm,Drummond:2007au,Dixon:2011pw,Dixon:2013eka,Dixon:2014voa}). 

It is convenient to expand perturbatively in the 't Hooft coupling,
\be 
g^2 \equiv \frac{g_{\rm YM}^2N}{16\pi^2}.
\ee
Note that we have normalised the operator (\ref{loopOp}) so that at weak coupling we have 
\be
W_n = 1 + O(g^2)\,.
\ee  
Note also that our choice of $F_n$ fully captures the order $g^2$ result and hence factor $R_n$ has a leading correction of order $g^4$ in perturbation theory.
\be
R_n = 1 + O(g^4)\,.
\ee
In the Abelian theory we simply have $R_n=1$ and the entire expression for the Wilson loop is given by exponentiating the $O(g^2)$ result (see equations (\ref{Di}), (\ref{GammasAb}) and (\ref{Fn})). We provide a review of the leading perturbative Wilson loop computations in Appendix A, including expressions for the divergent and finite integrals which arise at leading order.

Here we are interested in considering correlation functions of multiple such operators\footnote{These are no longer dual to amplitudes but we study them as interesting objects in their own right.}. Such objects are also UV divergent and the divergences should factorise over the corners on each loop. 
\be
W_{n_1,\ldots,n_m} = \langle \mathcal{L}(C_1) \ldots \mathcal{L}(C_m)\rangle = \prod_{r=1}^m \biggl[\Bigl[\prod_{i=1_r}^{n_r} D_{i_r} \Bigr] F_{n_r} \biggr]R_{n_1,\ldots,n_m}\,.
\label{Wnmfactors}
\ee
Again, our choice of normalisation of the loop operators means
\be
W_{n_1,\ldots,n_m} = 1 + O(g^2)\,.
\ee
In (\ref{Wnmfactors}) $D_{i_r}$ and $F_{n_r}$ are the same divergent and finite factors appearing in the equation (\ref{Wnfactors}) for the expectation value of the single light-like loop operator $\mathcal{L}(C_r)$. The function $R_{n_1,\ldots,n_m}$ is a new finite function of the data defining all of the loops $C_r$.
Alternatively, we can consider the ratio
\be
 \frac{\langle \mathcal{L}(C_1) \ldots \mathcal{L}(C_m)\rangle}{\langle \mathcal{L}(C_1) \rangle \ldots \langle \mathcal{L}(C_m) \rangle} = \frac{R_{n_1,\ldots,n_m}}{R_{n_1} \ldots R_{n_m}}
 \label{WLratiofn}
\ee
Objects such as $R_{n_1 \ldots n_r}$ or the ratio (\ref{WLratiofn}) should be finite and conformally invariant. 

Note that one contribution to the multi-loop correlator is just the product of the individual Wilson loops and in fact, at large $N$, this is the leading contribution. For example, for a correlator of two Wilson loops we can define a connected part as follows,
\be
W_{n_1,n_2} = W_{n_1} W_{n_2} +W_{n_1,n_2}^{\rm conn}\,.
\ee
The connected part is suppressed in the large $N$ limit by $1/N^2$ compared to the disconnected part. The factorisation of divergences and conformal anomalies implies that the connected part also obeys
\be
W_{n_1,n_2}^{\rm conn} = \prod_{r=1}^2 \biggl[\Bigl[\prod_{i_r=1}^{n_r} D_{i_r} \Bigr] F_{n_r} \biggr]R_{n_1,n_2}^{\rm conn}\,.
\ee
Equivalently, we can therefore write
\be
R_{n_1,n_2} = R_{n_1} R_{n_2} + R_{n_1,n_2}^{\rm conn}\,.
\ee

For correlation functions with more loop operators we can similarly define fully connected pieces, e.g. for three Wilson loops we have
\be
R_{n_1,n_2,n_3} = R_{n_1} R_{n_2} R_{n_3}  + R_{n_1} R_{n_2,n_3}^{\rm conn} + R_{n_2} R_{n_1,n_3}^{\rm conn} + R_{n_3} R_{n_1,n_2}^{\rm conn} + R_{n_1,n_2,n_3}^{\rm conn}  \,.
\ee
The fully connected piece $R_{n_1,n_2,n_3}^{\rm conn}$ is now suppressed by two powers of $1/N^2$ at large $N$. 

The connected parts of the correlators $W^{\rm conn}_{n_1,\ldots,n_m}$ require at a minimum $(m-1)$ propagators to connect all the loop operators.

We will write perturbative expansions in the following general form,
\begin{align}
&W_{n_1,\ldots,n_m} = \sum_l g^{2l} W_{n_1,\ldots,n_m}^{(l)}\,, \quad &R_{n_1,\ldots,n_m} = \sum_l g^{2l} R_{n_1,\ldots,n_m}^{(l)}\,, \notag \\
&W^{\rm conn}_{n_1,\ldots,n_m} = \sum_l g^{2l} W^{(l),{\rm conn}}_{n_1,\ldots,n_m}\,, &R^{\rm conn}_{n_1,\ldots,n_m} = \sum_l g^{2l} R^{(l),{\rm conn}}_{n_1,\ldots,n_m}\,.
\end{align}
Note that, although we use the 't Hooft coupling $g^2$ as the expansion parameter, we are not taking the large $N$ limit here, so that each coefficient in the expansion in $g^2$ is itself a function of $N$. The above statements about the low orders in the expansion translate into
\begin{align}
&W_{n_1,\ldots,n_m}^{(0)} = R_{n_1,\ldots,n_m}^{(0)} = 1\,, \qquad R_n^{(1)} = 0\,, \notag \\
&W_{n_1,\ldots,n_m}^{(l), {\rm conn}} = R_{n_1,\ldots,n_m}^{(l), {\rm conn}} = 0\,, \qquad (l=0,\ldots,m-2\,, \quad m\geq 2)\,.
\end{align}
As we will shortly see, the connected parts are even more suppressed for small $g^2$ in the $SU(N)$ theory for colour reasons and we then have in addition,
\be
W_{n_1,\ldots,n_m}^{(m-1), {\rm conn}} = R_{n_1,\ldots,n_m}^{(m-1), {\rm conn}} = 0\,, \qquad G=SU(N)\,.
\ee

\subsection{$R_{n_1,n_2}$ at order $g^2$}

\begin{figure}
\begin{center}
\begin{tikzpicture}[line width=1.0pt]
  \draw (0,0) rectangle (1,1);
  \draw (3,0) rectangle (4,1);
  \draw[decorate,decoration={snake,amplitude=1.5pt,segment length=6pt}] (1,0.5) -- (3,0.5);
\end{tikzpicture}
\caption{Feynman diagram contribution to the correlator of two squares in the Abelian theory.}
\label{Abelianleading}
\end{center}
\end{figure}
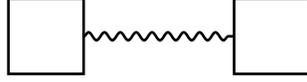

 For the case of two loop operators, the disconnected contribution, from diagrams where a gluon propagator begins and ends on the same loop, is well known \cite{Drummond:2007aua,Brandhuber:2007yx}. Therefore we just consider the extra piece from the connected diagram where the propagator $G^{1,2}_{\mu_1 \mu_2,a_1 a_2}$ (given in eq. (\ref{GandV})) crosses between the two loops,
 \begin{align}
    \langle \mathcal{L}(C_1) \mathcal{L}(C_2) \rangle^{\rm conn} = -\frac{1}{N^2} \tr(t_{a_1}) \tr(t_{a_2}) \int_0^1 dt_1 \dot{x}^{\mu_1}(t_1)  \int_0^1 dt_2 \dot{x}^{\mu_2}(t_2) G^{1,2}_{\mu_1 \mu_2,a_1 a_2} + O(g^4) \notag \,.
\end{align}
 The order $g^2$ contribution vanishes for $G=SU(N)$ as it comes with a colour factor of the form $\tr(t_a) \tr(t_a)$. It is non-vanishing for $G=U(N)$ however, in which case it is given as a sum over finite diagrams $I_{ij}$ where the propagator is between the edge $(x_i,x_{i+1})$ on loop $C_1$ and the edge $(x_j,x_{j+1})$ on $C_2$, as illustrated in Fig. \ref{Abelianleading},
 \begin{align}    
 \langle \mathcal{L}(C_1) \mathcal{L}(C_2) \rangle^{\rm conn} = W_{n_1,n_2}^{\rm conn} = \frac{g^2}{N^2} f_{n_1,n_2}+ O(g^4)\,, \qquad f_{n_1,n_2} = \sum_{i,j} I_{ij} \,.
 \label{W12connUN}
\end{align}
The finite diagram in question is given by
\begin{align}
I_{ij} = \quad &  
{\rm Li}_2\biggl[\frac{(x_{ij}^2-x_{i,j+1}^2)(x_{i,j+1}^2-x_{i+1,j+1}^2)}{x_{i,j+1}^2x_{i+1,j}^2-x_{ij}^2 x_{i+1,j+1}^2}\biggr] 
+ {\rm Li}_2\biggl[\frac{(x_{ij}^2-x_{i+1,j}^2)(x_{i+1,j}^2-x_{i+1,j+1}^2)}{x_{i,j+1}^2x_{i+1,j}^2-x_{ij}^2 x_{i+1,j+1}^2}\biggr] \notag \\
 - &{\rm Li}_2 \biggl[ \frac{(x_{ij}^2-x_{i,j+1}^2)(x_{ij}^2-x_{i+1,j}^2)}{x_{i,j+1}^2x_{i+1,j}^2-x_{ij}^2 x_{i+1,j+1}^2} \biggr]
-{\rm Li}_2 \biggl[ \frac{(x_{i,j+1}^2-x_{i+1,j+1}^2)(x_{i+1,j}^2-x_{i+1,j+1}^2)}{x_{i,j+1}^2x_{i+1,j}^2-x_{ij}^2 x_{i+1,j+1}^2} \biggr]\,,
\label{Iijsec2}
\end{align}
which is the same formula as found in \cite{Brandhuber:2007yx} and quoted in (\ref{Iij}) but the interpretation is different: here $x_i$ lies on loop $C_1$ while $x_j$ lies on $C_2$.

Since order $g^2$ is the leading order contribution to the connected part $W_{n_1,n_2}^{\rm conn}$, the same contribution can also be written as the leading contribution to the conformally invariant finite factor $R_{n_1,n_2}^{\rm conn}$,
\be
R_{n_1,n_2}^{\rm conn} = \frac{g^2}{N^2} f_{n_1,n_2} + O(g^4)\,.
\label{WWconnAb}
\ee
As explained in more detail in Appendix \ref{App-multi-WL}, by studying the total derivative $df_{n_1,n_1}$ one can see that $f_{n_1,n_2}$ is indeed conformally invariant. 

To write an expression for $f_{n_1,n_2}$ in a manifestly conformally invariant form we can define conformal cross-ratios via
\be
u_{i,j,k,l} = \frac{x_{ij}^2 x_{kl}^2}{x_{il}^2 x_{kj}^2}\,, \qquad v_{ij} = u_{i,j,i+1,j+1}\,.
\ee
The cross-ratios $v_{ij}$ are not all multiplicatively independent due to the relations
\be
\prod_i v_{ij} = 1\, \qquad \prod_j v_{ij} = 1\,.
\label{vrels}
\ee
In terms of these cross-ratios we have
\be
f_{n_1,n_2} =  \sum_{i,j}^{n_1,n_2} \, {\rm Li}_2(1-v_{ij}) +  \sum_{\substack{k \leq i \\ j \leq l }}  \log v_{ij}  \log v_{kl} \,.
\label{Gexpr2}
\ee
More details on the derivation of (\ref{Gexpr2}) are also given in Appendix \ref{App-multi-WL}.

Note that the function $f_{n_1,n_2}$ behaves as expected under collinear limits. For instance, we may introduce twistor variables to describe the geometry of both loops. Let us label them as $\{Z_1,\ldots,Z_{n_1}\}$ for the first loop and $\{\tilde{Z}_1,\ldots,\tilde{Z}_{n_2}\}$ for the second. If we take $f_{n_1,n_{2}+1}$
and send the momentum twistor \(\tilde{Z}_{n_2+1}\) collinear with \(\tilde{Z}_{n_2}\), e.g. via the limit
\be
\tilde{Z}_{n_2+1} \to \tilde{Z}_{n_2} - \epsilon \tilde{Z}_{\tilde{n}_2-1} + \frac{\langle \tilde{n}_2-1 \, \tilde{n}_2 \, \tilde{2} \, \tilde{3} \rangle}{\langle \tilde{n}_2 \, \tilde{1} \, \tilde{2} \, \tilde{3} \rangle} \epsilon \tau \tilde{Z}_{1} + \frac{\langle \tilde{n}_2-2\, \tilde{n}_2-1\, \tilde{n}_2\, \tilde{1} \rangle}{ \langle \tilde{n}_2-2\,\tilde{n}_2-1\,\tilde{2}\,\tilde{1} \rangle}\epsilon^2 \tilde{Z}_{2} 
\label{collinearlim}
\ee
where \(\epsilon \to 0\) parametrises the collinear limit and \(\tau\) is the longitudinal momentum fraction, the resulting function is simply \(f_{n_1,n_2}\) after sending \(\epsilon \to 0\). 

In the Abelian theory we obtain the full result from exponentiation of the order $g^2$ correction
\be
W^{U(1)}_{n_1,n_2} = W_{n_1} W_{n_2} \,{\rm exp}\,\Bigl\{g^2 f_{n_1,n_2}\Bigr\}\,.
\ee
This Abelian result generalises straightforwardly to a correlator of any number of Wilson loop operators,
\be
W^{U(1)}_{n_1,\ldots,n_m} = \Bigl[\prod_r W_{n_r} \Bigr] \, {\rm exp}\, \Bigl\{ g^2 \sum_{r>s} f_{n_r,n_s} \Bigr\}\,.
\ee

\subsection{Leading contributions for the $SU(N)$ theory}
\label{Sec-SU(N)Leading}
\begin{figure}
\begin{center}
\begin{tikzpicture}[line width=1.0pt]
  \draw (0,0) rectangle (1,1);
  \draw (3,0) rectangle (4,1);
  \draw[decorate,decoration={snake,amplitude=1.5pt,segment length=6pt}] (1,0.5) -- (3,0.5);
\def\r{0.8cm} 
  \draw  (1.65,-2.4) ++(0:\r)
    \foreach \x in {72,144,...,359} {
      -- ++(\x:\r)
    } -- cycle; 

 \draw[decorate,decoration={snake,amplitude=1.5pt,segment length=6pt}] (0.5,0) -- (1.6,-1.5);
 \draw[decorate,decoration={snake,amplitude=1.5pt,segment length=6pt}] (3.5,0) -- (2.5,-1.5);
\end{tikzpicture}
\caption{Feynman diagram contribution to the connected part of the correlator of two squares and one pentagon in the non-Abelian theory.}
\label{threeWLleading}
\end{center}
\end{figure}
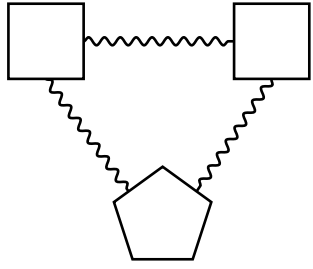

Although the result (\ref{W12connUN}) is for the $U(N)$ theory while the corresponding order $g^2$ term vanishes in the $SU(N)$ theory, we may still make use of our result for $f_{n_1,n_2}$ in the $SU(N)$ theory. 
In the $SU(N)$ theory one cannot have a single field contribution from any given loop operator due to the tracelessness of the generators $t_a$. The minimal contribution to the connected part of a correlator of multiple Wilson loops is therefore one with exactly two fields from the expansion of each loop operator. An example diagram is given in Fig. \ref{threeWLleading}.

We therefore find (recall $C_F = \frac{N^2-1}{2N}$)
\begin{align}
W_{n_1,\ldots,n_m}^{\rm conn} &= \langle \mathcal{L}(C_1) \ldots \mathcal{L}(C_m)\rangle^{\rm conn} \notag \\
&= \frac{(-1)^m}{N^m}\bigg\langle \prod_r \int_{t_{r,1} > t_{r,2} } \!\!\!\!\!\!\!\!\!\!\! dt_{r,1} dt_{r,2} \dot{x}_{r,1}^{\mu_r} \dot{x}_{r,2}^{\nu_r} A^{a_r}_{\mu_r}\bigl(x_r(t_{r,1})\bigr) A^{b_r}_{\nu_r}\bigl(x_r(t_{r,2})\bigr) \tr (t_{a_r} t_{b_r})\bigg\rangle^{\rm conn} + O(g^{2m+2}) \notag \\
&= \frac{2 N C_F}{2^m N^m} \Biggl[\biggl[ \prod_r \frac{1}{2} \frac{g_{\rm YM}^2}{4\pi^2} \int dt_{r+1,1} \int dt_{r,2} \frac{\dot{x}_{r+1,1} \cdot \dot{x}_{r,2}}{[-(x(t_{r+1,1}) - x(t_{r,2}))^2 + i \varepsilon]}\biggr]  + \ldots \Biggr] + O(g^{2m+2}) \notag \\
&= \frac{2 N C_F g^{2m}}{2^m N^{2m}} \Biggl[\biggl[ \prod_r  f_{r,r+1} \biggr]  + \ldots \Biggr] + O(g^{2m+2})\,,
\label{leadingcomb}
\end{align}
where the omitted terms are the other possible ways of making the contractions. There are $2^{m-1} (m-1)!$ such terms in total but only $\frac{(m-1)!}{2}$ are inequivalent for $m>2$. For $m=2$ there is one inequivalent set of contractions. We therefore find 
\be
W_{n_1,\ldots,n_m}^{\rm conn} = g^{2m} \frac{2 N C_F}{N^{2m}} \prod_{r=1}^m f_{n_r,n_{r+1}} + \text{(non-dihedral permutations)} + O(g^{2m+2})
\ee
for $m>2$. For example, for $m=3$ we have
\be
W_{n_1,n_2,n_3}^{\rm conn} = g^{6} \frac{2 N C_F}{N^{6}} f_{n_1,n_2} f_{n_2,n_3} f_{n_3,n_1} + O(g^{8})\,,
\ee
while for $m=4$ we have
\begin{align}
W_{n_1,\ldots,n_4}^{\rm conn} = &\,g^{8} \frac{2 N C_F}{N^{8}} [f_{n_1,n_2}f_{n_2,n_3}f_{n_3,n_4}f_{n_4,n_1} + f_{n_1,n_2}f_{n_2,n_4}f_{n_4,n_3}f_{n_3,n_1} + f_{n_1,n_3}f_{n_3,n_2}f_{n_2,n_4}f_{n_4,n_1}] \notag \\
&+ O(g^{10})\,.
\end{align}
For $m=2$ we have an extra factor of $\frac{1}{2}$ which can be thought of as a symmetry factor between the two propagators,
\be
W_{n_1,n_2}^{\rm conn} = g^{4} \frac{N C_F}{N^{4}} (f_{n_1,n_2})^2 + O(g^{2m+2})\,.
\ee

\subsection{Configurations involving triangles}

Note that, at least formally, one may consider expressions involving triangles. Since a light-like triangle must be collinear in real kinematics, this requires allowing complex kinematics. If one takes $f_{n_1,4}$ and takes the collinear limit described in (\ref{collinearlim}) (for which purpose one may replace the appearances there of $\tilde{Z}_2$ with e.g. $Z_1$) then the result does not vanish. Instead, looking at eq. (\ref{twistordW12}) we see that if $n_2=3$ and we label the three twistors describing the loop $C_2$ as $(\tilde{Z}_1,\tilde{Z}_2,\tilde{Z}_3)$ then we have
\begin{align}
d f_{n_1,3} &= \sum _{i,j} \log \frac{x_{i,j+1}^2 x_{i+1,j}^2}{x_{ij}^2 x_{i+1,j+1}^2} d \log \bigl[ \langle i-1\, i \, i+1 \, \tilde{\jmath} \rangle \langle i \, \tilde{1} \, \tilde{2} \, \tilde{3}\rangle\bigr] \notag \\
&= \sum _{i,j} \log \frac{x_{i,j+1}^2 x_{i+1,j}^2}{x_{ij}^2 x_{i+1,j+1}^2} d \log  \langle i-1\, i \, i+1 \, \tilde{\jmath} \rangle  \,,
\label{dfn13}
\end{align}
where the factor $\langle i \, \tilde{1} \, \tilde{2} \, \tilde{3}\rangle$ in the $d\log$ in the first line disappears via telescoping in the sum over $j$ as it no longer depends on $j$. If we also have $n_1=3$, the same happens for the remaining factor in the $d \log$ and we conclude that $df_{3,3}=0$, i.e. that $f_{3,3}$ is constant. Under a collinear limit on $C_1$, $f_{3,3}$ reduces to $f_{2,3}$ which vanishes as $C_1$ becomes a backtracking loop and we conclude that in fact $f_{3,3}=0$, even in complex kinematics. 

For \(n_1 > 3\), \(f_{n_1,3}\) is a non-zero function in complex kinematics. In real kinematics, \(f_{n_1,3}\) vanishes for all \(n_1\) owing to the fact that a closed light-like triangle in real Minkowski space is necessarily degenerate and the loop $C_1$ becomes a backtracking loop. To see this we may impose reality at the level of twistors by setting
\begin{align}
\tilde{Z}_2 = \tilde{Z}_1 + \epsilon(Z_{r_1} + aZ_{r_2} + bZ_{r_3}) \, \qquad
\tilde{Z}_3 = \tilde{Z}_1 + \epsilon(Z_{r_1} + cZ_{r_2} + dZ_{r_3})
\end{align}
and taking the \(\epsilon \to 0\) limit. Note that here \(Z_{r_i}\) are three twistors other than the three defining the triangle, which can be taken from $C_1$. It is then straightforward  to check that in this limit the derivative (\ref{dfn13}) vanishes and again reduction to $n_2=2$ implies $f_{n_1,3}=0$ in real kinematics. 
It is also worth noting that if one computes a one-loop integrand for a triangle-polygon correlator in the twistor Wilson loop formalism (as will be shortly reviewed), such functions vanish even at the level of the integrand, diagram by diagram, in real kinematics. 

\section{Correlators of Super Wilson loops}

We can extend the definition of multiple Wilson loop correlators to multiple super Wilson loop correlators $\mathcal{W}_{n_1,\ldots,n_m}$ following \cite{Mason:2010yk,Caron-Huot:2010ryg}. We can describe the kinematic dependence of these objects with supertwistors $\mathcal{Z}_{r,i_r}=(Z_{r,i_r}|\chi_{r,i_r})$ with $r=1,\ldots, m$ and $i_r = 1,\ldots,n_r$, where $\chi_{r,i_r}$ are Grassmann variables and $n_r$ is the number of supertwistors describing the contour $C_r$. We make use of the supertwistor Wilson loop formulation of Mason and Skinner \cite{Mason:2010yk} in order to perform computations for these objects. 

The super Wilson loops decompose into sectors of degree $4k$ in Grassmann variables,
\be
\mathcal{W}_{n_1,\ldots,n_m} = \sum_k \mathcal{W}^{(k)}_{n_1,\ldots,n_m}\,.
\ee
For a single loop operator, these sectors correspond to the N${}^k$MHV sectors of the corresponding super-amplitude. For correlators of multiple loop operators we will sometimes employ the same terminology here (MHV, NMHV etc.), although these sectors no longer correspond to helicity amplitudes.

The first term $\mathcal{W}^{(0)}_{n_1,\ldots,n_m} = W_{n_1,\ldots,n_m}$ is just the correlator of bosonic Wilson loops discussed above. Note that the factorisation of the divergent and anomalous conformal finite parts should be just as for the bosonic Wilson loop correlators,
\be
\mathcal{W}_{n_1,\ldots,n_m}  = \biggl[\prod_r \Bigl[\prod_{i=1}^n D_i \Bigr] F_{n_r}\biggr] \mathcal{R}_{n_1,\ldots,n_m}\,,.
\label{superWLfactors}
\ee
Here the finite conformally invariant factor $\mathcal{R}$ has an expansion into terms of Grassmann degree $4k$,
\be
\mathcal{R}_{n_1,\ldots,n_m} = \sum_k \mathcal{R}^{(k)}_{n_1,\ldots,n_m}
\ee
with $\mathcal{R}^{(0)}_{n_1,\ldots,n_m} = R_{n_1,\ldots,n_m}$ being the conformally invariant part of the bosonic Wilson loop correlator introduced in eq. (\ref{Wnmfactors}).

We will also use the same notation introduced in Sec. \ref{Sec-MWLcorrs} to denote connected contributions to correlators of multiple super Wilson loop operators, e.g.
\be
\mathcal{W}_{n_1,n_2} = \mathcal{W}_{n_1}\mathcal{W}_{n_2} + \mathcal{W}_{n_,n_2}^{\rm conn}\,, \qquad \mathcal{R}_{n_1,n_2} = \mathcal{R}_{n_1}\mathcal{R}_{n_2} + \mathcal{R}_{n_,n_2}^{\rm conn}\,.
\label{superWLconn}
\ee
Finally, each term in the expansion in Grassmann variables admits a perturbative expansion in $g^2$ which we write as follows,
\begin{align}
&\mathcal{W}^{(k)}_{n_1,\ldots,n_m} = \sum_l g^{2l} \mathcal{W}_{n_1,\ldots,n_m}^{(k,l)}\,, \quad &\mathcal{R}^{(k)}_{n_1,\ldots,n_m} = \sum_l g^{2l} \mathcal{R}_{n_1,\ldots,n_m}^{(k,l)}\,, \notag \\
&\mathcal{W}^{(k),\rm conn}_{n_1,\ldots,n_m} = \sum_l g^{2l} \mathcal{W}^{(k,l),{\rm conn}}_{n_1,\ldots,n_m}\,, &\mathcal{R}^{(k),\rm conn}_{n_1,\ldots,n_m} = \sum_l g^{2l} \mathcal{R}^{(k,l),{\rm conn}}_{n_1,\ldots,n_m}\,.
\label{Wklnotation}
\end{align}

\subsection{Review of the twistor formulation of $\mathcal{N}=4$ SYM}
Let us briefly recall the details of the reformulation of maximally supersymmetric Yang-Mills theory in twistor space. These developments were originally inspired by Witten's twistor string theory \cite{Witten:2003nn}, in which (at the perturbative level) the self-dual sector of \(\mathcal{N}=4\) Super-Yang Mills theory emerges from the open string sector of a topological B-model on supertwistor space \(\mathbb{CP}^{3|4}\). Historically, this gave rise to many important developments such as the MHV formalism for tree amplitudes \cite{Cachazo:2004kj}.

While Witten's original procedure of supplementing the action of a holomorphic Chern-Simons theory with \(D_1\)-instantons to recover the full theory also introduces conformal supergravity \cite{Berkovits:2004jj}, in \cite{Boels:2006ir}, it was shown that (at least at the perturbative level) \(\mathcal{N}=4\) Super-Yang Mills theory may be perfectly captured in twistor space using the twistor action
\be
S(\mathcal{A}) = S_1(\mathcal{A}) + S_2(\mathcal{A}) 
\ee
where \(S_1(\mathcal{A})\) is the action of a holomorphic Chern-Simons theory on \(\mathbb{CP}^{3|4}\), and \(S_2(\mathcal{A})\) provides the interaction terms which allow us to expand about the self-dual sector and recover the full theory. Explicitly, we have (here we choose a convenient normalisation to facilitate a precise match in the prefactors for amplitudes and single Wilson loop correlators in the large $N$ limit\footnote{In principle there remains a freedom of a change in the normalisation of the two terms in the action which becomes unity in the large $N$ limit. We have not written this explicitly here.})
\be
S_1(\mathcal{A}) = \frac{iN}{8\pi^3} \int D^{3|4} \mathcal{Z} \wedge \textrm{tr} \Bigl(\mathcal{A} \wedge \overline{\partial}\mathcal{A} + \frac{2}{3}\mathcal{A} \wedge \mathcal{A} \wedge \mathcal{A}\Bigr),
\ee
 and 
\be
S_2(\mathcal{A}) = \frac{g^2N}{\pi^2} \int d^{4|8} X \log \det\bigl(\overline{\partial} + \mathcal{A} \bigr)_X
\ee
where in \(S_2\), the integral is to be taken over all lines \(X\) in supertwistor space, and \(g^2\) is the t'Hooft coupling, \(\frac{g^2_{\rm YM}N}{16\pi^2}\). Here the $(0,1)$-form connection \(\mathcal{A}\) may be expanded in terms of components as
\begin{align}
\label{superA}
\mathcal{A}(Z,\chi) = a(Z,\overline{Z}) + \chi^{A'} \tilde{\gamma}_{A'}(Z,\overline{Z}) &+ \frac{1}{2!}\chi^{A'}\chi^{B'} \phi_{A'B'}(Z,\overline{Z}) \\
&+ \frac{1}{3!}\epsilon_{A'B'C'D'}\chi^{A'}\chi^{B'}\chi^{C'}
\gamma^{D'}(Z,\overline{Z}) + \frac{1}{4!}\chi^1\chi^2\chi^3\chi^4 g(Z,\overline{Z}). \notag 
\end{align}
The space-time component fields are then constructed from the components of the partial connection via the Penrose transform \cite{Penrose:1985bww,Ward:1990vs}.

Note that in order to obtain the self dual sector from \(S_1(\mathcal{A})\), a gauge choice must be made to reduce the symmetry to the spacetime gauge group, as explained in detail in \cite{Boels:2006ir}. For the purpose of perturbative calculations, the log-det term may be straightforwardly expanded as a power series in \(\mathcal{A}\) to give an infinite tower of interaction terms. Keeping only those terms which saturate the fermionic integration we have
\be
S_2(\mathcal{A}) = -\frac{g^2N}{\pi^2} \int d^{4|8}X \sum_{r=2}^{\infty}\frac{1}{r}  
\tr (-\bar{\partial}_X^{-1} \mathcal{A})^r\,.
\ee
\label{S2}
The operator $\bar{\partial}_X^{-1}$ acts on $(0,1)$-forms on the line $X$ (a copy of $\mathbb{CP}^1$, parametrised here by the complex variable $s'$) as follows,
\be
(\bar{\partial}_X^{-1} \omega)(s) = \int_{X} G(s,s') \wedge \omega(s')\,.
\ee
The Green's function $G$ is given by
\be
G(s,s') = -\frac{1}{2\pi i} \frac{ds'}{(s-s')}\,.
\ee
With these definitions we have
\be
\tr (\bar{\partial}^{-1} \mathcal{A})^r = \tr \biggl\{ \int_{X^r} G(s_r,s_1) \wedge \mathcal{A}(Z(s_1)) \ldots  G(s_{n-1},s_r) \wedge \mathcal{A}(Z(s_r)) \biggr\}\,.
\ee

Under these identifications, after  extracting the space-time field components via the Penrose transform and integrating out an auxiliary field, the action S aligns with the ordinary space-time formulation of \(\mathcal{N}=4\) super Yang-Mills theory at the perturbative level, 
\be
S = \frac{1}{g_{\rm YM}^2} \int d^4x \biggl[ -\frac{1}{2} \tr F^{\mu \nu} F_{\mu \nu}  + \ldots\biggr]\,.
\ee

\subsection{Review of twistor super Wilson loop calculations}
Here we provide a quick review of some of the formalism described in \cite{Mason:2010yk} for the computation of Wilson loop expectation values in twistor space. In twistor space, the contour can be represented by a sequence of intersecting lines $X_i$ (really each a copy of $\mathbb{CP}^1$) with the intersection points of $X_{i-1}$ and $X_i$ given by twistors $Z_i$. We can parametrise the line $X_i$ via
\be
Z_i(s) = s Z_{i-1} + Z_i\,,
\ee
so that $Z_i(0) = Z_i$ and $Z_i(\infty) = Z_{i-1}$.

Given a $(0,1)$-form Chern-Simons connection $a(Z)$ and given a line $X$, we can find a frame $H(x,\lambda)$ such that
\be
H^{-1} (\bar{\partial} + a)|_X H = \bar{\partial}_X\,,
\ee
or equivalently
\be
(\bar{\partial} + a)|_X H = 0\,.
\ee
The frame $H(x,\lambda)$ is unique up to multiplication by a gauge transformation $g(x)$ (independent of $\lambda$),
\be
H(x,\lambda) \rightarrow H(x,\lambda) g(x)\,.
\ee
For the sequence of intersecting lines $X_i$, with some choice of frame $H(x_i,\lambda)$ for each, we define (writing $\bar{\partial}_{X_i}^{-1}$ as $\bar{\partial}_i^{-1}$)
\be
H_i(\lambda) \equiv H(x_i,\lambda_i) H(x_i,\lambda_{i-1})^{-1} = \sum_{l_i=0}^{\infty} \bigl(-\bar{\partial}_i^{-1} a(Z_i(s))\bigr)^{l_i}
\ee
which then obeys the differential equation
\be
(\bar{\partial} + a)_{X_i} H_i(\lambda) = 0\,
\ee
on the line $X_i$ with the boundary condition that $H_i(\lambda_{i-1}) = 1$\,.

The twistor formulation of the loop operator (conventionally path-ordered so that $i$ increases from right to left) is then
\begin{align}
\mathcal{L} (C) =  \frac{1}{N} \tr \mathcal{P} \prod_{i=1}^n H_i(\lambda_i) &= \frac{1}{N} \tr \mathcal{P} \prod_{i=1}^n H(x_i,\lambda_i) H(x_i,\lambda_{i-1})^{-1} \notag \\
&= \frac{1}{N} \tr \mathcal{P} \prod_{i=1}^n \sum_{l_i=0}^{\infty} \bigl(-\bar{\partial}_i^{-1} a(Z_i(0))\bigr)^{l_i}\,.
\end{align}
This has a natural supersymmetric extension
\be
\mathcal{L} (C)=  \frac{1}{N} \tr \mathcal{P} \prod_{i=1}^n \sum_{l_i=0}^{\infty} \bigl(-\bar{\partial}_i^{-1} \mathcal{A}(Z_i(0))\bigr)^{l_i}\,,
\ee
where $\mathcal{A}$ is the supersymmetric extension of $a$ as given in (\ref{superA}).

When we expand we obtain from each edge
\begin{align}
H_i(\lambda_i) = \sum_{l_i=0}^{\infty} \bigl(-\bar{\partial}_i^{-1} \mathcal{A}(Z_i(0))\bigr)^{l_i} = 1 &- \int_{X_i} G(0,s') \wedge \mathcal{A}(Z_i(s')) \notag \\
&+ \int_{X_i} G(0,s') \wedge \mathcal{A}(Z_i(s')) \int_{X_i} G(s',s'') \wedge \mathcal{A}(Z_i(s'')) + \ldots\, \notag \\
= 1 &- \int \frac{1}{2 \pi i} \frac{ds' \wedge \mathcal{A}(Z_i(s'))}{s'} \notag \\
&- \int \frac{1}{2 \pi i} \frac{ds' \wedge \mathcal{A}(Z_i(s'))}{s'} \int \frac{1}{2 \pi i} \frac{ds'' \wedge \mathcal{A}(Z_i(s''))}{(s'-s'')} + \ldots
\end{align}
We now want to consider the expectation value of the loop operator. For now we consider the expectation value in holomorphic Chern-Simons theory without the interaction terms given by $S_2$. We will denote such an expectation value as $\mathcal{W}_n^{\rm CS} = \langle \mathcal{L}(C) \rangle^{\rm CS}$. These contributions to the Wilson loop expectation value correspond to tree-level amplitudes in $\mathcal{N}=4$ super Yang-Mills theory \cite{Mason:2010yk}. We recall that $\mathcal{A} = \mathcal{A}_a t_a$ and make use of the propagator 
\begin{align}
\langle \mathcal{A}_a(Z_i(s)) \mathcal{A}_b(Z_j(t)) \rangle^{\rm CS} &=  -\frac{8 \pi^2}{N}\delta_{ab} \Delta_*\bigl(Z_i(s),Z_j(t)\bigr) \notag \\ 
&= -\frac{8 \pi^2}{N} \delta_{ab} \bar{\delta}^{2|4}\bigl(Z_* , Z_i(s) , Z_j(t)\bigr)\notag \\
&= -\frac{8 \pi^2}{N} \delta_{ab} \int \frac{D^2c}{c_1 c_2 c_3} \bar{\delta}^{4|4}\bigl(c_1 Z_* + c_2 Z_i(s) + c_3 Z_j(t)\bigr)\,.
\end{align}
Here the measure is given by $D^2c = \frac{1}{3}(c_1 dc_2 \wedge dc_3 + \text{ cyc})$. Note that appearance of the \emph{reference twistor} \(Z_*\) is a consequence of our having chosen to work in an axial gauge in order to eliminate the cubic term in \(S_1\). While e.g. individual twistor Wilson loop diagrams can and do depend on the value of \(Z_*\), this should cancel out overall in a well-defined observable and this provides a very useful cross-check when performing computations. 

We then obtain
{\small
\begin{align}
\mathcal{W}_n^{\rm CS} = 1 &+ \sum_{i>j} \int \frac{ds}{s} \frac{dt}{t} \beta_1 \Delta_*^{ij}(s,t) \notag \\
&+ \sum_{i > j > k > l} \int \frac{ds}{s} \frac{dt}{t} \frac{du}{u} \frac{dv}{v} [\beta_2 \Delta_*^{ij}(s,t) \Delta_*^{kl}(u,v) + \beta_3 \Delta_*^{ik}(s,u) \Delta_*^{jl}(t,v) + \beta_2 \Delta_*^{il}(s,v) \Delta_*^{jk}(t,u)]\, \notag \\
&- \sum_{i>k>l} \int \frac{ds_1}{s_1} \frac{ds_2}{(s_1-s_2)}\frac{du}{u} \frac{dv}{v} [\beta_3 \Delta_*^{ik}(s_1,u) \Delta_*^{il}(s_2,v) + \beta_2 \Delta_*^{il}(s_1,v) \Delta_*^{ik}(s_2,u)]\notag \\
&- \sum_{i>j>l} \int \frac{ds}{s} \frac{dt_1}{t_1} \frac{dt_2}{(t_1-t_2)}\frac{dv}{v} [\beta_2 \Delta_*^{ij}(s,t_1) \Delta_*^{jl}(t_2,v) + \beta_3 \Delta_*^{ij}(s,t_2) \Delta_*^{jl}(t_2,v)] \notag \\
&- \sum_{i>j>k} \int \frac{ds}{s} \frac{dt}{t} \frac{du_1}{u_1} \frac{du_2}{(u_1-u_2)}[\beta_3 \Delta_*^{ik}(s,u_1) \Delta_*^{jk}(t,u_2) + \beta_2 \Delta_*^{ik}(s,u_2) \Delta_*^{jk}(t,u_1)] + \ldots
\label{TWLexp}
\end{align} 
}
Here we have factors $\beta_1 = (N^2-\alpha)/N^2$, $\beta_2= (N^2-\alpha)^2/{N^4}$ and $\beta_3 = (N^2 - 2\alpha N^2 + \alpha^2)/{N^4}$, where \(\alpha=0\) in the U($N$) theory and \(\alpha=1\) in the SU($N$) theory. Here, to obtain the \(\beta\)'s we have absorbed into the colour factors the powers of \(2\) which remain after cancelling powers of \((-\frac{1}{2 \pi i})\) from the Green's function with powers of \(-8\pi^2\) and \(\frac{1}{N}\) from the propagators. We also take care to remember the normalisation \(\frac{1}{N}\) of the loop operator.

In the $SU(N)$ theory in the large $N$ limit (in which case only $\beta_2$ survives and $\beta_3$ is suppressed), the boundary terms above are the origin of shifted N${}^2$MHV $R$-invariants which are Grassmann degree 8 Yangian invariants. In the Abelian case however we have $\beta_1=\beta_2=\beta_3 = 1$, and the third, fourth and fifth lines in (\ref{TWLexp}) simplify e.g.
\begin{align}
&-\int \frac{ds_1}{s_1} \frac{ds_2}{(s_1-s_2)}\frac{du}{u} \frac{dv}{v} [\Delta_*^{ik}(s_1,u) \Delta_*^{il}(s_2,v) + \Delta_*^{il}(s_1,v) \Delta_*^{ik}(s_2,u)] \notag \\
=&-\int \biggl[\frac{ds_1}{s_1} \frac{ds_2}{(s_1-s_2)} + \frac{ds_2}{s_2} \frac{ds_1}{(s_2-s_1)}\biggr]\frac{du}{u} \frac{dv}{v} [\Delta_*^{ik}(s_1,u) \Delta_*^{il}(s_2,v)] \notag \\
=&  \int\frac{ds_1}{s_1} \frac{ds_2}{s_2}\frac{du}{u} \frac{dv}{v} \Delta_*^{ik}(s_1,u) \Delta_*^{il}(s_2,v)\,.
\label{cancellingShifts}
\end{align}
Thus in the Abelian case these terms simply factorise and we find
\begin{align}
\mathcal{W}_n^{{\rm CS}} = 1 &+ \sum_{i>j} \int \frac{ds}{s} \frac{dt}{t}  \Delta_*^{ij}(s,t) \notag \\
&+ \sum_{i \geq j \geq k \geq l} \int \frac{ds}{s} \frac{dt}{t} \frac{du}{u} \frac{dv}{v} [\Delta_*^{ij}(s,t) \Delta_*^{kl}(u,v) +  \Delta_*^{ik}(s,u) \Delta_*^{jl}(t,v) + \Delta_*^{il}(s,v) \Delta_*^{jk}(t,u)]\, \notag \\
&+ \ldots\,,
\label{Abfactoring}
\end{align}
where the summation in the second line allows the boundary terms $i=j$ or $j=k$ or $k=l$, but not multiple boundary terms of the form $i=j=k$ or $i=j$ and $k=l$ which vanish. Note that the first line gives a sum over $R$-invariants since we have \cite{Mason:2010yk}
\be
\int \frac{ds}{s} \frac{dt}{t}  \Delta_*^{ij}(s,t) = [*,i-1,i,j-1,j]\,,
\label{NMHVtreeSingleWL}
\ee
with the usual notation for the superconformal invariant ($R$-invariant),
\be
[a,b,c,d,e] = \frac{\delta^{0|4}(\chi_a \langle bcde \rangle + \text{ cyc.})}{\langle abcd \rangle \langle bcde \rangle \langle cdea \rangle \langle deab \rangle \langle eabc \rangle}\,.
\ee
Here we make the usual definition 
\be
\langle ijkl \rangle = \det(Z_iZ_jZ_kZ_l).
\ee
Note also that the square of an $R$-invariant vanishes for Grassmann reasons. In the Abelian theory, we again have exponentiation of the expectation value (without including contributions from the $\log \det$ contribution to the action) of the loop operator,
\be
\mathcal{W}_n^{\rm CS}= {\rm exp}\biggl\{\sum_{i>j} [*,i-1,i,j-1,j]\biggr\}\,.
\label{Abtree}
\ee
The expansion (\ref{Abfactoring}) shows the first three orders in expanding out this exponential. Although it is not manifest term by term, the above expressions are independent of the choice of reference supertwistor (denoted by $*$) once the sum over $i$ and $j$ is performed.

Note that above we have omitted any terms where the propagator joins two adjacent lines. Such terms require a prescription and, depending on the prescription, could lead to divergences \cite{Belitsky:2011zm}. Here we take the prescription that, after evaluating all colour factors, we tilt the lines slightly so that they do not intersect. In other words we introduce a copy $\mathcal{Z}_i'$ of each supertwistor. Then we interpret line $i$ as being the line $(\mathcal{Z}_{i-1}, \mathcal{Z}_i')$. We then evaluate the diagrams before taking the take the limit $\mathcal{Z}_i' \rightarrow \mathcal{Z}_i$ back to the intersecting configuration in a way which respects $Q$-supersymmetry. The resulting R-invariants obtained then simply vanish due to the antisymmetry of their arguments.

Note that an alternative formulation of the chiral supersymmetric Wilson loop operators was presented in \cite{Chicherin:2016ybl} within the Lorentz Harmonic Chiral (LHC) superspace formalism of \cite{Chicherin:2016fac}. This formulation of the Wilson loop is quite close to the twistor formulation we have summarised here and a dictionary between the two formulations is given in \cite{Chicherin:2016ybl}. It was argued in \cite{Chicherin:2016ybl} that the twistor formulation of the Wilson loop omits certain \emph{edge factors} which are required in order to ensure gauge invariance. The diagrams contributing to the expectation value of a single Wilson loop in the LHC superspace formalism then fall into two classes: cusp diagrams, and edge diagrams. Cusp diagrams correspond precisely to the the diagrams which arise in Mason and Skinner's twistor formulation, while edge diagrams are omitted. While edge diagrams with external legs play a vital role for Wilson loop form factors (namely, in the cancellation of spurious poles), in the absence of external legs they simply evaluate to zero after evaluation in Euclidean signature and thus we believe they may safely be omitted from our calculations here.

\subsubsection{Interaction terms}

To expand the full theory around the self-dual sector, one must also include the interaction term \(S_2\). In terms of path integrals we have
\be
\langle \mathcal{L}(C) \rangle = \int [d\mathcal{A}] e^{- (S_1 + S_2)} \mathcal{L}(C)\,.
\ee
To treat the contributions from $S_2$ we expand perturbatively in $g$ and compute the coefficient at each order in the holomorphic Chern-Simons theory with action $S_1$,
\begin{align}
\langle \mathcal{L}(C) \rangle &= \int [d\mathcal{A}] e^{-S_1} \mathcal{L}(C) \biggl[1 + \frac{g^2N}{\pi^2} \int d^{4|8}X \sum_{r=2}^\infty \frac{1}{r} \tr (-\bar{\partial}_X^{-1} \mathcal{A})^r + O(g^4) \biggr] \notag \\
&= \langle \mathcal{L}(C) \rangle^{\rm CS} + \frac{g^2N}{\pi^2} \sum_{r=2}^\infty \frac{1}{r} \int d^{4|8}X \langle \mathcal{L}(C) \tr (-\bar{\partial}_X^{-1} \mathcal{A})^r \rangle^{\rm CS} + O(g^4)\,.
\label{gYMsqterm}
\end{align}

Note that contributions corresponding to diagrams which are related to each other only by a cyclic permutation on the order of the insertions on the Lagrangian line $X$ are clearly identical, and by choosing only one representative from each such class we may drop the factor of \(\frac{1}{r}\) in the integral expressions involving an \(r\)-vertex. 

Let us briefly remark on a simplification of the interaction terms in the Abelian case, namely that we only receive contributions from the two-vertex. This is to be expected since the Abelian theory should be a free theory, as in the case of non-supersymmetric loop operators discussed in Sec. \ref{Sec-MWLcorrs}. To see this, let us consider the relation between two diagrams which differ only by permuting the order of the insertions on the Lagrangian line $X$. For simplicity we will consider one-loop diagrams but the argument generalises in the obvious way.

\begin{figure}
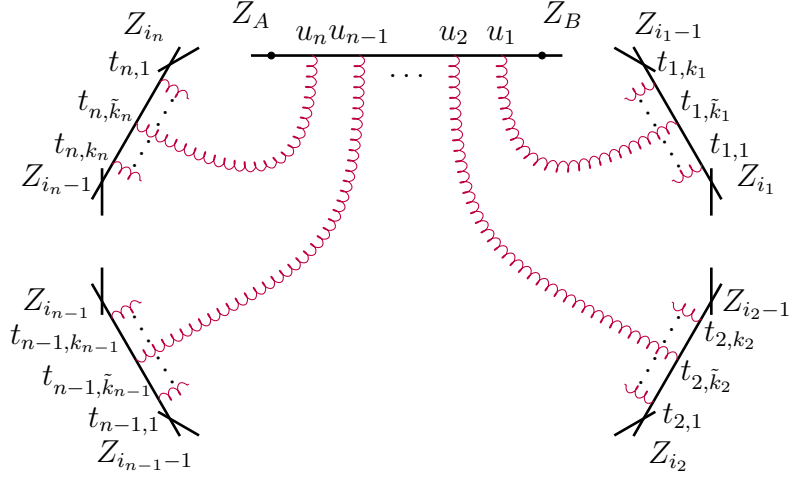

    \begin{center}
      \nvertexdiag
    \caption{A general twistor Wilson loop diagram with $n$ propagators attached to the Lagrangian line $X=(Z_A,Z_B)$; here we label the integration variables associated to the insertion point of the propagators on the twistor lines.}
    \label{nvertexdiag}      
    \end{center}
\end{figure}
Let us label by $w_{n-\textrm{vertex}}$ the general diagram given in Fig. \ref{nvertexdiag}, 
which involves $n$ propagators attached to the Lagrangian line $X$ (labelled by two twistors $Z_A$ and $Z_B$). 

We label the insertion positions on the line $X$ by $u_1,\ldots,u_n$. The $u$-dependent part of the integrand for this diagram is
\begin{align}
    &\textrm{Integrand}(w_{n-\textrm{vertex}}) \notag \\
    &\hspace{3mm}\sim\hspace{3mm}\frac{du_1}{u_1-u_2}\frac{du_2}{u_2-u_3}\hdots \frac{du_{n-1}}{u_{n-1}-u_n}\frac{du_n}{u_n-u_1} \frac{D^2a_{1}}{a_{1,1}a_{1,2}a_{1,3}}\hdots\frac{D^2a_{n}}{a_{n,1}a_{n,2}a_{n,3}}\notag\\
    &\hspace{14mm}\times \overline{\delta}^{4|4}(a_{n,1}Z_*+a_{n,2} Z_A+a_{n,2}u_nZ_B+a_{n,3}Z_{i_n-1}+a_{n,3}t_{n,\tilde{k}_n}Z_{i_n})\notag\\
    &\hspace{14mm}\times \overline{\delta}^{4|4}(a_{n-1,1}Z_*+a_{n-1,2} Z_A+a_{n-1,2}u_{n-1}Z_B+a_{n-1,3}Z_{i_{n-1}-1}+a_{n-1,3}t_{n-1,\tilde{k}_n}Z_{i_{n-1}})\notag\\
    &\hspace{14mm}\times \hdots \times \overline{\delta}^{4|4}(a_{1,1}Z_*+a_{1,2} Z_A+a_{1,2}u_1Z_B+a_{1,3}Z_{i_1-1}+a_{1,3}t_{1,\tilde{k}_1}Z_{i_1})
\end{align}

\begin{figure}
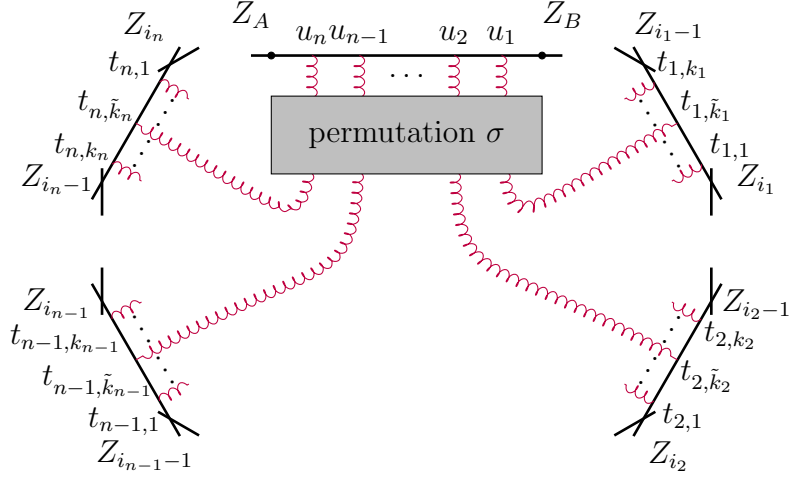

    \begin{center}
        \nvertexdiagwbox
    \end{center}
    \caption{A general twistor diagram with $n$ insertions on the Lagrangian line $X=(Z_A,Z_B)$, which differs from Fig. \ref{nvertexdiag} by some arbitrary permutation of the attachments to the Lagrangian line.}
    \label{nvertexdiagperm}
\end{figure}
Here, the integration measures associated with the propagators not attached to the Lagrangian line are omitted as they have no dependence on the $u$ variables. Now consider the same diagram but with an arbitrary permutation, $\sigma$, of the attachments of the propagators to the Lagrangian line, as given in Fig. \ref{nvertexdiagperm}. We label this as $w_{n-\textrm{vertex}}^\sigma$. Let us relabel the $u$ integration variables such that the propagator attached to $(i_m-1 \hspace{1mm} i_m)$ still attaches to the point on $\mathcal{L}$ associated to $u_m$, as is the case in  Fig. \ref{nvertexdiag}. Each propagator precisely matches those of Fig. \ref{nvertexdiag} but the measure associated to the $u$ variables will differ by a simple factor depending on the $u$'s, as a result of relabelling the integration variables. The rest of the integral will be the same as Fig. \ref{nvertexdiag} because the other propagators and the ordering of propagator attachments is unchanged. The $u$-dependent part of this permuted integral is
\begin{align}
     &\textrm{Integrand}(w_{n-\textrm{vertex}}^\sigma) \notag \\
     &\hspace{3mm}\sim\hspace{3mm}  \frac{du_{\sigma(1)}}{u_{\sigma(1)}-u_{\sigma(2)}}\frac{du_{\sigma(2)}}{u_{\sigma(2)}-u_{\sigma(3)}}\hdots \frac{du_{\sigma(n-1)}}{u_{\sigma(n-1)}-u_{\sigma(n)}}\frac{du_{\sigma(n)}}{u_{\sigma(n)}-u_{\sigma(1)}} \frac{D^2a_{1}}{a_{1,1}a_{1,2}a_{1,3}}\hdots\frac{D^2a_{n}}{a_{n,1}a_{n,2}a_{n,3}}\notag \\
    &\hspace{14mm}\times \overline{\delta}^{4|4}(Z_*+a_{n,2} Z_A+a_{n,2}u_nZ_B+a_{n,3}Z_{i_n-1}+a_{n,3}t_{n,\tilde{k}_n}Z_{i_n})\notag \\
    &\hspace{14mm}\times \overline{\delta}^{4|4}(Z_*+a_{n-1,2} Z_A+a_{n-1,2}u_{n-1}Z_B+a_{n-1,3}Z_{i_{n-1}-1}+a_{n-1,3}t_{n-1,\tilde{k}_n}Z_{i_{n-1}})\notag \\
    &\hspace{14mm}\times \hdots \times \overline{\delta}^{4|4}(Z_*+a_{1,2} Z_A+a_{1,2}u_1Z_B+a_{1,3}Z_{i_1-1}+a_{1,3}t_{1,\tilde{k}_1}Z_{i_1})\,.
\end{align}

In the Abelian theory, since all permutations come with the same colour factor, the full contribution comes from summing over all permutations modulo cyclic permutations\footnote{Cyclic permutations, as described above, merely cancel the factor of $\frac{1}{n}$ accompanying the $n$-vertex.}, which can be simply implemented by using only those permutations which keep \(u_1\) in a fixed position. For $n=2$ there are no non-cyclic permutations and we have a single contribution. For any \(n > 2\), we can see that such contributions cancel in the sum, although the mechanism of cancellation is different for odd versus even \(n\).

After having fixed \(u_1\), it is clear that for permutations which amount to reflecting the order of insertions from \(u_2\) to \(u_n\), the factor pulled out of the integral will reduce to \((-1)^n\). For odd \(n\) we therefore see immediately that the contributions from permutations which are related by reflection will cancel pairwise, and so the contribution for odd \(n\) cancels out to zero.

For even \(n > 2\), the contributions no longer cancel pairwise, but rather cancel out in cyclic classes i.e. the sum over all permutations which are related by cyclic permutations on labels \(2\) to \(n\) will cancel out. This amounts to the simple observation that the expression
\be
\frac{1}{(x_1-x_2)(x_2-x_3)(x_3-x_4) ... (x_{n}-x_1)}
\ee
when summed over cyclic permutations on labels \(2\) to \(n\) will be zero for all $n>2$ (though we needn't invoke this argument for odd \(n\) given the simpler pairwise cancellation observed). Although this is straightforward to algebraically verify for specific cases, it is also easy to see that this vanishes in general by considering residues.
In particular, if we consider the sum to be a function of complex variable \(x_1\), it is clear that the only possible residues are at \(x_i\) for \(i=2\) to \(n\), and that these residues cancel pairwise. For example, for the case \(n=4\) our sum reads
\begin{align}
\frac{1}{(x_1-x_2)(x_2-x_3)(x_3-x_4)(x_4-x_1)} &+ \frac{1}{(x_1-x_3)(x_3-x_4)(x_4-x_2)(x_2-x_1)} \notag \\
&+ \frac{1}{(x_1-x_4)(x_4-x_2)(x_2-x_3)(x_3-x_1)} 
\end{align}
and the poles are at \(x_1=x_2\) (coming from the first and second term), \(x_1=x_3\) (coming from the second and third terms) and \(x_1 = x_4\) (coming from the first and third terms). For each apparent pole, the residue in the two terms giving rise to it are equal except for a sign difference, and so in fact the residue on every pole is zero. Since there is also no pole at infinity, by Liouville's theorem the sum must be a constant with respect to \(x_1\). As \(|x_1| \to \infty\) the sum decays to zero, and thus in fact the sum must be zero as claimed. The pairwise cancellation of residues on each pole, and thus the argument, follows identically for larger \(n\). We therefore see that, as expected, only the two-vertex is able to contribute in the Abelian theory.

\subsection{Tree-level correlators of multiple super loop operators}

Now let us turn to the calculation of correlators of multiple super loop operators within the twistor formalism.
First we consider such correlators where the calculation is carried out in holomorphic Chern-Simons theory (i.e. with $g=0$). For such contributions the divergent and anomalous conformal factors in (\ref{superWLfactors}) are simply given by $D_i=1$ and $F_n=1$. For a single Wilson loop operator, such contributions are equivalent to the tree-level amplitudes of planar $\mathcal{N}=4$ super Yang-Mills theory. We will therefore refer to such contributions as `tree-level' even for correlators of multiple loop operators,
\begin{align}
\langle \mathcal{L}(C_1)\ldots\mathcal{L}(C_m)\rangle^{\rm CS} &= \mathcal{W}^{\rm tree}_{n_1,\ldots,n_m} = \mathcal{W}^{(0,0)}_{n_1,\ldots,n_m} + \mathcal{W}^{(1,0)}_{n_1,\ldots,n_m} + \ldots \notag \\
&= \mathcal{R}^{\rm tree}_{n_1,\ldots,n_m} = \mathcal{R}^{(0,0)}_{n_1,\ldots,n_m} + \mathcal{R}^{(1,0)}_{n_1,\ldots,n_m} + \ldots
\end{align}
Here we use the notation with superscript $(k,l)$ introduced in (\ref{Wklnotation}). 
The first (`MHV') term in the expansion in Grassmann variables is $\mathcal{W}^{(0,0)}_{n_1,\ldots,n_m} = \mathcal{R}^{(0,0)}_{n_1,\ldots,n_m} = 1$, while the second term $\mathcal{W}^{(1,0)}_{n_1,\ldots,n_m} = \mathcal{R}^{(1,0)}_{n_1,\ldots,n_m}$ is of Grassmann degree four and is the `NMHV' contribution, with later terms referred to as N\(^2\)MHV etc. Here, we present some explicit results at low MHV degree. Up to N\(^2\)MHV, all of the integrals which feature have already appeared in the literature, e.g. in \cite{Mason:2010yk}, and so we omit the details of their evaluation here.

Note that if we expand (\ref{superWLconn}) in Grassmann degree we have at tree level,
\begin{align}
\mathcal{W}_{n_1,n_2}^{\rm tree} &= \mathcal{W}_{n_1}^{\rm tree} \mathcal{W}_{n_2}^{\rm tree} + \mathcal{W}_{n_1,n_2}^{\rm tree, conn} \notag \\
&= (\mathcal{W}_{n_1}^{(0,0)} + \mathcal{W}_{n_1}^{(1,0)} + \ldots) (\mathcal{W}_{n_2}^{(0,0)} + \mathcal{W}_{n_2}^{(1,0)} + \ldots) + \mathcal{W}_{n_1,n_2}^{(0,0), {\rm conn}} +\mathcal{W}_{n_1,n_2}^{(1,0), {\rm conn}} + \ldots \notag \\
&= 1 + \mathcal{W}_{n_1,n_2}^{(0,0), {\rm conn}} + \mathcal{W}_{n_1}^{(1,0)} + \mathcal{W}_{n_2}^{(1,0)} + \mathcal{W}_{n_1,n_2}^{(1,0),{\rm conn}} + \ldots
\end{align}
and we conclude that 
\be
\mathcal{W}^{(0,0),{\rm conn}}_{n_1,n_2} = 0\,,
\ee
which is also apparent when considering the possible twistor diagrams.

\subsubsection{NMHV Contribution}

For NMHV tree-level correlators the answer is very simple, as it receives contributions only from a diagram with a single propagator between two distinct, non-adjacent twistor lines. In general we have 
\be
\mathcal{W}^{(1,0)}_{n_1,\ldots,n_m} = \sum_r \mathcal{W}^{(1,0)}_{n_r} +  \sum_{r < s} \mathcal{W}^{(1,0),{\rm conn}}_{n_r,n_s}
\ee
where the disconnected part is known from the literature and quoted in eq. (\ref{NMHVtreeSingleWL}) for the Abelian theory (as in Sec. \ref{Sec-MWLcorrs}, here one sets \(\alpha = 0\) in the case of the U($N$) theory, and \(\alpha = 1\) for SU$(N)$),
\be
\mathcal{W}^{(1,0)}_{n} = \frac{N^2 - \alpha}{N^2}\sum_{i<j} [*,i-1,i,j-1,j]
\ee
and the connected part is given similarly as (note that this colour factor vanishes in the SU($N$) theory)
\be
\mathcal{W}^{(1,0), \rm{conn}}_{n_r,n_s} = \frac{1-\alpha}{N^2}\sum_{i,j} [*,i-1,i,j-1,j];
\ee
where in this second sum, \(i\) runs over the indices of loop \(r\) and \(j\) runs over the indices of loop \(s\). Once again, the above expressions are independent of the choice of reference supertwistor (denoted by $*$) and this provides a useful sanity check on these calculations. 

Note also that (at least formally) the above expression allows non-zero expressions when the loops have three twistors defining them (but not two - due to antisymmetry). For example we could consider the case of two triangles. Let us denote the twistors defining the first loop by $\mathcal{Z}_{1,2,3}$ and those for the second by $\mathcal{Z}_{4,5,6}$. Using independence of the result on $\mathcal{Z}_*$ we can choose $\mathcal{Z}_* = \mathcal{Z}_1$. This leaves only three non-zero terms
\be
\mathcal{W}^{(1,0), \textrm{conn}}_{3,3} = \frac{1-\alpha}{N^2}\Bigl[[1,2,3,4,5] - [1,2,3,4,6] + [1,2,3,5,6]\Bigr] \,.
\ee
Note that symmetry between the two three-point loops is respected due to the basic identity obeyed by the R-invariants,
\be
[1,2,3,4,5]-[1,2,3,4,6]+[1,2,3,5,6]-[1,2,4,5,6]+[1,3,4,5,6]-[2,3,4,5,6]=0\,.
\ee

\subsubsection{N\(^2\)MHV Contribution}
\label{N2MHVtree}

We can also consider the contribution at N\(^2\)MHV, for which we consider diagrams with two propagators each running between a pair of non-adjacent twistor lines. For a correlator of two Wilson loops at N\(^2\)MHV we have
\be
\mathcal{W}_{n_1,n_2}^{(2,0)} = \mathcal{W}_{n_1}^{(2,0)} + \mathcal{W}_{n_2}^{(2,0)} + \mathcal{W}_{n_1}^{(1,0)} \mathcal{W}_{n_2}^{(1,0)} + \mathcal{W}_{n_1,n_2}^{(2,0),{\rm conn}}\,.
\label{Wn1n2decomp}
\ee

For three Wilson loops, we similarly have 
\begin{align}
\mathcal{W}_{n_1,n_2,n_3}^{(2,0)} = \,\,\,\,\, &\mathcal{W}_{n_1}^{(2,0)} + \mathcal{W}_{n_2}^{(2,0)} + \mathcal{W}_{n_3}^{(2,0)} \notag \\
+ &\mathcal{W}_{n_1}^{(1,0)} \mathcal{W}_{n_2}^{(1,0)} + \mathcal{W}_{n_1}^{(1,0)} \mathcal{W}_{n_3}^{(1,0)} + \mathcal{W}_{n_2}^{(1,0)} \mathcal{W}_{n_3}^{(1,0)} \notag \\
+ &\mathcal{W}_{n_1,n_2}^{(2,0),{\rm conn}} + \mathcal{W}_{n_1,n_3}^{(2,0),{\rm conn}} + \mathcal{W}_{n_2,n_3}^{(2,0),{\rm conn}} \notag \\
+ &\mathcal{W}_{n_1}^{(1)}\mathcal{W}_{n_2,n_3}^{(1,0),{\rm conn}} + \mathcal{W}_{n_2}^{(1,0)}\mathcal{W}_{n_1,n_3}^{(1,0),{\rm conn}} + \mathcal{W}_{n_3}^{(1,0)}\mathcal{W}_{n_1,n_2}^{(1,0),{\rm conn}} \notag \\
 + &\mathcal{W}_{n_1,n_2,n_3}^{(2,0),{\rm conn}}\,.
\end{align}
The disconnected contributions are of course known from the literature. Other than those given in the previous section, we have 
\begin{align}
\mathcal{W}_{n}^{(2,0)} = \frac{(N^2-\alpha)^2}{N^4}&\sum_{1 \leq i<j \leq k < l < n+i} [*,i-1,\widehat{i},j-1,j][*,k-1,\widehat{k},l-1,l] \notag \\
+ \, \frac{(1-2\alpha)N^2 + \alpha^2 }{N^4}&\sum_{1 \leq i<k \leq j < l < n+i} [*,i-1,\widehat{i},j-1,j][*,k-1,\widehat{k},l-1,l]
\end{align}

where we define
\be
\widehat{i} = \begin{cases} 
      \widehat{i}_k & i = l  \\
      i & i \neq l 
   \end{cases}
\ee
\be
\widehat{k} = \begin{cases} 
      \widehat{k}_i & k = j   \\ 
      k & k \neq j 
      \end{cases}
\ee

Here we make use of the notation
\be
\widehat{\mathcal{Z}}_{k,j} = (k-1 \, k) \cap (* \, j-1 \, j)
\label{shiftdef}
\ee
which can be written more explicitly as 
\be
\widehat{\mathcal{Z}}_{k,j} = \langle *\,j-1\,j\,k-1\rangle \mathcal{Z}_k - \langle *\,j-1\,j\,k\rangle \mathcal{Z}_{k-1}.
\ee

Here the first sum captures the contribution from planar diagrams and the second sum corresponds to the non-planar diagrams. 

Let us now focus our attention on the two types of connected contribution. Note that all of the integrals which arise here are the same as those appearing in the existing literature for \(\mathcal{W}_n^{(2,0)}\), and so we omit the details of the integration here. 

\subsubsection*{Connected part for two loop operators, \(\mathcal{W}_{n_1,n_2}^{(2,0), {\rm conn}}\)}

At N\(^2\)MHV, the connected parts of the correlator of two Wilson loops receives contributions from diagrams where both propagators cross from one Wilson loop to the other, and also those where one propagator crosses between the Wilson loops while another stays within the same Wilson loop.  
Diagrams in the former category all come with an identical factor (including here some powers of \(2\) from propagators and Green's functions, and the \(\frac{1}{N}\) factors from the propagators and normalisations of the loop operators) 
\be
\frac{4}{N^4}\delta_{ac}\delta_{bd}\textrm{tr}(t_at_b)\textrm{tr}(t_ct_d) = \frac{N^2 - \alpha}{N^4}
\ee
and supply the only contribution in the SU($N$) theory. 

Note that we must include those diagrams which include two propagators ending on the same twistor line (although diagrams where two propagators run between the same pair of twistor lines will vanish for Grassmann reasons). In the case of \(\mathcal{W}_n^{(2)}\), it is precisely such diagrams which give rise to the appearance of shifted twistors inside R-invariants; note that only one of the two orderings of the two insertions on one twistor line would be planar in that case. 

\begin{figure}
\begin{center}
\begin{tikzpicture}[scale=0.9]
   \path[use as bounding box] (-3.5,-2.5) rectangle (22.5,2.5);
  \defineline{lin1m}{originx=0.5, originy=2, angle=120, pointcount=3}
  \defineline{lin2m}{originx=0.5, originy=-1.5, angle=60, pointcount=3}
  \defineline{lin3m}{originx=3.5, originy=0.25, angle=270, pointcount=5}

 \defineline{lin4m}{originx=7.5, originy=2, angle=120, pointcount=3}
  \defineline{lin5m}{originx=7.5, originy=-1.5, angle=60, pointcount=3}
  \defineline{lin6m}{originx=10.5, originy=0.25, angle=270, pointcount=5}

  \drawline[showwings=true, showtwistors=true, fromlabel={\small $Z_{i_1-1}$},tolabel={\small $Z_{i_1}$}]{lin1m}
  \drawline[showwings=true, wingorientation=inner, showtwistors=true, fromlabel={\small $Z_{i_2-1}\,\,\,\,$},tolabel={\small $Z_{i_2}$}]{lin2m}
  \drawline[showwings=true, showtwistors=true, fromlabel={\small $Z_{j-1}$},tolabel={\small $\,\,Z_{j}$}]{lin3m}

 \drawline[showwings=true, showtwistors=true, fromlabel={\small $Z_{i_1-1}$},tolabel={\small $Z_{i_1}$}]{lin4m}
  \drawline[showwings=true, wingorientation=inner, showtwistors=true, fromlabel={\small $Z_{i_2-1}\,\,\,\,$},tolabel={\small $Z_{i_2}$}]{lin5m}
  \drawline[showwings=true, showtwistors=true, fromlabel={\small $Z_{j-1}$},tolabel={\small $\,\,Z_{j}$}]{lin6m}

  \drawpropagator{fromline=lin1m, toline=lin3m, fromip=2, toip=2}
  \drawpropagator{fromline=lin2m, toline=lin3m, fromip=2, toip=4}

    \drawpropagator{fromline=lin4m, toline=lin6m, fromip=2, toip=4}
  \drawpropagator{fromline=lin5m, toline=lin6m, fromip=2, toip=2}
    
\end{tikzpicture}
\end{center}
\caption{Two N\(^2\)MHV tree diagrams where the propagators run between the same twistor lines but with the order of insertion on the lines switched. While one of these diagrams would be colour suppressed in the SU$(N)$ theory for a single Wilson loop, for two Wilson loops both diagrams are planar and can be combined as a single integral, resulting in the disappearance of shifted twistors.}
\label{swappingInsertion}
\end{figure}
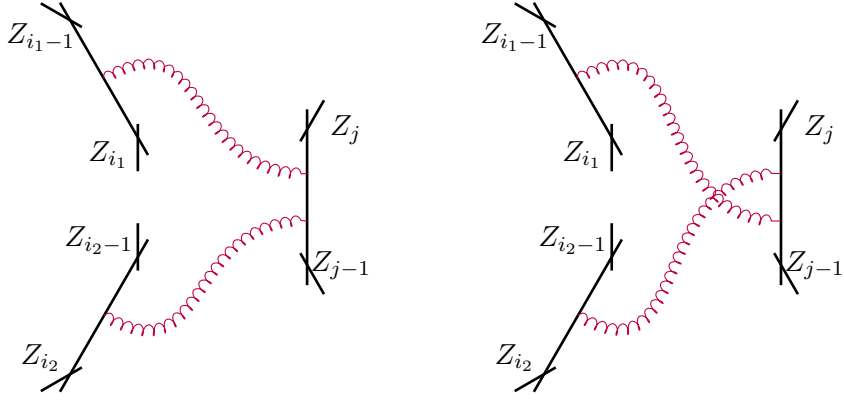

However, in the case of the connected contribution to the correlator of two loop operators, both the versions of a diagram with two field insertions on one twistor line come with an identical (leading) colour factor. Consider for example the diagram where we have propagators from \(X_{i_1}\) and \(X_{i_2}\) on one Wilson loop, both ending on the same line \(X_{j}\) on the other Wilson loop, as shown in Fig. \ref{swappingInsertion}. It is helpful to combine the contributions from these two diagrams into one integral, and in doing so we see that the integral simplifies exactly as we saw for a single Wilson loop in the Abelian case in (\ref{cancellingShifts}), giving an overall contribution of 
\be
\frac{N^2 - \alpha}{N^4}[*, i_1 - 1, i_1, j-1, j][*, i_2-1,i_2, j-1, j].
\ee

In the case where no twistor line has a double field insertion, and the propagators run from \(X_{i_1}\) to \(X_{j_1}\) and \(X_{i_2}\) to \(X_{j_2}\) (the \(i\)-labels being drawn from the first Wilson loop and the \(j\)-labels from the second), we simply have
\be
\frac{N^2-\alpha}{N^4} [*, i_1 - 1, i_1, j_1-1, j_1][*, i_2-1,i_2, j_2-1, j_2]\,.
\ee
Summing over all the contributions from this class of diagram, the disappearance of shifted twistors means we have the simple contribution
\be
\frac{N^2 - \alpha}{N^4}\sum_{i_1<i_2} \sum_{j_1<j_2} [*,i_1-1,i_1,j_1-1,j_1][*,i_2-1,i_2,j_2-1,j_2]\,,
\ee
which simplifies (using the nilpotence of the R-invariants) to
\be
\mathcal{W}_{n_1,n_2, (a)}^{(2,0), \textrm{conn}}=\frac{1}{2N^4} (N^2 - \alpha) \Bigl(\sum_{i,j} [*, i-1, i, j-1,j] \Bigr)^2\,.
\label{N2cont1}
\ee
The sum being squared is of course nothing other than the NMHV contribution in the Abelian theory.

In the U(N) theory, we must also account for diagrams in which one propagator stays within the same Wilson loop, with only one propagator crossing over. These all come with a factor of the form 
\be
\frac{4}{N^4}\delta_{ab} \delta_{cd}\textrm{tr}(t_at_bt_c)\textrm{tr}(t_d) = (1-\alpha)\frac{N^2 - \alpha}{N^4} = (1-\alpha)/N^2 =
\begin{cases}
0\, \quad G=SU(N)\,, \\
1/N^2 \, \quad G=U(N)\,,
\end{cases},
\ee
which of course vanishes entirely in the SU$(N)$ theory ($\alpha=1$). The overall contribution from such diagrams is given by  
\begin{align}
\mathcal{W}_{n_1,n_2, (b)}^{(2,0), \textrm{conn}} =\frac{1-\alpha}{N^2}\sum_{j}^{n_2} \sum_{i_1<i_2 \leq i_3 \leq i_1  + n_1} [*,i_1-1,i_1,i_2-1,i_2] &[*,i_3-1,i_3,j-1,j]
+ \notag \\
&+(n_1 \leftrightarrow n_2) \,.
\label{N2cont2}
\end{align}
Note that shifted twistors cancel out via exactly the same mechanism spelled out for the previous class of diagram. The full expression for \(W_{n_1,n_2}^{(2),\textrm{conn},\textrm{tree}}\) is then given by adding the contributions given in equations (\ref{N2cont1}) and (\ref{N2cont2}),
\be
\mathcal{W}_{n_1,n_2}^{(2,0), \textrm{conn}} = \mathcal{W}_{n_1,n_2, (a)}^{(2,0), \textrm{conn}} + \mathcal{W}_{n_1,n_2, (b)}^{(2,0), \textrm{conn}}\,.
\ee

\subsubsection*{Connected part for three loop operators, \(\mathcal{W}^{(2,0),{\rm conn}}_{n_1,n_2,n_3}\)}

For a correlator of three Wilson loops, the N\(^2\)MHV contribution also includes diagrams where all three Wilson loops are connected, which means there are two propagators ending on one Wilson loop, each running to a different Wilson loop. All such diagrams come with identical factor of the form
\begin{equation}
\frac{4}{N^4}\delta_{ac}\delta_{bd}\textrm{tr}(t_at_b)\textrm{tr}(t_c)\textrm{tr}(t_d) = \frac{1-\alpha}{N^4}
\end{equation}
which of course vanishes in the SU($N$) theory. Shifted twistors from diagrams with two propagators ending on the same twistor line cancel out via exactly the same mechanism as we saw for \(\mathcal{W}^{(2,0),\textrm{conn}}_{n_1,n_2}\) and overall we find
\begin{align}
\mathcal{W}^{(2,0),\textrm{conn}}_{n_1,n_2,n_3} = \frac{1-\alpha}{N^4}\sum_{i_1 \leq i_2}^{n_1} \sum_j^{n_2} \sum_k^{n_3} [*,i_1-1,i_1,j-1,j][*,i_2-1,i_2,k-1,k]
+ {\text{cyc}}(n_1,n_2,n_3)\,,
\end{align}
where \(i\)-, \(j\)-, and \(k\)-labels are drawn from the first, second and third Wilson loops respectively.

\subsubsection{N${}^3$MHV Contribution: \(W^{(3,0),{\rm conn}}_{n_1,n_2}\) in the SU(N) theory}
\label{N3MHVtree}

We may perform the same analysis for the N\(^3\)MHV contribution, which will similarly decompose into various connected and disconnected parts. For simplicity, let us focus our attention on the contribution \(\mathcal{W}^{(3,0),\textrm{conn}}_{n_1,n_2}\) which provides the most non-trivial piece of the answer, although the integrals which we will present suffice for the computation of any N\(^3\)MHV correlator. Moreover, let us restrict our attention to the SU(N) theory, so that diagrams in which one Wilson loop has only a single propagator ending on it will come with vanishing colour factor. 

Subject to these constraints, the diagrams which arise naturally partition into two categories: those for which all three propagators cross from one Wilson loop to the other, which we will refer to as `Class 1', and those for which one propagator stays within the same Wilson loop, which we will refer to as `Class 2'.

\subsubsection*{Colour factors}

Within each class, diagrams can come with one of two colour factors, only one of which contributes to the planar limit. The simple heuristic to determine whether a diagram is planar is to imagine a cylinder for which the Wilson loops are the plane caps; planar diagrams are those for which the propagators can be drawn without crossing \emph{on the cylinder}. 

For Class 1 diagrams, colour dominant diagrams come with a factor of the form (note that these formulae are specifically for the SU($N$) theory)
\be
\frac{8}{N^3}\delta_{ad}\delta_{be}\delta_{cf}\textrm{tr}(t_at_bt_c)\textrm{tr}(t_dt_et_f) = \frac{N^4 - 3N^2 + 2}{N^4}
\ee
and colour subdominant diagrams come with a colour factor of the form
\be
\frac{8}{N^3}\delta_{ae}\delta_{bd}\delta_{cf}\textrm{tr}(t_at_bt_c)\textrm{tr}(t_dt_et_f)  = -\frac{2(N^2-1)}{N^4}.
\ee

Similarly, for Class 2 diagrams we have colour factors of the form
\be
\frac{8}{N^3}\delta_{ad}\delta_{be}\delta_{cf} \textrm{tr}(t_at_bt_ct_d)\textrm{tr}(t_et_f) = \frac{(N^2-1)^2}{N^4}
\ee
for dominant diagrams and 
\be
\frac{8}{N^3}\delta_{ac}\delta_{be}\delta_{df} \textrm{tr}(t_at_bt_ct_d)\textrm{tr}(t_et_f) = -\frac{N^2-1}{N^4}
\ee
for subdominant diagrams. Note that, at leading order in \(N\), dominant diagrams in Class 1 and Class 2 both come with the same colour factor of \(1\). 

\subsubsection*{Integrals for N\(^3\)MHV computations}
While the cases investigated thus far have only required the very simple integrals which had already appeared in e.g. \cite{Mason:2010yk}, for N\(^3\)MHV calculations we now have a number of more complicated integrals which feature. Here we provide a catalogue of their evaluations in terms of R-invariants; note that as the integrals themselves are unaffected by which twistor line is on which Wilson loop (this only affects the trace structure), the same integrals suffice for N\(^3\)MHV calculations even beyond the present case of \(\mathcal{W}^{(3),\textrm{conn}, \textrm{tree}}_{n_1,n_2}\) in the SU(N) theory. 

Here we simply quote the results for each integral; in Appendix B, we provide explicit details on the evaluation of some illustrative examples.

Let us use the notation\footnote{Of course, if we swap the order of the three propagators in the argument, we still refer to the same integral.}
\be
I\bigl(\Delta_*^{ij}(s,t),\Delta_*^{kl}(u,v),\Delta_*^{mn}(w,x)\bigr)
\ee
\label{colourStrippedIntegral}
to denote an integral stripped of its numerical prefactors (including the colour trace and also e.g. powers of \(N\) and \(2\) coming from the propagators) where the superscripts on the \(\Delta_*\)'s label the twistor lines on which the associated propagator ends, and explicitly giving the variables associated to each end of the propagator allows us to distinguish between integrals where the order of the insertions on a twistor line has been switched. In the case that multiple propagators end on the same twistor line, we use the same letter for the associated variable, with a subscript. So for instance an example with a single double insertion would be
\be
I\bigl(\Delta_*^{ij}(s_1,t),\Delta_*^{ik}(s_2,u),\Delta_*^{lm}(v,w)\bigr)
\ee
and an example with a triple insertion would be
\be
I\bigl(\Delta_*^{ij}(s_1,t),\Delta_*^{ik}(s_2,u),\Delta_*^{il}(s_3,v)\bigr).
\ee

Recall that of course the dependence on the integration variables associated to a multi-insertion in the measure is asymmetric; for instance for a double insertion we have
\be
\frac{ds_1}{s_1}\frac{ds_2}{s_2-s_1}
\ee
and for a triple insertion we have
\be
\frac{ds_1}{s_1}\frac{ds_2}{s_2-s_1}\frac{ds_3}{s_3-s_2}.
\ee
This pattern would continue for e.g. a quadruple insertion, but these will not arise for the case we presently consider. 

\subsubsection*{No multi-insertion}
The simplest integral which may arise is the one where there are no multi-insertions on any twistor line. The result for such a case is simply
\be
I\bigl(\Delta_*^{ij}(s,t), \Delta_*^{kl}(u,v), \Delta_*^{mn}(u,v)\bigr) = [*,i-1,i,j-1,j][*,k-1,k,l-1,l][*,m-1,m,n-1,n].
\ee

\subsubsection*{One double-insertion}
\begin{figure}
\begin{center}
\begin{tikzpicture}[scale=0.9]
   \path[use as bounding box] (-3.5,-3.5) rectangle (3.5,1.5);
  \defineline{lin1m}{originx=-3, originy=2, angle=0, pointcount=5}
  \defineline{lin2m}{originx=-1, originy=-0.2, angle=90, pointcount=3}
  \defineline{lin3m}{originx=-3, originy=-2.5, angle=180, pointcount=3}

  \defineline{lin5m}{originx=2, originy=1.6, angle=240, pointcount=3}
  \defineline{lin6m}{originx=2, originy=-2.1, angle=300, pointcount=3}

  \drawline[showwings=true, wingorientation=outer, showtwistors=true, fromlabel={\small $Z_{i}$},tolabel={\small $Z_{i-1}$}]{lin1m}
  \drawline[showwings=true, wingorientation=inner, showtwistors=true, fromlabel={\small $Z_{m-1}$ \hspace{2mm}},tolabel={\small $Z_{m}\hspace{1mm}\,\,\,$}]{lin2m}
  \drawline[showwings=true, wingorientation=outer, showtwistors=true, fromlabel={\small $Z_{j}$},tolabel={\small $Z_{j-1}$}]{lin3m}

  \drawline[showwings=true, wingorientation=inner, showtwistors=true, fromlabel={\small $Z_{k-1}$},tolabel={\small $Z_{k}$}]{lin5m}
  \drawline[showwings=true, showtwistors=true, fromlabel={\small $Z_{l-1}$},tolabel={\small $\,\,Z_{l}$}]{lin6m}

  \drawpropagator{fromline=lin1m, toline=lin3m, fromip=2, toip=2}

  \drawpropagator{fromline=lin1m, toline=lin5m, fromip=4, toip=2,looseval=0.7}

  \drawpropagator{fromline=lin2m, toline=lin6m, fromip=2, toip=2}

\end{tikzpicture}
  \end{center}

\caption{An example of a diagram with a single double-insertion which is relevant for \(\mathcal{W}^{(3), \textrm{conn}}_{n_1,n_2}\) in the SU(N) theory. In our nomenclature, this would be a `Class 2' diagram and moreover would be colour-dominant. As an integral, this corresponds to \(I\bigl(\Delta_*^{ij}(s_2,t), \Delta_*^{ik}(s_1,u), \Delta_*^{lm}(v,w)\bigr)\).}
\label{oneDouble}
\end{figure}
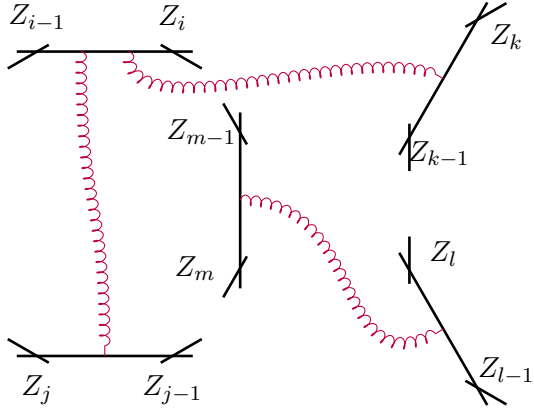

Next, we have the case where two propagators end on one twistor line, with all other twistor lines involved in the integral only having a single propagator ending on them. Such a case is shown in Fig. \ref{oneDouble}; for its evaluation, we have
\be
I\bigl(\Delta_*^{ij}(s_2,t), \Delta_*^{ik}(s_1,u), \Delta_*^{l,m}(v,w)\bigr) = [*,l-1,l,m-1,m][*,i-1,i,k-1,k][*,i-1,i,j-1,\widehat{j}_{k}].
\ee

If we swap the order of the insertions on the line \(X_i\), i.e. take the above integral with \(s_1 \leftrightarrow s_2\), the result is obtained by simply swapping over \(j\) and \(k\) in the given expression. 

\subsubsection*{Two double-insertions}
\begin{figure}
\begin{center}

\begin{tikzpicture}[scale=0.9]
    \path[use as bounding box] (-3.5,-3) rectangle (3.5,2.0);
  \defineline{lin1m}{originx=-3, originy=2.75, angle=120, pointcount=5}
  \defineline{lin2m}{originx=-3, originy=-0.75, angle=60, pointcount=5}
  \defineline{lin3m}{originx=0, originy=1, angle=270, pointcount=3}
  \defineline{lin4m}{originx=2, originy=-1, angle=0, pointcount=3}

  \drawline[showwings=true, showtwistors=true, fromlabel={\small $Z_{i-1}$ \hspace{2mm}},tolabel={\small $Z_{i}\hspace{1mm}$}]{lin1m}
  \drawline[showwings=true, wingorientation=inner, showtwistors=true, fromlabel={\small $Z_{j-1}$ \, },tolabel={\small $Z_{j}$}]{lin2m}
  \drawline[showwings=true, showtwistors=true, fromlabel={\small \hspace{2mm} $Z_{k-1}$},tolabel={\small $\,\,Z_{k}$}]{lin3m}
  \drawline[showwings=true, showtwistors=true, fromlabel={\small $Z_{l-1}$},tolabel={\small $\,\,\,Z_{l}$}]{lin4m}

  \drawpropagator{fromline=lin1m, toline=lin3m, fromip=4, toip=2}

  \drawpropagator{fromline=lin1m, toline=lin2m, fromip=2, toip=4}

  \drawpropagator{fromline=lin4m, toline=lin2m, fromip=2, toip=2,looseval=0.8}

\end{tikzpicture}
\end{center}
\caption{A diagram with two double insertions. In our nomenclature, this would be a colour dominant Class 2 diagram, corresponding to \(I\bigl(\Delta_{*}^{ij}(s_1,t_2), \Delta_{*}^{ik}(s_2,u), \Delta_{*}^{jl}(t_1,v)\bigr)\).}
\label{twoDoub}
\end{figure}
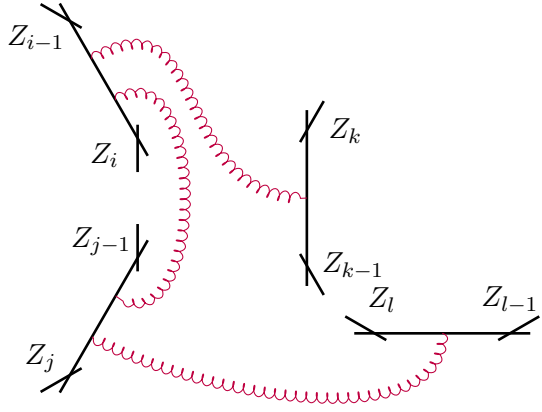

Next, we have the case where there are \emph{two} double-insertions; an example of such a diagram is shown in Fig. \ref{twoDoub}.

For a fixed choice of twistor lines, such an integral comes in four varieties due to the freedom to switch the order of the two pairs of double insertions. The four results which arise are
{\small
\begin{align}
&I\bigl(\Delta_{*}^{ij}(s_1,t_1), \Delta_{*}^{ik}(s_2,u), \Delta_{*}^{jl}(t_2,v)\bigr) = [*,i-1,i,j-1,j][*,i-1,\widehat{i}_{j},k-1,k][*,j-1,\widehat{j}_i,l-1,l] \notag \\
&I\bigl(\Delta_{*}^{ij}(s_2,t_1), \Delta_{*}^{ik}(s_1,u), \Delta_{*}^{jl}(t_2,v)\bigr) = [*,i-1,i,k-1,k][*,j-1,\widehat{j}_{i},l-1,l][*,i-1,\widehat{i}_k,j-1,j] \notag \\
&I\bigl(\Delta_{*}^{ij}(s_1,t_2), \Delta_{*}^{ik}(s_2,u), \Delta_{*}^{jl}(t_1,v)\bigr) = [*,j-1,j,l-1,l][*,i-1,i,j-1,\widehat{j}_l][*,i-1,\widehat{i}_j,k-1,k] \notag \\
&I\bigl(\Delta_{*}^{ij}(s_2,t_2), \Delta_{*}^{ik}(s_1,u), \Delta_{*}^{jl}(t_1,v)\bigr) = [*,j-1,j,l-1,l][*,i-1,i,k-1,k][*,i-1,\widehat{i}_k,j-1,\widehat{j}_l] \notag
\end{align}
}

\subsubsection*{Three double-insertions}
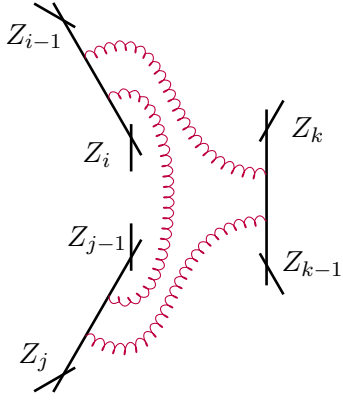
\begin{figure}
\begin{center}

\begin{tikzpicture}[scale=0.9]
     \path[use as bounding box] (-3.5,-2.5) rectangle (3.5,3.0);

  \defineline{lin1m}{originx=-1.5, originy=2, angle=120, pointcount=5}
  \defineline{lin2m}{originx=-1.5, originy=-1.5, angle=60, pointcount=5}
  \defineline{lin3m}{originx=1, originy=0.25, angle=270, pointcount=5}

  \drawline[showwings=true, showtwistors=true, fromlabel={\small $Z_{i-1}$},tolabel={\small $Z_{i}\,\,$}]{lin1m}
  \drawline[showwings=true, wingorientation=inner, showtwistors=true, fromlabel={\small $Z_{j-1}$ \hspace{2mm}},tolabel={\small $Z_{j}$}]{lin2m}
  \drawline[showwings=true, showtwistors=true, fromlabel={\small \hspace{2mm} $Z_{k-1}$},tolabel={\small $\,\,\,Z_{k}$}]{lin3m}

  \drawpropagator{fromline=lin1m, toline=lin3m, fromip=4, toip=2}

  \drawpropagator{fromline=lin1m, toline=lin2m, fromip=2, toip=4,looseval=1.2}

  \drawpropagator{fromline=lin2m, toline=lin3m, fromip=2, toip=4}

\end{tikzpicture}

\end{center}
\caption{A diagram with three double insertions. This would be a colour-dominant Class 2 diagram and would specifically correspond to the integral \(I\bigl(\Delta_{*}^{ij}(s_1,t_2), \Delta_*^{jk}(t_1,u_2), \Delta_*^{ki}(u_1,s_2)\bigr)\).}
\label{threeDoub}
\end{figure}

We can also have the case where there are three twistor lines with a double-insertion; note that this means every twistor line which enters into the integral has a double-insertion on it. Such an example is depicted in Fig. \ref{threeDoub}.

For this case, we really have eight different integrals as a result of the freedom to swap the order within the three pairs of double-insertions. The cases which arise are
{\small
\begin{align}
&I\bigl(\Delta_{*}^{ij}(s_2,t_1), \Delta_*^{jk}(t_2,u_1), \Delta_*^{ki}(u_2,s_1)\bigr) = [*,j-1,\widehat{j}_i,k-1,k][*,k-1,\widehat{k}_j,i-1,i][*,i-1,\widehat{i}_k,j-1,j] \notag \\
&I\bigl(\Delta_{*}^{ij}(s_1,t_1), \Delta_*^{jk}(t_2,u_1), \Delta_*^{ki}(u_2,s_2)\bigr) = [*,j-1,\widehat{j}_i,k-1,k][*,k-1,\widehat{k}_j,i-1,\widehat{i}_j][*,i-1,i,j-1,j] \notag \\
&I\bigl(\Delta_{*}^{ij}(s_2,t_2), \Delta_*^{jk}(t_1,u_1), \Delta_*^{ki}(u_2,s_1)\bigr) = [*,j-1,j,k-1,k][*,k-1,\widehat{k}_j,i-1,i][*,i-1,\widehat{i}_k,j-1,\widehat{j}_k] \notag \\
&I\bigl(\Delta_{*}^{ij}(s_2,t_1), \Delta_*^{jk}(t_2,u_2), \Delta_*^{ki}(u_1,s_1)\bigr) = [*,j-1,\widehat{j}_i,k-1,\widehat{k}_i][*,k-1,k,i-1,i][*,i-1,\widehat{i}_k,j-1,j] \notag \\
&I\bigl(\Delta_{*}^{ij}(s_1,t_2), \Delta_*^{jk}(t_1,u_1), \Delta_*^{ki}(u_2,s_2)\bigr) = [*,j-1,j,k-1,k][*,k-1,\widehat{k}_j,i-1,\widehat{i}_j][*,i-1,i,j-1,\widehat{j}_k] \notag \\
&I\bigl(\Delta_{*}^{ij}(s_1,t_1), \Delta_*^{jk}(t_2,u_2), \Delta_*^{ki}(u_1,s_2)\bigr) = [*,j-1,\widehat{j}_i, k-1,\widehat{k}_i][*,k-1,k,i-1,\widehat{i}_j][*,i-1,i,j-1,j] \notag \\
&I\bigl(\Delta_{*}^{ij}(s_2,t_2), \Delta_*^{jk}(t_1,u_2), \Delta_*^{ki}(u_1,s_1)\bigr) = [*,j-1,j,k-1,\widehat{k}_i][*,k-1,k,i-1,i][*,i-1,\widehat{i}_k,j-1,\widehat{j}_k] \notag \\
&I\bigl(\Delta_{*}^{ij}(s_1,t_2), \Delta_*^{jk}(t_1,u_2), \Delta_*^{ki}(u_1,s_2)\bigr) = [*,j-1,j,k-1,\widehat{k}_i][*,k-1,k,i-1,\widehat{i}_j][*,i-1,i,j-1,\widehat{j}_k]. \notag 
\end{align}
}

\subsubsection*{One triple insertion}

\begin{figure}
\begin{center}
\begin{tikzpicture}[scale=0.9]
  \path[use as bounding box] (-3.5,-3) rectangle (3.5,2.5);

  \defineline{lin1m}{originx=-4, originy=0, angle=90, pointcount=5}
  \defineline{lin4m}{originx=2, originy=2, angle=180, pointcount=3}
  \defineline{lin5m}{originx=0, originy=0, angle=270, pointcount=3}
  \defineline{lin6m}{originx=2, originy=-2, angle=0, pointcount=3}
  
  \drawline[showwings=true, showtwistors=true, fromlabel={\small $Z_{i-1}$},tolabel={\small $Z_{i}\,$}]{lin1m}
  \drawline[showwings=true, showtwistors=true, fromlabel={\small $Z_{j-1}$},tolabel={\small $Z_{j}$}]{lin4m}
  \drawline[showwings=true, wingorientation=inner, showtwistors=true, fromlabel={\small \,  $Z_{k-1}$},tolabel={\small $\,\,\,Z_{k}$}]{lin5m}
  \drawline[showwings=true, showtwistors=true, fromlabel={\small $Z_{l-1}$},tolabel={\small $\,\,Z_{l}$}]{lin6m}

 \drawpropagator{fromline=lin1m, toline=lin6m, fromip=2, toip=2,looseval=0.8}
 \drawpropagator{fromline=lin1m, toline=lin5m, fromip=3, toip=2}
 \drawpropagator{fromline=lin1m, toline=lin4m, fromip=4, toip=2,looseval=0.8}
\end{tikzpicture}
\end{center}

\caption{A diagram with a triple-insertion on the twistor line \(X_i\). In our language, this would be a colour-dominant Class 1 diagram corresponding to the integral \(I\bigl(\Delta_*^{ij}(s_3,t),\Delta_*^{ik}(s_2,u),\Delta_*^{il}(s_1,v)\bigr)\). }
\label{tripleInsertion}
\end{figure}
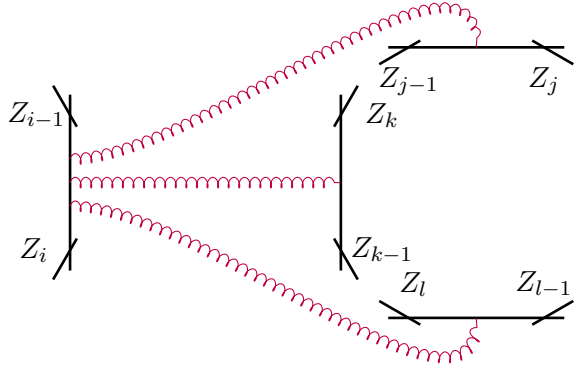

Finally, we can have the case with a single triple insertion, with each propagator on the line with a triple insertion running to a distinct twistor line. An example is depicted in Fig. \ref{tripleInsertion}. For such diagrams we have
\be
I\bigl(\Delta_*^{ij}(s_3,t),\Delta_*^{ik}(s_2,u),\Delta_*^{il}(s_1,v)\bigr) = [*,i-1,i,l-1,l][*,\widehat{i}_{l}, k-1,k][*,i-1,\widehat{i}_k,j-1,j]
\ee
Cases which amount to permuting the orders of the insertions, i.e. permuting \(\{s_1,s_2,s_3\}\), are generated by subjecting \(\{j,k,l\}\) to the same permutation that has been applied to \(\{s_1,s_2,s_3\}\). 

\subsubsection*{Generating the correlator}
\begin{figure}
\centering
\begin{tabular}{ccccccc} 
    \includegraphics[scale=0.17]{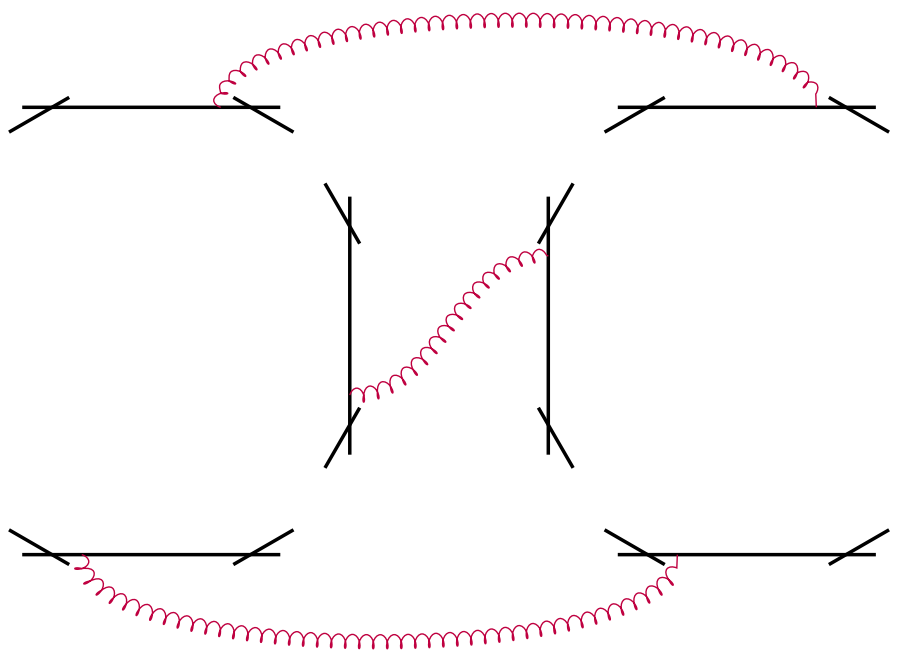} & \includegraphics[scale=0.17]{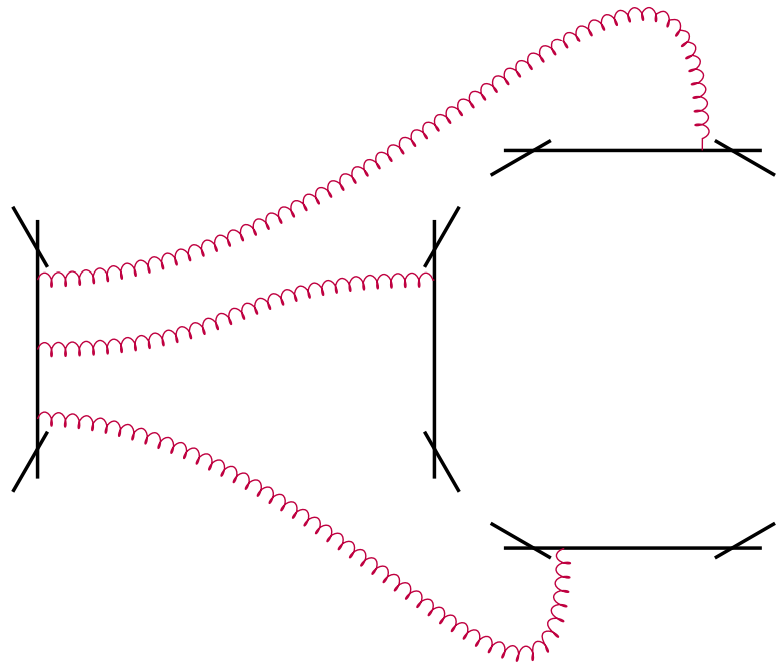} & 
    \includegraphics[scale=0.17]{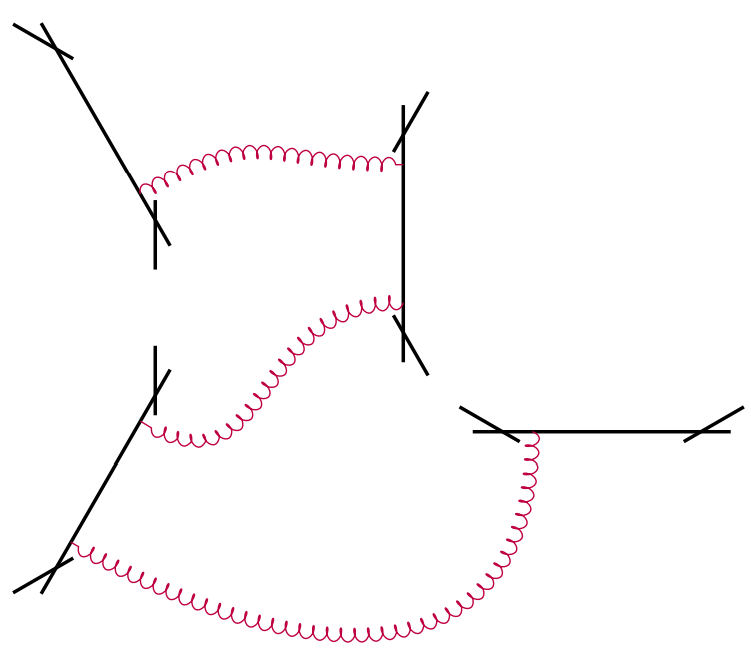} & 
        \includegraphics[scale=0.17]{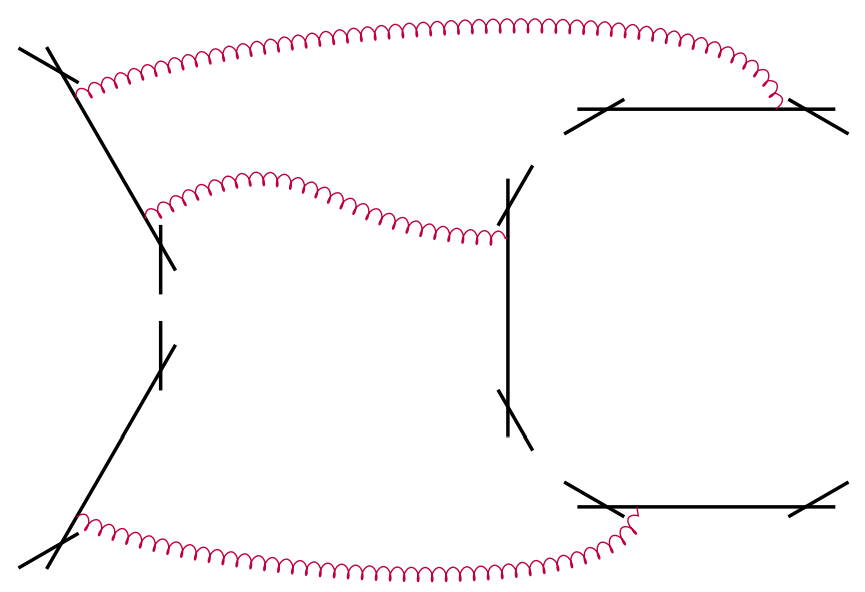} & 
     \includegraphics[scale=0.17]{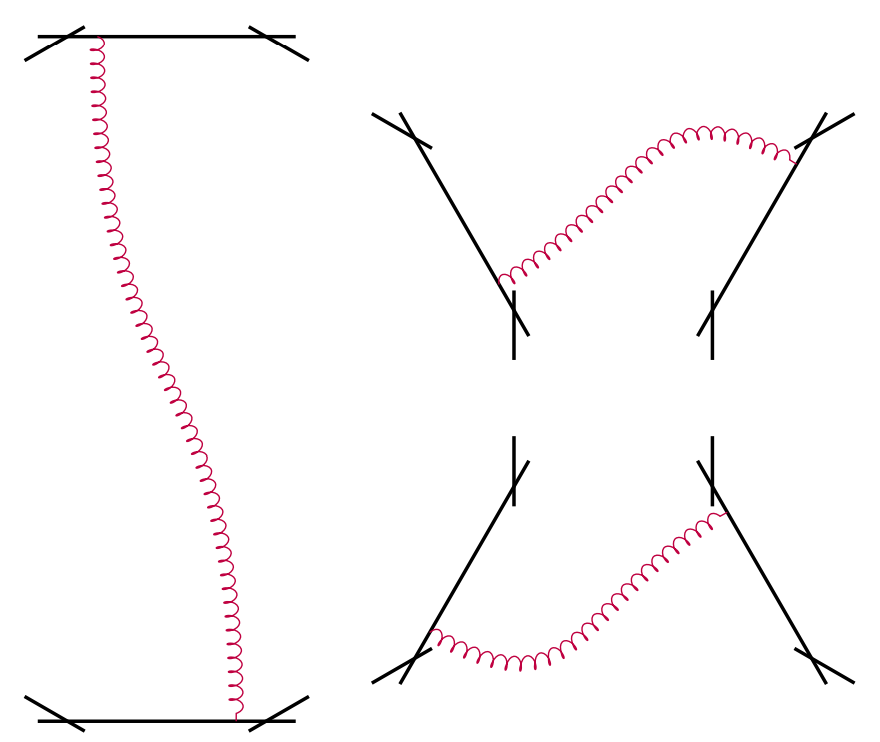} & \includegraphics[scale=0.17]{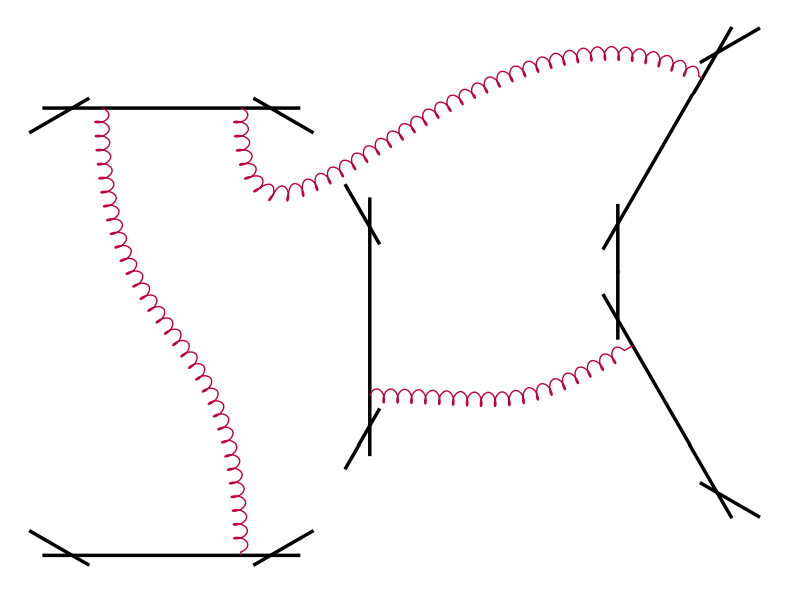} & \\ 
    \includegraphics[scale=0.17]{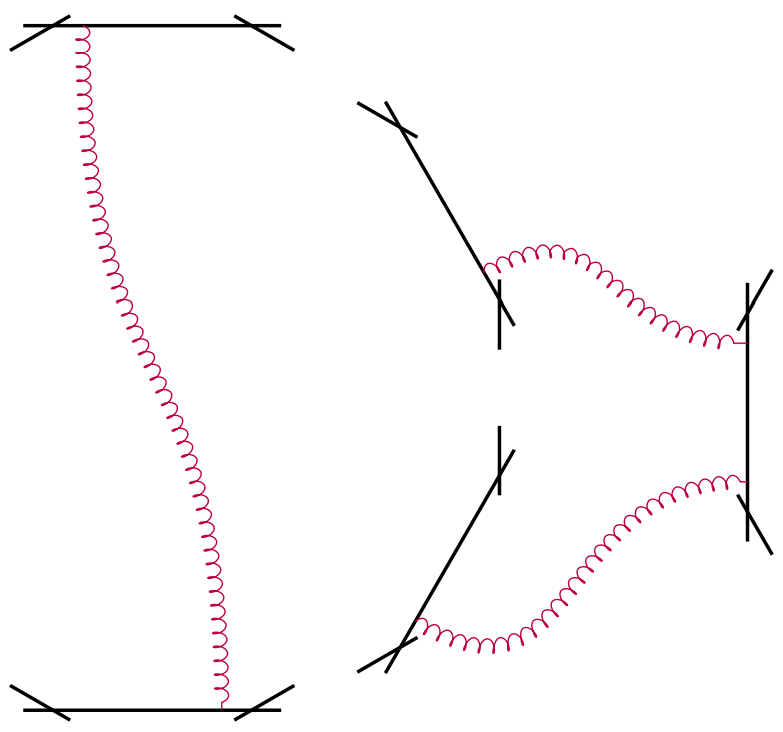} &
    \includegraphics[scale=0.17]{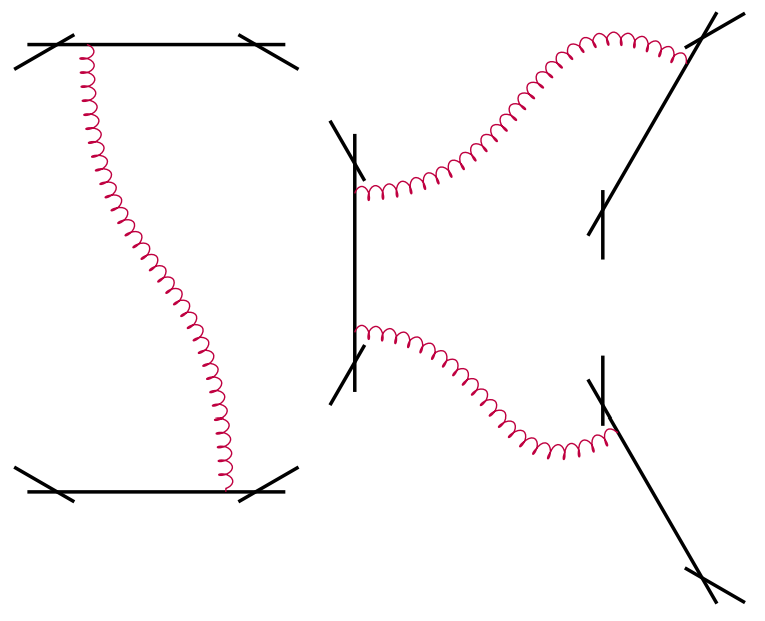} & 
\includegraphics[scale=0.15]{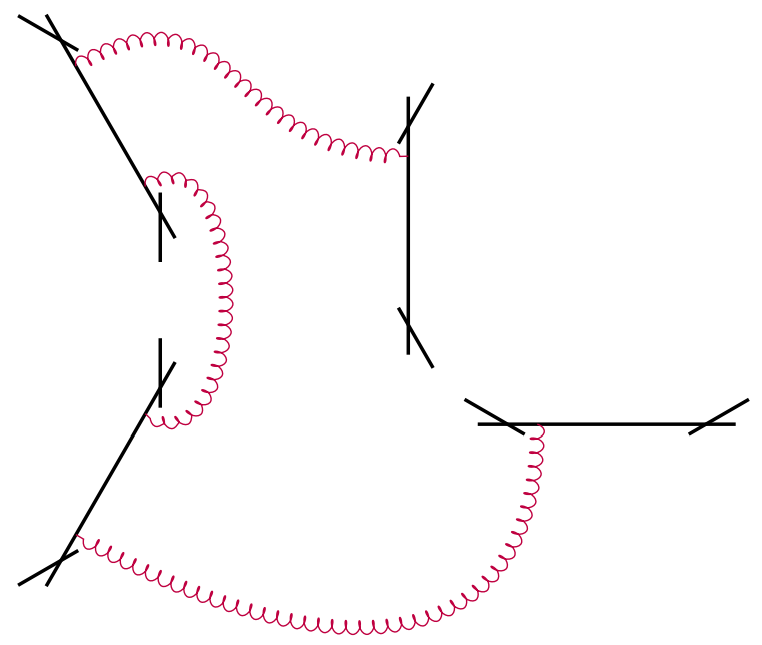} & 
\includegraphics[scale=0.17]{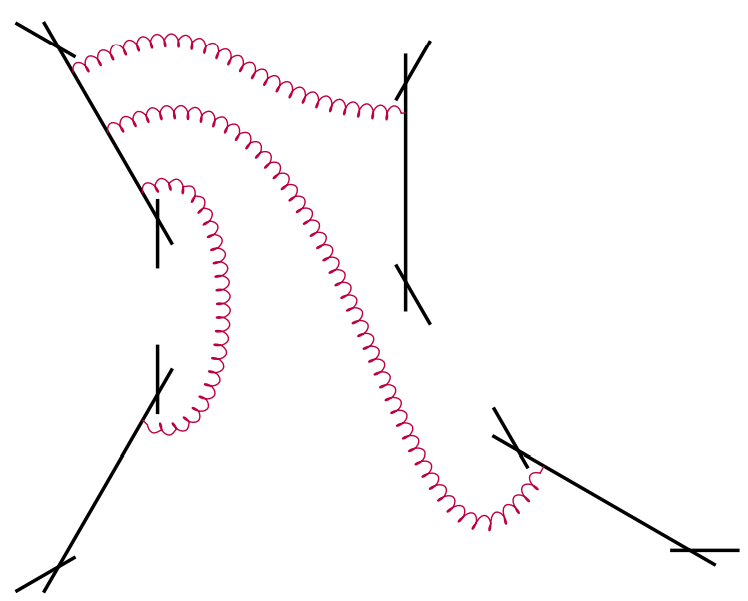} & 
    \includegraphics[scale=0.17]{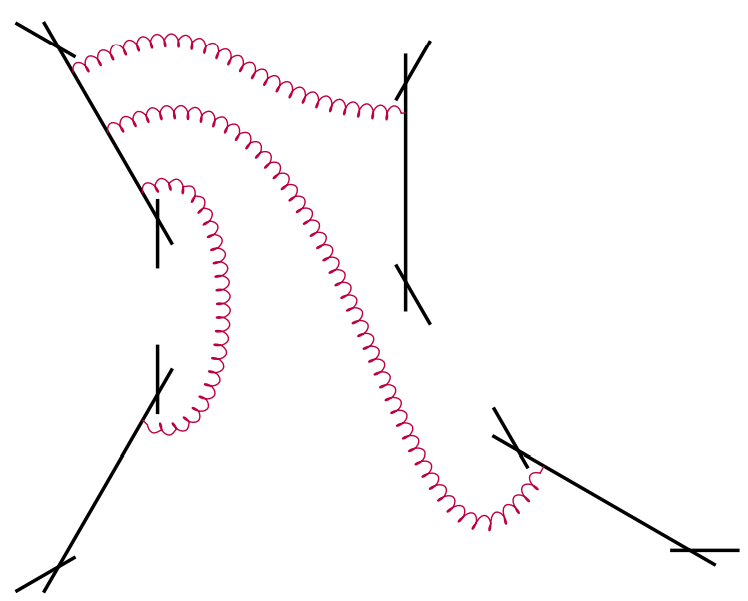} &
    \includegraphics[scale=0.1]{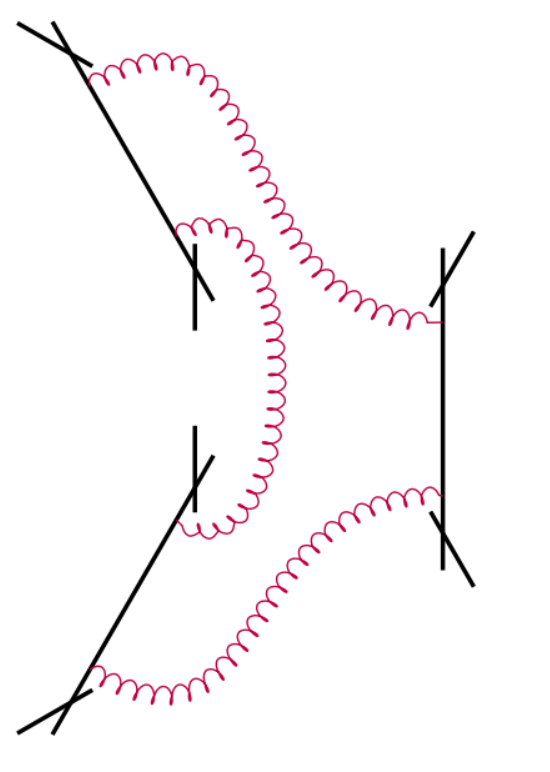} &
\end{tabular}
\caption{All integral topologies which enter into the N\(^3\)MHV, connected contribution to the tree-level correlator of two Wilson loops in the SU(N) theory. For each class of diagram, we must account for all ways of choosing which twistor lines to involve in each Wilson loop, and moreover all possibilities for re-ordering the multiple insertions of the propagators on any twistor lines with a double or triple insertion.}
\label{allDiags}
\end{figure}

All of the integrals which arise for N\(^3\)MHV correlators are of one of the forms given above. To generate \(\mathcal{W}^{(3,0), \textrm{conn}}_{n_1,n_2}\), one simply generates all diagrams (which amounts to all the possible choices of three distinct propagators between non-adjacent lines, such that at least two cross from one Wilson loop to the other, and accounting for the different possible orderings of propagators in cases with multiple insertions on the same twistor line), dresses them with the appropriate powers of \(2\) and \(N\) from the propagators and also the colour factor as spelled out in the above section, and evaluates them using the above formulae. In practice this is a very straightforward process to automate. In the ancillary files, we provide a Mathematica-readable expression for \(\mathcal{W}^{(3,0),\textrm{conn}}_{n_1,n_2}\) for all \(4 \leq n_1 \leq n_2 \leq 6\). All of the integral topologies which arise for \(\mathcal{W}^{(3,0), \textrm{conn}}_{n_1,n_2}\) are shown in Fig. \ref{allDiags}. 

\subsection{Feynman rules in R-invariant form}
\label{Feynman}
While we have enumerated all of the integrals which arise in the case of tree-level N\(^3\)MHV computations, it is helpful to automate the evaluation of these integrals for the purpose of higher MHV tree-level computations, and also the computation of loop integrands. Wilson loop diagrams may be straightforwardly evaluated in terms of products of R-invariants by relating to the dual MHV diagram, using a procedure spelled out in \cite{Mason:2010yk}. In particular, it was identified in that reference that the momentum twistor space MHV diagrams of an n-gluon scattering amplitude in planar $\mathcal{N}=4$ SYM are in one-to-one correspondence with the twistor diagrams of an n-sided light-like Wilson loop correlator. The rule for a propagator in a Wilson loop diagram is the same as the rule for the equivalent propagator in the corresponding MHV diagram of the dual amplitude. For an MHV diagram and Wilson loop (colour-stripped) Feynman diagram with a single propagator, we have the following correspondence \cite{Bullimore:2013jma},
\begin{center}
\begin{tikzpicture}[scale=0.9]
   \filldraw[black] (2,1.75) circle (1.5pt) node[anchor=west]{$x_i$};
   \filldraw[black] (2,0.25) circle (1.5pt) node[anchor=west]{$x_j$};
   \filldraw[black] (0,2) node[anchor=south]{$i-1$};
   \filldraw[black] (4,2) node[anchor=south]{$i$};
   \filldraw[black] (0,0) node[anchor=north]{$j$};
   \filldraw[black] (4,0) node[anchor=north]{$j-1$};
   \filldraw[black] (4,1) node[anchor=west]{\hspace{9mm}$=$\hspace{52mm}$=$\hspace{8mm}$[*,i-1,i,j-1,j]\,.$};
   
   \filldraw[black] (-0.1,1) circle (0.75pt);
   \filldraw[black] (0,1.3) circle (0.75pt);
   \filldraw[black] (0,0.7) circle (0.75pt);
   
   \filldraw[black] (4.1,1) circle (0.75pt);
   \filldraw[black] (4,1.3) circle (0.75pt);
   \filldraw[black] (4,0.7) circle (0.75pt);

   \filldraw[black] (8-1.7,1) circle (0.75pt);
   \filldraw[black] (8-1.6,1.3) circle (0.75pt);
   \filldraw[black] (8-1.6,0.7) circle (0.75pt);

   \filldraw[black] (8+1.7,1) circle (0.75pt);
   \filldraw[black] (8+1.6,1.3) circle (0.75pt);
   \filldraw[black] (8+1.6,0.7) circle (0.75pt);
   
   \draw[line width=1] (0,0+2) -- (1,-1+2);
   \draw[line width=1] (1,-1+2) -- (0,-2+2);
   \draw[line width=1] (1,-1+2) -- (3,-1+2);
   \draw[line width=1] (3,-1+2) -- (4,0+2);
   \draw[line width=1] (3,-1+2) -- (4,-2+2);
   
  \defineline{lin1}{originx=8, originy=1.5+1, angle=0, pointcount=3}
  \defineline{lin2}{originx=8, originy=-1.5+1, angle=180, pointcount=3}

  \drawline[showwings=true,showtwistors=true,wingorientation=outer,fromlabel=$Z_{i}$,tolabel=$Z_{i-1}$]{lin1}
  \drawline[showwings=true,wingorientation=outer,showtwistors=true, fromlabel=$Z_{j}$,tolabel=$Z_{j-1}$]{lin2}
  
  \drawpropagator{fromline=lin1, toline=lin2, fromip=2, toip=2}
\end{tikzpicture}
\end{center}
Here $Z_{i}$ is the momentum twistor associated to the $i$-th external particle with momentum $p_i$, and, along with the $(i-1)$-th external particle and the propagator, forms the external region $x_i$. For loop diagrams there will also be internal regions formed by internal propagators (see section 5.3 of \cite{Bullimore:2013jma} for more examples of dual diagrams). The rule for the general propagator separating regions $x_i$ and $x_k$ in the following MHV diagram \cite{Bullimore:2013jma} is
\begin{equation}
    \generalMHVrule\hspace{10mm} = \hspace{12.5mm}[*,i-1,\widehat{i}_j,k-1,\widehat{k}_l]\,.  
\end{equation}
where $\widehat{i}_j$ is as defined in \ref{shiftdef}. $x_i$, $x_j$, $x_k$, $x_l$ may be internal regions (loop regions) or external regions. If $x_i$ does not have an adjacent region in the clockwise direction, then $\widehat{i}_j$ is replaced by $i$ and if $x_k$ does not have an adjacent region in the clockwise direction, then $\widehat{k}_l$ is replaced by $k$. $[*,i,j,k,l]$ is antisymmetric in the twistors, so $[*,i-1,\widehat{i}_j,k-1,\widehat{k}_l]$ is invariant under a reordering of the pairs $\{i-1,\widehat{i}_j\}$ and $\{k-1,\widehat{k}_l\}$.\\

Using the amplitude-Wilson loop duality (detailed in \cite{Adamo:2011pv,Bullimore:2013jma}), this rule is equivalent to that of the general propagator connecting the lines $(i-1 \hspace{0.75mm} i)$ and $(k-1 \hspace{0.75mm} k)$ in the following Wilson loop diagram,
\begin{equation}
    \GeneralPropRule\hspace{5mm}=\hspace{5mm}[*,i-1,\widehat{i}_j,k-1,\widehat{k}_l]
    \label{Wilsonlooprule}
\end{equation}

If $(i-1 \hspace{0.75mm} i)$ is a line on a Wilson loop and there is no other propagator in the direction of $i$ on the line $(i-1 \hspace{0.75mm} i)$, then $\widehat{i}_j$ is replaced by $i$ and similarly, if $(k-1 \hspace{0.75mm} k)$ is a line on a Wilson loop and there is no other propagator in the direction of $k$ on the line $(k-1 \hspace{0.75mm} k)$, then $\widehat{k}_l$ is replaced by $k$. If $(i-1 \hspace{0.75mm} i)$ is a loop line (Lagrangian line), then there is cyclic symmetry in the ordering of propagators on that line, so if there is no other propagator in the direction of $i$ on the line $(i-1 \hspace{0.75mm} i)$, then $j$ in $\widehat{i}_j$ is the twistor in $(j-1 \hspace{0.75mm} j)$ which the next propagator in the cyclic ordering of propagators attached to $(i-1 \hspace{0.75mm}i )$ goes to.\\

Although the rule above originates from the amplitude-Wilson loop duality, which holds for the expectation value of a single Wilson loop, the expression in (\ref{Wilsonlooprule}) does not depend on the Wilson loop which the lines are part of. Thus, this rule can be used to obtain the kinematic part of diagrams appearing in our more general Wilson loop correlators involving multiple Wilson loops.\\

As an illustrative example, consider the following diagram appearing in the one-loop, N$^2$MHV contribution to a pentagon-square correlator. By following the rules we have spelled out, its evaluation is given by\\\\
\begin{equation}
    \begin{split}
    &\hspace{-2mm}\PentSquareOneloopDiagram\\\\
        &=[*,5,1,3,4][*,3,\widehat{4}_1,A,\widehat{B}_9][*,3,\widehat{4}_B,9,6][*,A,\widehat{B}_4,8,9]
    \end{split}
\end{equation}\\\\
where $(AB)$ is the Lagrangian line and so $B-1=A$.\\

In contrast to the computations presented in Appendix \ref{N3MHVintegrals}, the expression for each diagram obtained using (\ref{Wilsonlooprule}) does not require one to manually integrate out the bosonic parts of the delta functions in order to arrive at the simple, R-invariant form for the diagram. Using (\ref{Wilsonlooprule}), any Wilson loop correlator loop integrand/tree level expression (any number of Wilson loops, number of sides, loop order, MHV degree) can be expressed in terms of a sum over products of R-invariants (dressed with the appropriate colour factors).

\section{Super Wilson loop correlators at order $g^2$}

\subsection{Review for a single loop operator}
We can also apply the twistor super Wilson loop formalism to the computation of $g^2$ corrections which, in the case of a single Wilson loop, can be related to MHV diagrams \cite{Bullimore:2010pj}. If we now include the interactions from the second term in the action, including the MHV vertices we have contribution to a single Wilson loop given by the sum over terms of the form 
\begin{align}
D_{ij} = &\frac{4}{N^3}\tr(t_{a_1}t_{a_2})\tr(t_{b_1}t_{b_2}) \delta_{a_1 b_1} \delta_{a_2 b_2} \frac{g^2N}{\pi^2} \times \notag \\
&\int d^{4|8}x_{AB} \int \frac{ds}{s} \frac{dt}{t} \frac{du_1 du_2}{(u_1-u_2)(u_2-u_1)} \Delta_*^{ix}(s,u_1) \Delta_*^{jx}(t,u_2).
\label{Dij}
\end{align}

\begin{figure}
\begin{center}
\begin{tikzpicture}[scale=0.9]
   \path[use as bounding box] (-3.5,-2.5) rectangle (22.5,2.5);
  \defineline{lin1m}{originx=4.5, originy=2, angle=120, pointcount=3}
  \defineline{lin2m}{originx=4.5, originy=-1.5, angle=60, pointcount=3}
  \defineline{lin3m}{originx=7.5, originy=0.25, angle=270, pointcount=2}

  \drawline[showwings=true, showtwistors=true, fromlabel={\small $Z_{i-1}$},tolabel={\small $Z_{i}$}]{lin1m}
  \drawline[showwings=true, wingorientation=inner, showtwistors=true, fromlabel={\small $Z_{j-1}\,\,\,\,$},tolabel={\small $Z_{j}$}]{lin2m}
  \drawline[showwings=false, showtwistors=true, tolabel={\small $\hspace{2mm}(AB)$}]{lin3m}

  \drawpropagator{fromline=lin1m, toline=lin3m, fromip=2, toip=1}
  \drawpropagator{fromline=lin2m, toline=lin3m, fromip=2, toip=2}

\end{tikzpicture}
\end{center}
\caption{The single type of twistor diagram which contributes to the one-loop, MHV contribution to a single Wilson loop. Here, \((AB)\) is the Lagrangian line.}
\label{kermitDiagram}
\end{figure}
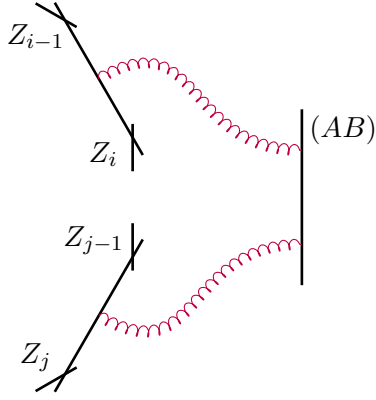

Diagrammatically, these contributions come from those diagrams where two propagators run from the Lagrangian to two distinct lines of the Wilson loop, as shown in Fig. \ref{kermitDiagram}. Note that an \(x\) in the superscript of a \(\Delta_*\) denotes that the correpsonding end of that propagator is on the Lagrangian line. Note that we omit the factor of \(\frac{1}{2}\) which appears in the expansion of \(S_2\) (\ref{S2}) because we have two identical diagrams corresponding to interchanging \(u_1 \leftrightarrow u_2\), which diagrammatically means cycling through the order of field insertions on the Lagrangian line $X_{AB}$. In performing diagrammatic computations, we will simply omit the \(\frac{1}{n}\) factors but identify diagrams differing by such cyclic permutations as indistinct. 

Simplifying, the contribution $D_{ij}$ in (\ref{Dij}) reduces to \cite{Mason:2010yk}
\begin{align}
D_{ij} &= g^2 \frac{(N^2-\alpha)}{N^2}    \int \frac{d^{4|8}x_{AB}}{\pi^2}  \int \frac{ds}{s} \frac{dt}{t} \frac{du_1 du_2}{(u_1-u_2)(u_2-u_1)}\Delta_*^{ix}(s,u_1) \Delta_*^{jx}(t,u_2) \notag \\
&= g^2 \frac{(N^2-\alpha)}{N^2} K_{ij}\,,
\end{align}
where we define the `Kermit' diagram 
\begin{align}
K_{ij} = &\int \frac{d^{4|8}x_{AB}}{\pi^2} \int \frac{ds}{s} \frac{dt}{t} \frac{du_1 du_2}{(u_1-u_2)(u_2-u_1)}  \Delta_*^{ix}(s,u_1) \Delta_*^{jx}(t,u_2) \notag \\
= &\int \frac{d^{4|8}x_{AB}}{\pi^2} [*,i-1,i,A,B'][*,j-1,j,A,B''].
\label{Kermit}
\end{align}
This expression can be derived either by integrating out the support of the bosonic delta functions, or using the `Feynman rules in R-invariant form' outlined in Section \ref{Feynman}. 

In the expression (\ref{Kermit}) we have
\begin{align}
\mathcal{Z}^{\prime}_{B} &= \mathcal{Z}_A \langle B\, * \, j-1\,j \rangle - \mathcal{Z}_B \langle A\, *\, j-1\, j\rangle\,, \notag \\
 \mathcal{Z}^{\prime\prime}_{B} &= \mathcal{Z}_A \langle B\, * \, i-1\,i \rangle - \mathcal{Z}_B \langle A\, *\, i-1\, i\rangle\,,
\end{align}
We can write each contribution after performing the fermionic integration as 
\be
K_{ij} = -\int \frac{d^4 x_{AB}}{\pi^2} \frac{(\langle *\, i-1\, i \,[A\rangle \langle B] \,j-1 \, j\, * \rangle)^2}{\langle A\,B\, i-1\, i\rangle \langle A\,B\, j-1\, j\rangle \langle A\, B\, i-1 \,* \rangle \langle A\, B\, i \,* \rangle \langle A\, B\, j-1 \,* \rangle \langle A\, B\, j \,* \rangle}\,.
\ee

Each contribution $K_{ij}$ exhibits spurious poles involving the reference twistor $Z_*$ but these cancel in the sum over $i,j$. One way to make this explicit is to rewrite the full expression at the level of the integrand (i.e. before performing any integration over $x_{AB}$) in terms of chiral pentagon integrals \cite{Arkani-Hamed:2010pyv},
\be
\mathcal{W}_n^{(0,1)} = \frac{N^2-\alpha}{N^2} \sum_{i<j} K_{ij}  = -\frac{N^2-\alpha}{N^2} \sum_{i<j} P^X_{ij}\,.
\label{superW01}
\ee
The chiral pentagon integral $P^X_{ij}$ is given by
\be
P^X_{ij} = \int \frac{d^4 x_{AB}}{\pi^2} \frac{\langle A\, B\, \bar{i}\, \bar{j} \rangle \langle i\, j\, X \rangle}{\langle i-1\, i\, A\, B \rangle \langle i\, i+1\, A\, B \rangle \langle j-1\, j\, A\, B \rangle \langle j\, j+1\, A\, B \rangle \langle A\, B\, X\rangle}\,. 
\ee
Here $X$ is an auxiliary spacetime point (or twistor line). The equivalence of the Kermit and pentagon expressions shows that the final answer depends on neither the reference twistor $Z_*$ nor the auxiliary point $X$.

The expression (\ref{superW01}) is formal in the sense that the remaining integral over $x_{AB}$ is divergent. In the case of the pentagon expression this can be seen to be the case when $i=j-1$. Such divergences are expected since this result of the twistor Wilson loop calculation should reproduce the result at order $g^2$ presented in Sec. \ref{Sec-MWLcorrs} and described in detail in Appendix \ref{App-lightlikeloop} for the bosonic (non-supersymmetric) Wilson loop operators.

In the Abelian case exponentiation still occurs, and the tree-level NMHV and order $g^2$ MHV results can be combined to give the full Abelian result for the super Wilson loop,
\be
\langle \mathcal{L}(C) \rangle = \mathcal{W}_n = {\rm exp}\biggl\{\sum_{i<j} \bigl[g^2K_{ij} + [*,i-1,i,j-1,j]\bigr]\biggr\}\,, \qquad G=U(1)\,.
\label{U1expsuperWL}
\ee
We can identify the Kermit contribution with the regularised contribution from the spacetime formulation of the Abelian Wilson loop, responsible for the exponentiated divergent and anomalous conformal finite factors $D_i$ and $F_n$,
\be
{\rm exp} \biggl\{ \sum_{i<j} g^2 K_{ij} \biggr\} = \biggl[\prod_i D_i\biggr] F_n\,.
\ee
The remaining conformally invariant factor for $G=U(1)$ is then simply given by (\ref{Abtree}),
\be
\mathcal{R}_n = {\rm exp}\biggl\{\sum_{i<j} [*,i-1,i,j-1,j]\biggr\} = {\rm exp}\, \mathcal{R}^{(1,0)}_n = \sum_{k=0}^{n-4} \frac{1}{k!}\biggl[\sum_{i<j} [*,i-1,i,j-1,j] \biggr]^k \,
\ee
and we see that it is independent of $g$. At degree $4k$ in the Grassmann variables we have
\be
\mathcal{R}^{(k,0)}_n = \frac{1}{k!}\biggl[\sum_{i<j} [*,i-1,i,j-1,j] \biggr]^k = \frac{1}{k!} \bigl(\mathcal{R}^{(1,0)}_n\bigr)^k\,.
\ee

\subsection{Multiple loop operators}

Generalising to the correlators of multiple Wilson loops, the simplest one-loop case is the MHV correlator of two Wilson loops where we can compute the connected part. We obtain a sum over diagrams very similar to (\ref{Dij}). The diagram is the same as the one shown in Fig. \ref{kermitDiagram}, except that the edges $i$ and $j$ are now interpreted as parts of different loops. The resulting diagram differs only in the colour structure,
\begin{align}
\tilde{D}_{ij} = &\frac{4}{N^4}\tr(t_{a_1})\tr(t_{a_2})\tr(t_{b_1}t_{b_2}) \delta_{a_1 b_1} \delta_{a_2 b_2} \frac{g^2N}{\pi^2} \times \notag \\
&\int d^{4|8}x_{AB} \int \frac{ds}{s} \frac{dt}{t} \frac{du_1 du_2}{(u_1-u_2)(u_2-u_1)} \Delta_*^{ix}(s,u_1) \Delta_*^{jx}(t,u_2).
\label{tildeDij}
\end{align}
The colour factor vanishes for the $SU(N)$ theory. For the $U(N)$ theory, the expression thus obtained is again a sum over Kermit diagrams,
\begin{align}
\tilde{D}_{ij} = \frac{g^2}{N^2} K_{ij}\,.
\end{align}
Summing over diagrams we then have
\be
\mathcal{W}^{(0,1),{\rm conn}}_{n_1,n_2} = \frac{1}{N^2} \sum_{i,j} K_{ij}\,,
\label{superW01connKermit}
\ee
where now \(i\) runs over the indices of the first Wilson loop and \(j\) runs over the indices of the second Wilson loop.

Note that, as is the case a for a single Wilson loop, only the first two factors in the denominator of the Kermit expression are physical poles. The remaining four factors are spurious poles which must cancel in the sum over terms (which must also be independent of the reference twistor $Z_*$). For example, if we consider the sum of two terms $K_{i,j} + K_{i+1,j}$ we find after ignoring some common factors and the prefactor of \(\pi^2\),
\be
-\frac{(\langle *\, i-1\, i \,[A\rangle \langle B] \,j-1 \, j\, * \rangle)^2}{\langle A\,B\, i-1\, i\rangle  \langle A\, B\, i-1 \,* \rangle \langle A\, B\, i \,* \rangle} - \frac{(\langle *\, i\, i+1 \,[A\rangle \langle B] \,j-1 \, j\, * \rangle)^2}{\langle A\,B\, i\, i+1\rangle  \langle A\, B\, i \,* \rangle \langle A\, B\, i+1 \,* \rangle}\,.
\ee
Combining denominators we find a numerator of the form
\begin{align}
\text{num} = \quad -&(\langle *\, i-1\, i \,[A\rangle \langle B] \,j-1 \, j\, * \rangle)^2 \langle A\,B\, i\, i+1\rangle \langle A\, B\, i+1 \,* \rangle \notag \\
- &(\langle *\, i\, i+1 \,[A\rangle \langle B] \,j-1 \, j\, * \rangle)^2 \langle A\,B\, i-1\, i\rangle  \langle A\, B\, i-1 \,* \rangle\,.
\end{align}
If we now consider $Z_i, Z_A, Z_B ,Z_*$ to be coplanar we may write
\be
Z_i = Z_* + \alpha Z_A + \beta Z_B
\ee
We then find
\begin{align}
\text{num} = (\beta& \langle * \, i-1\, B A \rangle \langle B \, j-1 \, j\, * \rangle - \alpha \langle * \, i-1\, A B \rangle \langle A \, j-1 \, j\, * \rangle)^2 \langle A \, B \, * \, i+1 \rangle^2 \notag \\
- (\beta& \langle *\, B\, i+1 \, A \rangle \langle B \,j-1 \, j\, * \rangle - \alpha \langle *\, A \, i+1 \, B \rangle \langle A \,j-1 \, j\, * \rangle )^2 \langle A\, B\, i-1\, *\rangle^2 \,,
\end{align}
which vanishes identically. We therefore see that the pole at $\langle i \, A\, B\, * \rangle=0$ is absent in the sum of the two terms. This is sufficient to show that all such spurious poles in the Kermit expression cancel in the sum over $i$ and $j$. The resulting sum therefore has only the local poles of the form $\langle A\,B\, i-1\, i\rangle$ and $\langle A\,B\, j-1\, j\rangle$. The argument above shows that by using Pl\"ucker relations on the minors one can cancel the spurious poles and obtain an expression which manifestly has only local poles. Since the reference twistor cannot appear in the denominator of such an expression, it also cannot appear in the numerator, and hence the sum over diagrams $K_{ij}$ above is also correctly independent of $Z_*$.

Just as for the case of a single loop operator, the expression for the integrand obtained above is equal to a purely local expression in terms of chiral pentagon integrands \cite{Arkani-Hamed:2010pyv}. 

Each of the chiral pentagons has unit residue on a single physical quad-cut 
\be
\langle AB i-1 \, i \rangle = \langle AB j-1 \, j \rangle = \langle AB i \, i+1 \rangle = \langle AB j \, j+1 \rangle = 0\,,
\ee
which should therefore match the residue on the same singularity from the sum of the four Kermit diagrams $K_{i,j} + K_{i+1,j} + K_{i,j+1} + K_{i+1,j+1}$. Indeed it is straightforward to check that the associated residue for that sum is \(-1\) and thus we conclude that indeed
\be
-\frac{1}{N^2} \sum_{i,j} P^X_{ij} 
\label{proposedPXijexpr}
\ee
has all the right physical poles to match the expression (\ref{superW01connKermit}). We still need to show that it does not depend on the choice of $X$. We may make use of the explicit expression for the finite pentagon integral \cite{Arkani-Hamed:2010pyv},
\be
P^X_{ij} = \log u \log v + {\rm Li}_2 (1-u) + {\rm Li}_2 (1-v) + {\rm Li}_2 (1-w) - {\rm Li}_2 (1-uw) - {\rm Li}_2 (1-vw)\,,
\label{PXij}
\ee
where
\begin{align}
u = \tfrac{\langle X \, i-1\,i \rangle \langle i\, i+1\, j\, j+1\rangle}{\langle X \, i \, i+1 \rangle \langle i-1\, i\, j\, j+1\rangle}\,,  \quad
v = \tfrac{\langle X \, j\, j+1 \rangle \langle i-1\, i\, j-1\, j \rangle}{\langle X j-1\, j \rangle \langle i-1 \, i\, j\, j+1\rangle}\,,  \quad 
w = \tfrac{\langle i-1\,i\,j\,j+1\rangle \langle i\, i+1\, j-1\, j\rangle}{\langle i-1\, i\, j-1\, j \rangle \langle i\, i+1\, j\, j+1\rangle}\,.
\end{align}
Taking the total derivative we have
\begin{align}
d P^X_{ij} = \quad &\log u [d \log v - d \log (1-u) + d \log (1-uw)] \notag \\
+ &\log v [d \log u - d \log(1-v) + d \log (1-vw)] \notag \\
+ & \log w [d \log (1-uw) + d \log (1-vw)] \notag \\
= \quad &\log u \, d \log \frac{v(1-uw)}{1-u} + \log v \, d \log \frac{u(1-vw)}{1-v} + \log w \, d \log [(1-uw)(1-vw)]
\end{align}
Note that
\begin{align}
uw &= \frac{\langle X \, i-1\,i \rangle  \langle i\, i+1\, j-1\, j\rangle}{\langle X \, i \, i+1 \rangle  \langle i-1\, i\, j-1\, j \rangle } = u|_{j \rightarrow j-1}\notag \\
vw &= \frac{\langle X \, j\, j+1 \rangle \langle i\, i+1\, j-1\, j \rangle}{\langle X j-1\, j \rangle \langle i \, i+1\, j\, j+1\rangle} = v|_{i\rightarrow i+1}
\end{align}
Thus the second and fifth terms in (\ref{PXij}) cancel in the sum over $j$ while the third and sixth terms cancel in the sum over $i$. The fourth term is independent of $X$ and can be written as 
\be
{\rm Li}_2(1-w) = {\rm Li}_2 (1-1/v_{ij}) = - \frac{1}{2} (\log v_{ij})^2 - {\rm Li}_2(1-v_{ij}) \,.
\ee
The first term in (\ref{PXij}) can be written
\begin{align}
&\bigl[\log \langle X \, i-1\,i \rangle + \log \langle i\, i+1\, j\, j+1\rangle - \log \langle X \, i \, i+1 \rangle - \log \langle i-1\, i\, j\, j+1\rangle \bigr] \notag \\
\times & \bigl[ \log \langle X \, j\, j+1 \rangle + \log \langle i-1\, i\, j-1\, j \rangle - \log \langle X j-1\, j \rangle - \log \langle i-1 \, i\, j\, j+1\rangle \bigr]\,.
\end{align}
All terms involving $X$ cancel in the sums over $i$ or $j$ and we see that the expression (\ref{proposedPXijexpr}) is indeed independent of $X$. We can write the resulting expression as
\begin{align}
\mathcal{W}^{(0,1),{\rm conn}}_{n_1,n_2} &= \frac{1}{N^2} \sum_{i,j} K_{ij} = -\frac{1}{N^2} \sum_{i,j} P^X_{ij}\,,  \\
&= -\frac{1}{N^2} \sum_{i,j} \biggl[ \frac{1}{2} (\log v_{ij})^2 - {\rm Li}_2(1-v_{ij}) + \log \frac{\langle i\, i+1\, j\, j+1\rangle}{\langle i-1\, i\, j\, j+1\rangle} \log \frac{\langle i-1\, i\, j-1\, j \rangle}{\langle i-1 \, i\, j\, j+1\rangle} \biggr]\,, \notag
\end{align}
which is conformally invariant but not manifestly homogeneous. Alternatively, choosing $X$ as the infinity bitwistor, we can write it as
\be
\mathcal{W}^{(0,1),{\rm conn}}_{n_1,n_2} = -\frac{1}{N^2}\sum_{i,j} \biggl[ \frac{1}{2} (\log v_{ij})^2 - {\rm Li}_2(1-v_{ij}) + \log \frac{x_{i+1,j+1}^2}{x_{i,j+1}^2} \log \frac{x_{ij}^2}{x_{i,j+1}^2} \biggr]\,.
\ee
If we now eliminate $x_{ij}^2$ in favour of $v_{ij}$ and $x_{kl}$ with $k>i$ or $l>j$ for $i\leq n_1 -1$ and $j \leq n_2 -1$, all remaining $x_{kl}$ cancel and we are left with a manifestly conformally invariant expression which agrees exactly with the expression (\ref{Gexpr2}) for $f_{n_1,n_2}$,
\be
-\sum_{i,j} P^X_{ij} =  f_{n_1,n_2}\,.
\ee
Hence we have
\be
\mathcal{W}^{(0,1),{\rm conn}}_{n_1,n_2} = \frac{1}{N^2} f_{n_1,n_2}\,,
\label{superWL12connUN}
\ee
exactly as it should be, since the twistor calculation at Grassmann degree zero (MHV level) should match the purely spacetime calculation presented in Sec. \ref{Sec-MWLcorrs} in eq. (\ref{W12connUN}). We remind the reader that both (\ref{W12connUN}) and (\ref{superWL12connUN}) are for the $U(N)$ theory, while the corresponding quantities both vanish for the $SU(N)$ theory.

If we consider the Abelian theory and look at the correlator of multiple loop operators we again expect an exponentiation of the form
\begin{align}
\langle \mathcal{L}(C_1) \ldots \mathcal{L}(C_m) \rangle = \mathcal{W}_{n_1,\ldots,n_m} =  {\rm exp}\biggl\{&\sum_r \sum_{i_r<j_r} \bigl[g^2K_{i_r j_r} + [*_r,i_r-1,i_r,j_r-1,j_r]\bigr] \notag \\
 + &\sum_{r<s} \sum_{i_r,j_s} \bigl[g^2K_{i_r j_s} + [*_{rs},i_r-1,i_r,j_s-1,j_s]\bigr]  \biggr\}\,.
 \label{superMWLexp}
\end{align}
The first line here is a product over the expression (\ref{U1expsuperWL}) for each individual loop operator. The second line accounts for the connected parts. We have used the notation $*_r$ and $*_{rs}$ for the reference twistor here since each part is gauge invariant individually and so one can choose different reference twistors for each part if wanted.

As above, the first term in the first line in (\ref{superMWLexp}) accounts for the divergent and anomalous conformal factors $D_{i_r}$ and $F_{n_r}$ in (\ref{superWLfactors}). Therefore we have for the conformally invariant remainder,
\begin{align}
 \mathcal{R}_{n_1,\ldots,n_m} =  {\rm exp}\biggl\{&\sum_r \sum_{i_r<j_r} [*_r,i_r-1,i_r,j_r-1,j_r] \notag \\
 + &\sum_{r<s} \sum_{i_r,j_s} \bigl[g^2K_{i_r j_s} + [*_{rs},i_r-1,i_r,j_s-1,j_s]\bigr]  \biggr\}\, \notag \\
 =  {\rm exp}\biggl\{&\sum_r \sum_{i_r<j_r} [*_r,i_r-1,i_r,j_r-1,j_r] \notag \\
 + &\sum_{r<s}\Bigl[g^2 f_{n_r,n_s} + \sum_{i_r,j_s} [*_{rs},i_r-1,i_r,j_s-1,j_s]\Bigr]  \biggr\}\,.
 \label{superMWLexpR}
\end{align}

Let us note that, as already discussed for the spacetime calculation in Section \ref{Sec-SU(N)Leading}, the one-loop Abelian MHV result \(f_{n_1,n_2}\) supplies the basic ingredient for the first non-zero contributions in the SU(N) theory. For instance, we depict the leading contributions for two Wilson loops at MHV (O($g^4$)) in Fig. \ref{leadingOrderDiagramsMHV}; diagrammatically, it is clear from the twistor diagrams that (after computing the correct colour factors)
\be
\mathcal{W}^{(0,2),{\rm conn}}_{n_1,n_2} = \frac{NC_F}{N^4}(f_{n_1,n_2})^2
\label{superW02conn}
\ee
exactly as we deduced in the case of the space-time calculation in Section \ref{Sec-SU(N)Leading}, because each relevant twistor diagram is essentially the product of two of the one-loop diagrams which enter into \(f_{n_1,n_2}\).

\begin{figure}
\begin{center}
\begin{tikzpicture}[scale=0.9]
   \path[use as bounding box] (-3.5,-2.5) rectangle (22.5,2.5);

    \defineline{lin4m}{originx=3.5, originy=2.5, angle=120, pointcount=3}

    \defineline{lin4bm}{originx=3.5, originy=-1, angle=60, pointcount=3}

    \defineline{lin5m}{originx=6, originy=2.5, angle=90, pointcount=4}
  \defineline{lin6m}{originx=5, originy=-1, angle=90, pointcount=4}
    \defineline{lin7m}{originx=8.3, originy=2.5, angle=240, pointcount=3}

   \defineline{lin8m}{originx=8.3, originy=-1, angle=300, pointcount=3}
  \drawpropagatortwo{fromline=lin4m, toline=lin5m, fromip=2, toip=3,looseval=0.5}

\drawpropagator{fromline=lin5m, toline=lin7m, fromip=2, toip=2,looseval=0.5}

\drawpropagator{fromline=lin4bm, toline=lin8m, fromip=2, toip=2,looseval=0.5}

  \drawline[showwings=true, showtwistors=true, fromlabel={\small $Z_{i_1-1} \, \, $},tolabel={\small $Z_{i_1} \hspace{2mm}$}]{lin4m}

  \drawline[showwings=true, showtwistors=true, fromlabel={\small $Z_{i_2-1} \, \, \,  $},tolabel={\small $Z_{i_2}\hspace{1mm}$}]{lin4bm}

  \drawline[showwings=false, showtwistors=true, fromlabel={\small $(AB)\hspace{2mm}$}]{lin5m}

  \drawline[showwings=true, showtwistors=true, fromlabel={\small $\, \, \,  \,  Z_{j_1-1}$},tolabel={\small $\, \, \hspace{1mm}Z_{j_1}$}]{lin7m}

  \drawline[showwings=true, showtwistors=true, fromlabel={\small $Z_{j_2-1}$},tolabel={\small $Z_{j_2}$}]{lin8m}

\end{tikzpicture}
\end{center}
\caption{The type of diagram which contributes for NMHV at leading order (O($g^2$)) in the SU(N) theory for the correlator of two Wilson loops.}
\label{leadingOrderDiagramsNMHV}
\end{figure}
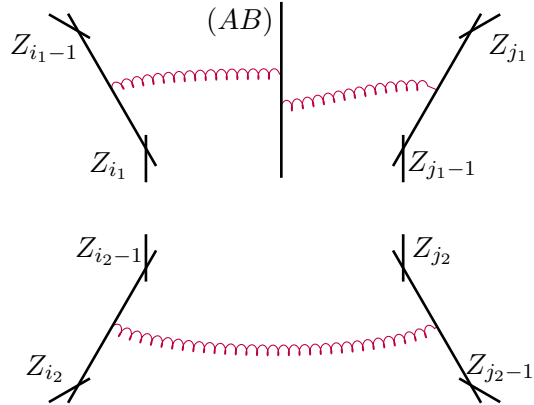

We can apply similar logic beyond the MHV case, e.g. noting that the leading order contribution in the SU(N) theory for two Wilson loops (for which we depict the diagrams in Fig. \ref{leadingOrderDiagramsNMHV}) is given by
\be
\mathcal{W}^{(1,1),{\rm conn}}_{n_1,n_2} = \frac{2NC_F}{N^4} f_{n_1,n_2}\sum_{i,j}[*,i-1,i,j-1,j]. 
\label{superW11conn}
\ee
Note that the sum is nothing more than the Abelian NMHV tree-level result for the connected part of the correlator of two Wilson loops. 

Finally we have the connected tree-level contribution at N${}^2$MHV level quoted in (\ref{N2cont1}) which we repeat here in the case of the $SU(N)$ theory,
\be
\mathcal{W}_{n_1,n_2}^{(2,0),{\rm conn}}=\frac{NC_F}{N^4}  \Bigl(\sum_{i,j} [*, i-1, i, j-1,j] \Bigr)^2\,.
\label{superW20conn}
\ee

\begin{figure}
\begin{center}
\begin{tikzpicture}[scale=0.9]
   \path[use as bounding box] (-3.5,-2.5) rectangle (22.5,2.5);
  \defineline{lin1m}{originx=9.5, originy=1, angle=90, pointcount=3}
  \defineline{lin2m}{originx=11.5, originy=1, angle=90, pointcount=4}
  \defineline{lin3m}{originx=13.5, originy=1, angle=270, pointcount=3}

    \defineline{lin4m}{originx=-1.5, originy=2.5, angle=120, pointcount=3}

    \defineline{lin4bm}{originx=-1.5, originy=-1, angle=60, pointcount=3}

    \defineline{lin5m}{originx=1, originy=2.5, angle=90, pointcount=4}
  \defineline{lin6m}{originx=1, originy=-1, angle=90, pointcount=4}
    \defineline{lin7m}{originx=3.3, originy=2.5, angle=240, pointcount=3}

   \defineline{lin8m}{originx=3.3, originy=-1, angle=300, pointcount=3}

  \drawline[showwings=true, showtwistors=true, fromlabel={\small $Z_{i-1}$},tolabel={\small $Z_{i}\hspace{1mm}$}]{lin1m}
  \drawline[showwings=false, showtwistors=true,  fromlabel={\small $(AB)\hspace{2mm}$}]{lin2m}
  \drawline[showwings=true, showtwistors=true, fromlabel={\small $Z_{j-1}$},tolabel={\small $\,\,Z_{j}$}]{lin3m}
  \drawpropagatortwo{fromline=lin1m, toline=lin2m, fromip=2, toip=3}
  \drawpropagator{fromline=lin2m, toline=lin3m, fromip=2, toip=2}

  \drawpropagatortwo{fromline=lin4m, toline=lin5m, fromip=2, toip=3,looseval=0.5}

  \drawpropagator{fromline=lin5m, toline=lin7m, fromip=2, toip=2,looseval=0.5}

  \drawpropagatortwo{fromline=lin4bm, toline=lin6m, fromip=2, toip=2,looseval=0.5}

  \drawpropagator{fromline=lin6m, toline=lin8m, fromip=3, toip=2,looseval=0.5}

  \drawline[showwings=true, showtwistors=true, fromlabel={\small $Z_{i_1-1}$ \, \, \, },tolabel={\small $Z_{i_1} \hspace{2mm}$}]{lin4m}

  \drawline[showwings=true, showtwistors=true, fromlabel={\small $Z_{i_2-1}$ \, \, \, },tolabel={\small $Z_{i_2}\hspace{1mm}$}]{lin4bm}

  \drawline[showwings=false, showtwistors=true, fromlabel={\small $(A_1B_1)\hspace{5mm}$}]{lin5m}

  \drawline[showwings=false, showtwistors=true, fromlabel={\small $(A_2B_2)\hspace{5mm}$}]{lin6m}

  \drawline[showwings=true, showtwistors=true, fromlabel={\small \, \, $Z_{j_1-1}$},tolabel={\small $\hspace{1mm}Z_{j_1}$}]{lin7m}

  \drawline[showwings=true, showtwistors=true, fromlabel={\small $Z_{j_2-1}$},tolabel={\small $\hspace{3mm}Z_{j_2}$}]{lin8m}

\end{tikzpicture}
\end{center}
\caption{On the left, we depict the type of diagram which supplies the leading order in contribution in \(g\), for the correlator of two Wilson loops computed in the SU(N) theory and for the MHV sector. By comparison with the diagram on the right which depicts the diagram entering into the U(N) (including Abelian) case, it is clear that (up to colour factors) we essentially have the square of the O(\(g^2\)) Abelian, MHV result.}
\label{leadingOrderDiagramsMHV}
\end{figure}
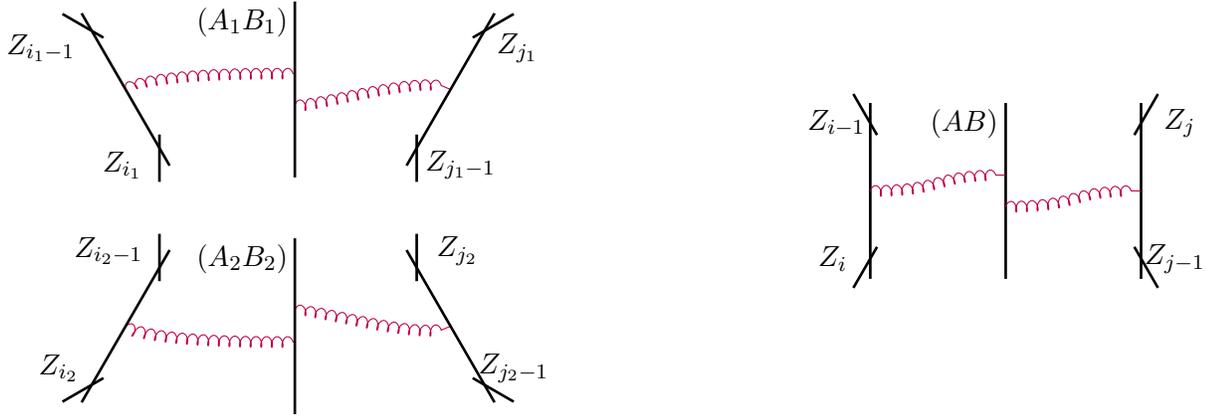

We see that the three contributions (\ref{superW02conn}), (\ref{superW11conn}) and (\ref{superW20conn}) to $\mathcal{W}_{n_1,n_2}^{\rm conn}$ can be packaged together into a perfect square,
\be
\frac{NC_F}{N^4}\Bigl( g^2 f_{n_1,n_2} + \sum_{i,j}[*,i-1,i,j-1,j]\Bigr)^2\,.
\label{SUNleadgincombined}
\ee
The object in parentheses is the sum of the Abelian MHV order $g^2$ and NMHV tree-level results. Indeed, the combination
\be
g^2K_{i j} + [*,i-1,i,j-1,j]
\ee
which, when summed over $i$ and $i$, produces the combined Abelian result, can be thought of as a combined propagator which includes the standard twistor propagator together with the two-insertion vertex from the non self-dual interaction Lagrangian in $S_2$.

\section{$\bar{Q}$ equation:}
Just as for a single Wilson loop \cite{Caron-Huot:2011dec,Bullimore:2011kg}, there should be an equation relating the action of the operator
\be
\bar{Q}_A^{A'} = \sum_i \chi_i^{A'}\frac{\partial}{\partial Z_i^A}
\ee
to the integral over a correlator of Wilson loops with an additional point inserted on each edge of each loop. The equation should be a generalisation of the equation for a single Wilson loop presented in \cite{Caron-Huot:2011dec} which reads,
\be
\bar{Q}_A^{A'} \mathcal{R}^{(k)}_n = \frac{1}{4} \Gamma_{\rm cusp} \int [d^{2|3} \mathcal{Z}_{n+1}]^{A'}_A \bigl[ \mathcal{R}^{(k+1)}_{n+1} - \mathcal{R}^{(k)}_n \mathcal{R}^{(1,0)}_{n+1}\bigr] + \text{ cyc}\,.
\label{QbarOriginal}
\ee
This equation has been extensively tested and used to calculate new results in the planar limit, with particular focus on the relevance to planar scattering amplitudes (see e.g. \cite{Caron-Huot:2011dec,Caron-Huot:2013vda,He:2019jee,Li:2021bwg}).
The natural generalisation of (\ref{QbarOriginal}) equation for correlators of $m$ light-like loop operators would be 
\be
\label{Qbargeneralm}
\bar{Q}_A^{A'} \mathcal{R}^{(k)}_{n_1,\ldots,n_m} = \frac{1}{4}\Gamma_{\rm cusp}\sum_r \biggl[   \int \bigl[d^{2|3} \mathcal{Z}_{n_r+1} \bigr]^{A'}_{A} \Bigl[ \mathcal{R}^{(k+1)}_{n_1,\ldots,n_r+1,\ldots,n_m} - \mathcal{R}^{(k)}_{n_1,\ldots,n_m} \mathcal{R}^{(1,0)}_{n_r+1}\Bigr] + \text{cyc}_r\biggr]\,.
\ee
Here, as for the \(\bar{Q}\) equation for a single Wilson loop, we define the integration measure as
\be 
\int d^{2|3}\mathcal{Z}_{{n_r}+1} = C(n_r-1 n_r 1_r)_a  \oint_{\epsilon=0} \epsilon d \epsilon \int_0^{\infty} d\tau (d^3 \chi_{n_r+1})^A 
\ee
where by \(1_r\) we mean the first twistor on loop \(C_r\) i.e. the twistor which follows \(n_r\) in the cyclic ordering. 

As in the case of a single Wilson loop, since at each order in \(g^2\) the \(\bar{Q}\) equation relates (for the SU(N) theory) the \(\bar{Q}\) of the \(l\)-loop N\(^k\)MHV result to a collinear integral of the \(l-1\)-loop N\(^{k+1}\)MHV result, this equation should in principle provide a useful means for obtaining results at high order in powers of \(g^2\). While we leave a more detailed exploration of this point to forthcoming work, let us now verify the \(\bar{Q}\) equation (\ref{Qbargeneralm}) in the simple case of the Abelian theory.  

\subsection{$\bar{Q}$ equation in the Abelian case}

\subsubsection{Review for a single Wilson loop}

We can consider the $\bar{Q}$ equation in the Abelian case. The equation for a single Wilson loop reads (note that as \(\Gamma_{\textrm{cusp}} = 4 g^2\) in the Abelian theory, the factor of \(\frac{1}{4}\Gamma_{\textrm{cusp}}\) is simply $g^2$)
\be
\bar{Q}_A^{A'} \mathcal{R}^{(k)}_n = g^2 \int [d^{2|3} \mathcal{Z}_{n+1}]^{A'}_A \bigl[ \mathcal{R}^{(k+1)}_{n+1} - \mathcal{R}^{(k)}_n \mathcal{R}^{(1,0)}_{n+1}\bigr] + \text{ cyc}\,.
\label{QbarAbelian}
\ee
Note that the LHS is manifestly zero since $\bar{Q}$ annihilates each $R$-invariant (upon setting the reference twistor to e.g. $\mathcal{Z}_1$).

As we have just seen, in the Abelian case we have $\mathcal{R}^{(1,0)}_{n+1} = \mathcal{R}^{(1)}_{n+1}$. Furthermore we may exploit the independence on the reference supertwistor $\mathcal{Z}_*$ to write (setting $\mathcal{Z}_*=\mathcal{Z}_1$)
\begin{align}
\mathcal{R}^{(1)}_{n+1} &= \sum^{n+1}_{i>j} [*,i-1,i,j-1,j] \notag \\
&= \sum^{n}_{i<j} [1,i-1,i,j-1,j] + \sum_{i=3}^{n-1} [1,i-1,i,n,n+1] = \mathcal{R}^{(1)}_n + X_{n+1}\,,
\end{align}
where the final step defines $X_{n+1}$ as the second term in the penultimate expression. It follows that if we write the integrand on the RHS of the $\bar{Q}$ equation in terms of $\mathcal{R}^{(1)}_n$ and $X_{n+1}$, the term linear in $X_{n+1}$ precisely cancels,
\be
\label{Abexpansion}
\mathcal{R}^{(k+1)}_{n+1} - \mathcal{R}^{(k)}_n \mathcal{R}^{(1)}_{n+1} = \biggl[\frac{1}{(k+1)!} - \frac{1}{k!}\biggr]\bigl(R^{(1)}_n\bigr)^{k+1}  + \frac{1}{2(k-1)!} X_{n+1}^2 \bigl(\mathcal{R}^{(1)}_{n}\bigr)^{k-1} + O(X_{n+1}^3)\,.
\ee
Now we recall that the measure of integration on the RHS of (\ref{QbarAbelian}) contains an integral $[d^{0|3} \chi_{n+1}]^{A'}$ and the factor $\epsilon d\epsilon$ with instruction to keep the residue of the simple pole $1/\epsilon$. The first term in the expansion (\ref{Abexpansion}) then integrates to zero as it does not depend on $\chi_{n+1}$ at all. Poles in $\epsilon$ can only arise from vanishing denominator factors in the $R$-invariants in $X_{n+1}$. Each such $R$-invariant has three denominator factors which vanish linearly as $\epsilon$ is taken to zero. However each of the four Grassmann odd factors from the delta function in the numerator vanishes linearly. The threefold Grassmann integral removes three of these zeros. Accounting for the factor of $\epsilon$ from the measure we find that terms linear in $X_{n+1}$ would indeed have a residue upon integration. However, terms with higher powers of $X_{n+1}$ have no residue. Since the linear terms in $X_{n+1}$ cancelled in (\ref{Abexpansion}), we find the whole of the expression on the RHS of (\ref{QbarAbelian}) integrates to zero, matching the vanishing of the LHS.

More compactly, we can consider the $\bar{Q}$ equation obtained by summing over $k$ on both sides,
\be
\bar{Q}_A^{A'} \mathcal{R}_n = g^2 \int [d^{2|3} \mathcal{Z}_{n+1}]^{A'}_A \bigl[ \mathcal{R}_{n+1} - \mathcal{R}_n \mathcal{R}^{(1), {\rm tree}}_{n+1}\bigr] + \text{ cyc}\,.
\label{QbarAbelianExp}
\ee
The integrand on the RHS of (\ref{QbarAbelianExp}) simplifies as follows
\begin{align}
\mathcal{R}_{n+1} - \mathcal{R}_n \mathcal{R}^{(1), {\rm tree}}_{n+1} &= {\rm exp} \bigl\{ \mathcal{R}^{(1)}_{n+1} \bigr\} - {\rm exp} \bigl\{ \mathcal{R}^{(1)}_n \bigr\} \mathcal{R}^{(1)}_{n+1} \notag \\
&= {\rm exp} \bigl\{ \mathcal{R}^{(1)}_{n} + X_{n+1} \bigr\} - {\rm exp} \bigl\{ \mathcal{R}^{(1)}_n \bigr\} \bigl[ \mathcal{R}^{(1)}_{n} + X_{n+1} \bigr] \notag \\
&= {\rm exp} \bigl\{ \mathcal{R}^{(1)}_{n} \bigr\} \Bigl[  {\rm exp} \bigl\{ X_{n+1} \bigr\} - X_{n+1} - \mathcal{R}^{(1)}_{n}  \Bigr]\,.
\end{align}
Again we see that the terms linear in $X_{n+1}$ have cancelled and we may repeat the argument as above for the vanishing of the integral.

\subsubsection{Multiple Wilson loops}

If we consider the contribution to a correlator of multiple loop operators we have the exponentiation given in (\ref{superMWLexpR}). The $\bar{Q}$ equation we wish to understand is as follows (again written summing over all Grassmann degrees and noting that \(\Gamma_{\textrm{cusp}} = 4g^2\) in the Abelian theory),
\be
\bar{Q}_A^{A'} \mathcal{R}_{n_1,\ldots,n_m} = g^2 \sum_r \biggl[  \int \bigl[d^{2|3} \mathcal{Z}_{n_r+1} \bigr]^{A'}_{A} \Bigl[ \mathcal{R}_{n_1,\ldots,n_r+1,\ldots,n_m} - \mathcal{R}_{n_1,\ldots,n_m} \mathcal{R}^{(1,0)}_{n_r+1}\Bigr] + \text{cyc}_r\biggr]\,.
\label{ManyWQbarAb}
\ee
Recall that in the Abelian case we have $\mathcal{R}_{n_r+1}^{(1,0)} = \mathcal{R}^{(1)}_{n_r+1}$ and that $\mathcal{R}_{n_1,\ldots,n_m}$ is given in eq. (\ref{superMWLexpR}).
The LHS of (\ref{ManyWQbarAb}) is given by
\be
\bar{Q}_A^{A'} \mathcal{R}_{n_1,\ldots,n_m} = g^2 \mathcal{R}_{n_1,\ldots,n_m} \sum_{r>s} \bar{Q}_A^{A'} f_{n_r n_s}  \,,
\ee
with
\be
\bar{Q}_A^{A'} f_{n_r n_s} =  \sum_{\{ i,j \}}^{\{n_r,n_s\}} \log \frac{x_{ij}^2 x_{i+1,j+1}^2}{x_{i,j+1}^2 x_{i+1,j}^2} \bar{Q}_A^{A'} \log \bigl[ \langle i-1\, i \, i+1 j \rangle \langle i \, j-1 \, j \, j+1\rangle\bigr]\,,
\label{Qbarf}
\ee
which follows from (\ref{twistordW12}).
Now let us note that we can write (choosing $\mathcal{Z}_{*r} = \mathcal{Z}_{*rs} = \mathcal{Z}_{1_r}$ in eq. (\ref{superMWLexpR})),
\be
\mathcal{R}_{n_1,\ldots,n_r+1,\ldots,n_m} = \mathcal{R}_{n_1,\ldots,n_m} {\rm exp}\biggl\{X_{n_r+1} + g^2 f_\epsilon + \sum_{s \neq r}  \tilde{X}_{s,n_r+1} \biggr\}
\ee
where
\be
\tilde{X}_{s,n_r+1}  = \sum_{i_s} [1_r,i_s-1,i_s,n_r,n_r+1]\,, \qquad f_\epsilon \equiv \sum_{s \neq r} (f_{n_r+1, n_s} - f_{n_r n_s})\,.
\ee
Note that the collinear limit of $f_{n_r+1,n_s}$ is smooth and hence we have $f_\epsilon = O(\epsilon)$.

By a similar argument to the one above for a single Abelian Wilson loop, the only terms which can contribute to the integral on the RHS of (\ref{ManyWQbarAb}) are those which are linear in $X$ or $\tilde{X}$. The term linear in $X$ again cancels between the two terms in the integrand, leaving only the term linear in $\tilde{X}$. The term $f_\epsilon$ never contributes since it is higher order in $\epsilon$. Thus the $\bar{Q}$ equation holds if
\be
\sum_{r<s} \bar{Q}_A^{A'} f_{n_r n_s} =    \sum_r \biggl[ \int [d^{2|3} \mathcal{Z}_{n_r+1}]^{A'}_{A} \sum_{s \neq r}  \tilde{X}_{s,n_r+1} + \text{cyc}_r\biggr]\,.
\label{QbarrelAb}
\ee
The integral is over $R$-invariants of the form $[1_r,i_s-1,i_s,n_r,n_r+1]$. The result of integrating such $R$-invariants is given in \cite{Caron-Huot:2011dec}. We have
\be
\int [d^{2|3} \mathcal{Z}_{n_r+1}]^{A'}_{A} [1_r,i_s-1,i_s,n_r,n_r+1] = \int d \log \frac{\langle X i_s-1 i_s \rangle}{\langle X 1_r 2_r \rangle} \bar{Q}^{A'}_{A} \log \frac{\langle \bar{n}_r \, i_s  \rangle }{\langle \bar{n}_r \, i_s -1 \rangle}\,,
\ee
where
\be
X(\tau) = Z_{n_r} \wedge (Z_{n_r-1} - C \tau Z_{1_r}), \qquad C = \frac{\langle n_r-1\, n_r\, 2_r\, 3_r \rangle}{\langle n_r \, 1_r \, 2_r \, 3_r \rangle}\,.
\ee
In total for the RHS of (\ref{QbarrelAb}) we have
\begin{align}
& \sum_{\substack{r,s\\ r \neq s}} \sum_{j_r,i_s} \log \frac{\langle j_r -1\, j_r \, j_r +2 \, j_r +3 \rangle \langle j_r\, j_r +1\, i_s -1 \, i_s  \rangle}{\langle j_r \, j_r +1 \, j_r +2\, j_r +3 \rangle \langle j_r -1 \, j_r \, i_s -1 \, i_s \rangle} \bar{Q}^{A'}_A \log \frac{\langle j_r -1\, j_r \, j_r +1\, i_s \rangle}{\langle j_r -1 \, j_r \, j_r +1\, i_s -1 \rangle} \notag \\
=&  \sum_{\substack{r,s\\ r \neq s}} \sum_{j_r,i_s} \log \frac{\langle j_r\, j_r +1\, i_s -1 \, i_s  \rangle \langle j_r -1 \, j_r \, i_s  \, i_s+1 \rangle}{\langle j_r -1 \, j_r \, i_s -1 \, i_s \rangle \langle j_r\, j_r +1\, i_s  \, i_s+1  \rangle} \bar{Q}^{A'}_A \log \langle j_r -1\, j_r \, j_r +1\, i_s \rangle\notag \\
=&  \sum_{\substack{r,s\\ r \neq s}} \sum_{j_r,i_s} \log \frac{x_{j_r+1,i_s}^2  x_{j_r,i_s+1}^2}{x_{j_r,i_s}^2  x_{j_r+1,i_s+1}^2} \bar{Q}^{A'}_A \log \langle j_r -1\, j_r \, j_r +1\, i_s \rangle\notag \\
=&  \sum_{r<s} \sum_{j_r,i_s} \log \frac{x_{j_r+1,i_s}^2  x_{j_r,i_s+1}^2}{x_{j_r,i_s}^2  x_{j_r+1,i_s+1}^2} \bar{Q}^{A'}_A \log \bigl[ \langle j_r -1\, j_r \, j_r +1\, i_s \rangle \langle i_s -1\, i_s \, i_s +1\, j_r \rangle\bigr]\,,
\end{align}
which precisely matches the LHS of (\ref{QbarrelAb}) upon making use of (\ref{Qbarf}). 

Note that, by having proved the validity of the \(\bar{Q}\) equation in the Abelian theory, we have also proved the validity at MHV and leading order in \(g\) in the SU(N) theory, as a consequence of the analysis presented in Section \ref{Sec-SU(N)Leading}. Specifically, the equations (\ref{QbarOriginal}) and (\ref{Qbargeneralm}) together can be used to write an equation for the connected part of the correlator of multiple loop operators. We then have
\begin{align}
\bar{Q}_{A}^{A'} \mathcal{R}_{n_1,n_2}^{(k),{\rm conn}} = \frac{1}{4} &\Gamma_{\rm cusp} \biggl[   \int \bigl[d^{2|3} \mathcal{Z}_{n_1+1} \bigr]^{A'}_{A} \Bigl[ \mathcal{R}^{(k+1),{\rm conn}}_{n_1+1,n_2} - \mathcal{R}^{(k),{\rm conn}}_{n_1,n_2} \mathcal{R}^{(1,0)}_{n_r+1}\Bigr] + \text{cyc}_1\biggr] \notag \\
&+ (n_1 \leftrightarrow n_2)\,.
\end{align}
Taking the large $N$ limit, we can see that this equation is compatible with the results combined in (\ref{SUNleadgincombined}). The steps essentially follow the analysis of the Abelian case presented above. We will present more sophisticated checks in the SU(N) theory in forthcoming work. 

\section{Conclusion}
The correlators of multiple light-like Wilson loop provide a rich class of observable to explore. Although these are kinematically complicated objects, we have demonstrated that in simple cases the calculations are tractable. We have also verified that the symmetry properties of individual Wilson loops carry over to the correlators of multiple loop operators in the form of the $\bar{Q}$ equation.

Having introduced these objects and the basic tools for their computation, we will turn in a series of forthcoming papers to the study of the properties and mathematical structure enjoyed by these objects. Firstly, the story of \emph{holomorphic linking} presented in \cite{Bullimore:2011ni} generalises very naturally to more general nodal curves, including the case of multiple Wilson loops, and by following the same procedure as in that reference we are able to derive a version of the BCFW recursion relation (for tree level contributions and also for loop integrands) for multiple Wilson loops. This generalised BCFW equation, its derivation, and the verification of the equation in simple cases, will appear in forthcoming work. 

The study of O($g^2$) correlators in the SU(N) theory also warrants further exploration. It is straightforward to write down loop-level integrands in the twistor Wilson loop formulation by using the Feynman rules we have outlined here, and by probing the leading singularities of these integrands it should be possible to use the \emph{chiral box expansion} \cite{Bourjaily:2013mma} for one-loop local integrands in order to perform the loop integration and arrive at integrated answers. In work to appear, we carry out these calculations for loop-level correlators of multiple Wilson loop and use the integrated answers to perform more sophisticated checks on the \(\bar{Q}\)-equation in the SU(N) theory. We will also show that the O(\(g^2\)) leading singularities may be expressed quite generally in terms of tree-level objects, as is the case for a single Wilson loop. 

There are a range of other exciting questions to ask about these objects which we have only begun to explore. For instance, in the case of the expectation value of a single Wilson loop there is a Grassmannian integral which generates the tree-level answer as well as leading singularities \cite{Arkani-Hamed:2009ljj,Arkani-Hamed:2009nll,Mason:2009qx}; to what extent can this integral be modified to generalise to our case here? An answer in the affirmative may hint at some version of Yangian-type symmetry for these objects \cite{Drummond:2009fd,Drummond:2010qh,Drummond:2010uq} and might plausibly lead to a generalisation of the amplituhedron \cite{Arkani-Hamed:2013jha} to the case of correlators of multiple loop operators.  

The \(\bar{Q}\)-equation which we have presented here should provide a useful tool for calculating correlators at higher loop orders, and it would be fascinating to follow the procedure used in \cite{Li:2021bwg} to e.g. integrate up from N\(^2\)MHV one-loop expression to an MHV three-loop expression. There are many interesting questions to ask about the analytic structure of such objects, such as their symbol alphabets, and whether there any interesting constraint on the consecutive branch cuts such as Steinmann-like relations \cite{Caron-Huot:2016owq} or a link to cluster algebras \cite{Golden:2013xva,Drummond:2017ssj}. 

\section*{Acknowledgements}
The authors thank Paul Heslop and Gregory Korchemsky for helpful discussions. All authors are supported by the STFC consolidated grant ST/X000583/1. \"OCG is also supported by the Royal Society University Research Fellowship URF\textbackslash R1\textbackslash221236.

\appendix

\newpage

\section{Perturbative Wilson loop computations}
\label{app-YMandWL}

Here we set out our conventions and collect some standard textbook facts about Wilson loops in (super) Yang-Mills theory. We then briefly recap some known results on expectation values of a single light-like loop.

\subsection{Preliminaries}

We consider four-dimensional $\mathcal{N}=4$ super Yang-Mills theory with gauge group $G$. The theory has the action
\be
S = \frac{1}{g_{\rm YM}^2} \int d^4x \biggl[ -\frac{1}{2} \tr F^{\mu \nu} F_{\mu \nu}  + \ldots\biggr]\,,
\ee
where we have omitted the fermionic and scalar fields. The gauge connection $A_\mu = A_{\mu}^a t^a$ has curvature $F_{\mu \nu} = F^a_{\mu \nu} t^a$ with $F^a_{\mu \nu} = \partial_\mu A^a_\nu - \partial_\nu A^a_\mu + f^{abc} A^b_{\mu} A^c_{\nu}$. The $t^a$ are the generators of the gauge group in some representation and obey
\be
\tr (t^a t^b) = \frac{1}{2} \delta_{ab}\,, \quad [t^a,t^b] = i f^{abc} t^c\,.
\ee
Note that we have normalised the fields so that the Yang-Mills coupling only appears as a prefactor of $1/g_{\rm YM}^2$ in front of the action. Here we are mostly interested in the case of $G=SU(N)$, but it is also instructive to consider $G=U(N)$ and in particular the Abelian case $G=U(1)$, in which case the theory is really a free theory although we will not rescale the field to remove $g_{\rm YM}$.

Recall that a Wilson line operator between two points $x_0$ and $x$ can be defined via a path-ordered exponential. We define a curve from $x_0$ to $x_1$ via a map $x^\mu : [0,1] \rightarrow \mathbb{R}^{1,3}$ with $x(0)=x_0$ and $x(1)=x_1$. We then have
\begin{align}
U(x_1,x_0,A) &= \mathcal{P}\,{\rm exp} \,i \int_{x_0}^{x_1} A \notag \\
&= 1\!\!1 + i \int_0^1 dt_1 \dot{x}^\mu(t_1) A_\mu\bigl(x(t_1)\bigr) - \int_0^1 dt_1 \dot{x}^\mu(t_1) A_\mu \bigl(x(t_1)\bigr) \int_0^{t_1} dt_2 \dot{x}^\nu(t_2) A_\nu \bigl(x(t_2)\bigr) + \ldots \notag
\end{align}
Recall that under a gauge transformation $V(x) \in SU(N)$ we have
\be
A_\mu(x) \mapsto  A^V_\mu(x) = V(x) \bigl[A_\mu(x) +i \partial_\mu \bigr] V^\dagger(x)\,.
\ee
If we then consider $U(x(s),x_0,A)$, we should consider a reparametrisation $\tilde{x}$ of the curve described by a map $t : [0,1] \rightarrow [0,s]$ (with $t(0)=0$ and $t(1) = s$) such that $\tilde{x}(\tilde{t}) = x(t(\tilde{t}))$. Then
\begin{align}
U(x(s),x_0,A) &= 1\!\!1 + i \int_0^1 d\tilde{t}_1 \dot{\tilde{x}}^\mu(\tilde{t}_1) A_\mu\bigl(\tilde{x}(\tilde{t}_1)\bigr) - \int_0^1 d\tilde{t}_1 \dot{\tilde{x}}^\mu(\tilde{t}_1) A_\mu \bigl(\tilde{x}(\tilde{t}_1)\bigr) \int_0^{\tilde{t}_1} d\tilde{t}_2 \dot{\tilde{x}}^\nu(\tilde{t}_2) A_\nu \bigl(\tilde{x}(\tilde{t}_2)\bigr) + \ldots \notag \\
&= 1\!\!1 + i \int_0^s dt_1 \dot{x}^\mu(t_1) A_\mu\bigl(x(t_1)\bigr) - \int_0^s dt_1 \dot{x}^\mu(t_1) A_\mu \bigl(x(t_1)\bigr) \int_0^{t_1} dt_2 \dot{x}^\nu(t_2) A_\nu \bigl(x(t_2)\bigr) + \ldots \,.\notag
\end{align}
We then find
\be
\frac{d}{ds} U(x(s),x_0,A) = i\dot{x}^\mu(s) A_\mu(x(s)) U(x(s),x_0,A)
\ee
and hence
\be
\bigl[\partial_\mu - iA_\mu(x)\bigr] U(x,x_0,A) = 0\,.
\label{Udiffeq}
\ee
Now consider $V(x)U(x,x_0,A) V(x_0)^\dagger$. We have
\begin{align}
&\bigl[\partial_\mu - iA^V_\mu(x)\bigr] \bigl[V(x)U(x,x_0,A) V(x_0)^\dagger\bigr] \notag \\
= &\biggl[\partial_\mu - i V(x) \bigl[A_\mu(x) + i \partial_\mu \bigr] V^\dagger(x)\biggr] \bigl[V(x)U(x,x_0,A) V(x_0)^\dagger\bigr] \notag \\
= & \partial_\mu V(x) U(x,x_0,A) V(x_0)^\dagger + V(x)\bigl[\partial_\mu - i A_\mu(x)\bigr] U(x,x_0,A)V(x_0)^\dagger \notag \\
& +V(x) \partial_\mu V(x)^\dagger V(x) U(x,x_0,A) V(x_0)^\dagger\,.
\end{align}
The first and third terms cancel while the second vanishes due to (\ref{Udiffeq}). On the other hand, simply rewriting (\ref{Udiffeq}) we know that
\be
\bigl[\partial_\mu - i A^V_\mu(x)\bigr] U(x,x_0,A^V) = 0\,.
\ee
Thus we find
\be
U(x,x_0,A^V) = V(x)U(x,x_0,A) V(x_0)^\dagger
\ee
since both sides obey the same first order differential equation with same boundary condition,
\be
U(x_0,x_0,A^V) = V(x_0)U(x_0,x_0,A) V(x_0)^\dagger = 1\!\!1\,.
\ee
It follows that if we take a closed loop $C$ and then take the trace, the resulting loop operator,
\be
\mathcal{L}(C)= \frac{1}{N} \tr \mathcal{P}\,{\rm exp} \, i \oint_C A\,,
\ee
is gauge invariant. The analysis holds for any gauge group and choice of representation, but here we are interested in $G=SU(N)$ or $G=U(N)$ and the trace taken in the fundamental representation, hence our normalisation so that $\mathcal{L}(C) = 1 + \ldots$.

Let us remind ourselves how to expand the Wilson loop operator to perform perturbative computations. We parametrise the curve $C$ via a map $x^\mu : [0,1] \rightarrow \mathbb{R}^{1,3}$ (with $x^\mu(0) = x^\mu(1)$ for a closed loop),
\begin{align}
\mathcal{L}(C) 
= \frac{1}{N}\tr\biggl[1\!\!1 + i\!\int_0^1 dt_1 \dot{x}^\mu(t_1) A_\mu\bigl(x(t_1)\bigr) - \!\int_0^1 dt_1 \dot{x}^\mu(t_1) A_\mu \bigl(x(t_1)\bigr) \int_0^{t_1} dt_2 \dot{x}^\nu(t_2) A_\nu \bigl(x(t_2)\bigr) + \ldots \biggr] \notag
\end{align}
If we compute the expectation value of this operator we find the term linear in $A$ gives nothing as there is nothing to contract with the field. The remaining powers in $A$ all contribute however,
\begin{align}
\langle \mathcal{L}(C) \rangle = 1 &- \frac{1}{N}\int_{t_1>t_2} \!\!\!\!\!\!\! \, \dot{x}_1^{\mu} \dot{x}_2^{\nu} \,\bigl[G_{\mu \nu, ab}^{1,2}  \tr (t_a t_b ) + \ldots\bigr] \notag \\
&- \frac{i}{N}\int_{t_1>t_2>t_3} \!\!\!\!\!\!\!\!\!\!\!\!\!\!  \dot{x}_1^{\mu} \dot{x}_2^{\nu} \dot{x}_3^{\rho}  \,\bigl[V^{1,2,3}_{\mu \nu \rho, abc} \tr (t_a t_b t_c) +\ldots \bigr]\notag \\
&+  \frac{1}{N}\int_{t_1>t_2>t_3>t_4} \!\!\!\!\!\!\!\!\!\!\!\!\!\!\!\!\!\!\!\!\!  \, \dot{x}_1^{\mu} \dot{x}_2^{\nu} \dot{x}_3^{\rho} \dot{x}_4^{\sigma} \Bigl[\bigl[ G^{1,2}_{\mu\nu,ab} G^{3,4}_{\rho \sigma,cd} + G^{1,3}_{\mu\rho,ac} G^{2,4}_{\nu \sigma,bd} + G^{1,4}_{\mu\sigma,ad} G^{2,3}_{\nu \rho,bc} \bigr] \tr (t_a t_b t_c t_d) +\ldots \Bigr] \notag \\
&+ \ldots\,.
\label{WLnonAbexp}
\end{align}
In perturbation theory the diagrams typically exhibit UV divergences. Here we employ dimensional regularisation with $d=4-2\epsilon$ with the Green's function and vertex given by
\begin{align}
G^{i,j}_{\mu\nu,ab} &=  -\frac{g_{\rm YM}^2 }{4 \pi^2} \eta_{\mu \nu}\delta_{ab}  \frac{(\pi \tilde{\mu}^2)^\epsilon \Gamma(1-\epsilon)}{[-(x_i - x_j)^2 + i \varepsilon]^{(1-\epsilon)}}\,, \notag \\
V^{i,j,k}_{\mu \nu \rho ,abc} &= \frac{1}{g_{\rm YM}^2} f_{a'b'c'} \tilde{\mu}^{-2\epsilon}  \int d^{4-2\epsilon} x_0 \mathcal{D}^{ijk}_{\mu' \nu' \rho'} [G_{\mu\mu',aa'}^{i,0} G_{\nu\nu',bb'}^{j,0}G_{\rho\rho',cc'}^{k,0}]\,, \notag \\
\mathcal{D}^{ijk}_{\mu \nu \rho} &= \eta_{\mu \nu} (\partial_{i \rho} - \partial_{j \rho}) + \eta_{\nu \rho} (\partial_{j \nu} - \partial_{k \nu}) + \eta_{\rho \mu} (\partial_{k \mu} - \partial_{i \mu})\,.
\label{GandV}
\end{align}
In (\ref{WLnonAbexp}) we have written the leading contributions in $g_{\rm YM}$ to each power in $A$ appearing in the path-ordered exponential up to $A^4$. Higher order terms in $g_{\rm YM}$ will come from the various interaction vertices in the Lagrangian, including the coupling of the Yang-Mills field to the matter content of the theory.

The colour factors from the various terms can be deduced from the relations
\begin{align}
[t_a,t_b] &= i f_{abc} t_c\,,  &&(t_a)_i^j (t_a)_k^l = \frac{1}{2}\biggl(\delta_i^l \delta_k^j - \frac{\alpha}{N} \delta_i^j \delta_k^l\biggr)\,, \notag \\
t_a t_b t_a &= \frac{1}{2}\biggl(\tr(t_b) 1\!\!1 -\frac{\alpha}{N}t_b\biggr)\,, &&\tr(t_a)\tr (t_a) = \frac{N}{2}(1-\alpha)\,,
\end{align}
where $\alpha=1$ for $G=SU(N)$ and $\alpha=0$ for $G=U(N)$.  
The various terms in (\ref{WLnonAbexp}) are then as follows,
\begin{align}
\delta_{ab} \tr (t_a t_b) &= \frac{1}{2}\delta_{ab} \delta_{ab} = \frac{1}{2}(N^2-\alpha) \equiv N C_F\,, \notag \\
\delta_{ab} \delta_{cd} \tr (t_a t_b t_c t_d) &= \frac{1}{4N}(N^2-\alpha)^2  \notag \\
\delta_{ac} \delta_{bd} \tr (t_a t_b t_c t_d) &= \frac{1}{4N}(N^2-2\alpha N^2+\alpha^2)  \notag \\
\delta_{ad} \delta_{bc} \tr (t_a t_b t_c t_d) &= \frac{1}{4N}(N^2-\alpha)^2   \notag \\
i f_{abc} \tr(t_a t_b t_c) &= \tr (t_a t_b [t_a,t_b]) = -\frac{1}{4}N(N^2-1) \,.
\end{align}
Note that the third term above (which contributes to the crossed diagram) is suppressed with respect to the second and the fourth at large $N$.

In the Abelian theory only even powers of $A$ contribute as we only have free propagators to use to contract the field insertions. Note we have a single generator $t$ but we will keep our convention that $\tr (t t) = C_F = \frac{1}{2}$. In the Abelian case we therefore find
\begin{align}
\langle \mathcal{L}(C) \rangle = 1 &- \frac{1}{2}\int_0^1 dt_1 \int_0^{t_1} dt_2\, \dot{x}_1^{\mu} \dot{x}_2^{\nu} \,G_{\mu \nu}^{1,2} \notag \\
&+  \frac{1}{4}\int_0^1 dt_1 \int_0^{t_1} dt_2  \int_0^{t_2} dt_3 \int_0^{t_3} dt_4 \, \dot{x}_1^{\mu} \dot{x}_2^{\nu} \dot{x}_3^{\rho} \dot{x}_4^{\sigma} \bigl[ G^{1,2}_{\mu\nu} G^{3,4}_{\rho \sigma} + G^{1,3}_{\mu\rho} G^{2,4}_{\nu \sigma} + G^{1,4}_{\mu\sigma} G^{2,3}_{\nu \rho} \bigr] + \ldots
\end{align}
with the Green's function $G^{ij}_{\mu \nu}$ given by the first line of (\ref{GandV}) with the factor of $\delta_{ab}$ removed.
The combinatorics of the expansion rearrange the series into the form
\be
\langle \mathcal{L}(C) \rangle = {\rm exp} \biggl\{ - \frac{1}{2}\int_0^1 dt_1 \int_0^{t_1} dt_2\, \dot{x}_1^{\mu} \dot{x}_2^{\nu} \,G_{\mu \nu}^{1,2} \biggr\} \equiv {\rm exp} \biggl\{ -\frac{1}{2}\int_{t_1>t_2} \!\!\!\!\!\!\! \dot{x}_1^{\mu} \dot{x}_2^{\nu} \,G_{\mu \nu}^{1,2} \biggr\} \,,
\label{WLAbexp}
\ee
i.e. the first non-zero correction actually exponentiates to give the whole result. 

\subsection{Light-like loops}
\label{App-lightlikeloop}

We review here the computations at one-loop (order $g^2$) for light-like Wilson loops presented in \cite{Drummond:2007aua,Brandhuber:2007yx} as well as the anomalous conformal Ward identity of \cite{Drummond:2007cf,Drummond:2007au}.

For a piecewise light-like loop with $x_{i+1,i}^2=0$ we parametrise each edge of the loop via
\be
x_i^\mu(t) = x_i^\mu (1-t) + x_{i+1}^\mu t = x_i^{\mu} + t x_{i+1,i}^\mu\,.
\ee
It is convenient to expand the Wilson loop in the 't Hooft coupling,
\be
g^2 = \frac{g_{\rm YM}^2 N}{16 \pi^2}\,.
\ee
The $O(g^2)$ contribution in (\ref{WLnonAbexp}) decomposes into a sum of contributions $I_{ij}$,
\be
\langle \mathcal{L}(C) \rangle = 1 + g^2 \cdot \frac{2 C_F}{N} \sum_{i\leq j} I_{ij} + O(g^4)\,.
\ee
There are three distinct types of contributions. Those with the propagator beginning and ending on the same segment vanish due to the light-like nature of the edges,
\be
I_{ii} = 2 (\pi \mu^2)^\epsilon \Gamma(1-\epsilon)  \int_0^1 dt_1 \int_0^{t_1} dt_2 \frac{x_{i+1,i} \cdot x_{i+1,i}}{[-(x_i(t_1) - x_i(t_2))^2 + i \varepsilon]^{(1-\epsilon)}} = 0\,.
\ee
Those diagrams with the propagator crossing between adjacent segments (including the first and the last) are divergent,
\begin{align}
I_{j-1,j} &= 2 (\pi \tilde{\mu}^2)^\epsilon \Gamma(1-\epsilon)  \int_0^1 dt_1 \int_0^1 dt_2 \frac{x_{j,j-1} \cdot x_{j+1,j}}{[-(x_{j-1}(t_1) - x_j(t_2))^2 + i \varepsilon]^{(1-\epsilon)}} \notag \\
&=- (\pi \tilde{\mu}^2)^\epsilon \Gamma(1-\epsilon)  \int_0^1 dt_1\int_0^1 dt_2 \frac{(- x_{j-1,j+1}^2)}{[-(x_{j-1,j}(1-t_1) - x_{j+1,j}t_2)^2 + i \varepsilon]^{(1-\epsilon)}} \notag \\
&= -  \bigl(\pi \tilde{\mu}^2  (-x_{j-1,j+1}^2)\bigr)^\epsilon \frac{\Gamma(1-\epsilon)}{\epsilon^2}\, \notag \\
&= -  \bigl(\mu^2  (-x_{j-1,j+1}^2)\bigr)^\epsilon \biggl[\frac{1}{\epsilon^2} + \frac{1}{2} \zeta_2 + O(\epsilon)\biggr]\,,
\label{Idiv}
\end{align}
where we have made a slight redefinition of the dimensional regularisation scale $\mu^2 = \tilde{\mu}^2 \pi e^\gamma$ with $\gamma$ the Euler gamma constant.
Finally, the diagrams with the propagator between two well-separated edges are finite and can be evaluated for $\epsilon=0$,
\begin{align}
I_{ij} =\quad &2 \int_0^1  dt_1 \int_0^1 dt_2 \frac{x_{i+1,i} \cdot x_{j+1,j}}{[-(x_{i}(t_1) - x_j(t_2))^2 + i \varepsilon]} \notag \\
= \quad&  
{\rm Li}_2\biggl[\frac{(x_{ij}^2-x_{i,j+1}^2)(x_{i,j+1}^2-x_{i+1,j+1}^2)}{x_{i,j+1}^2x_{i+1,j}^2-x_{ij}^2 x_{i+1,j+1}^2}\biggr] 
+ {\rm Li}_2\biggl[\frac{(x_{ij}^2-x_{i+1,j}^2)(x_{i+1,j}^2-x_{i+1,j+1}^2)}{x_{i,j+1}^2x_{i+1,j}^2-x_{ij}^2 x_{i+1,j+1}^2}\biggr] \notag \\
 - &{\rm Li}_2 \biggl[ \frac{(x_{ij}^2-x_{i,j+1}^2)(x_{ij}^2-x_{i+1,j}^2)}{x_{i,j+1}^2x_{i+1,j}^2-x_{ij}^2 x_{i+1,j+1}^2} \biggr]
-{\rm Li}_2 \biggl[ \frac{(x_{i,j+1}^2-x_{i+1,j+1}^2)(x_{i+1,j}^2-x_{i+1,j+1}^2)}{x_{i,j+1}^2x_{i+1,j}^2-x_{ij}^2 x_{i+1,j+1}^2} \biggr]  \,.
\label{Iij}
\end{align}
Recall the factorised form of the Wilson loop given in (\ref{Wnfactors}),
\be
W_n = \langle \mathcal{L}(C) \rangle = \Bigl[\prod_{i=1}^n D_i \Bigr] F_n R_n\,.
\ee
From (\ref{Idiv}) we have the first contribution to the divergent factor
\begin{align}
\label{Di}
D_i ={\rm exp}  \Biggl\{ -\frac{1}{4}\sum_{l=1}^\infty  g^{2l} (-\mu^2 x_{i-1,i+1}^2)^{l\epsilon} \biggl[ \frac{\Gamma_{\rm cusp}^{(l)}}{(l \epsilon)^2} +  \frac{\Gamma_{\rm sub}^{(l)}}{l \epsilon}\biggr]\Biggr\}\,.
\end{align}
We see that we have
\be
\label{GammasAb}
\Gamma_{\rm cusp}^{(1)} = 4 \cdot \frac{2 C_F}{N}\,, \qquad \Gamma_{\rm sub}^{(1)} = 0\,.
\ee
Note that for $G=U(N)$, and hence in both the large $N$ limit and the Abelian case $G=U(1)$ we have $\frac{2 C_F}{N} = 1$. In the Abelian case, all higher coefficients $\Gamma^{(l)}_{\rm cusp}$ and $\Gamma^{(l)}_{\rm sub}$ are vanishing and we have simply
\be
\Gamma_{\rm cusp} = 4 g^2\,, \qquad \bigl(G= U(1)\bigr)\,.
\ee
In the non-Abelian case, both $\Gamma_{\rm cusp} = \sum_l g^{2l} \Gamma_{\rm cusp}^{(l)}$ and $\Gamma_{\rm sub} = \sum_l g^{2l} \Gamma_{\rm sub}^{(l)}$ are non-trivial functions of the coupling $g^2$ and $N$. In the large $N$ limit they are given by
\begin{align}
\Gamma_{\rm cusp}(g^2) &= 4g^2 - \frac{4\pi^2}{3} g^4 + O(g^6) \,, \notag \\
\Gamma_{\rm sub}(g^2) &=  - 28 \zeta_3 g^4 + O(g^6)\,.
\label{Gammas}
\end{align}

Writing $f_n$ for the leading correction to $F_n$, we have
\begin{equation}
    F_n = 1 + g^2 \cdot \frac{2 C_F}{N} f_n + O(g^4)\,, \qquad f_n = \sum_{\{i,j\}} I_{ij} - \frac{n}{2} \zeta_2\,, \label{Fn1loop}
\end{equation}
where the sum is over all non-adjacent pairs of edges $\{i,j\}$. 

By inspecting the total derivative of $I_{ij}$, we may verify that the function $f_n$ obeys the anomalous conformal Ward identity of \cite{Drummond:2007cf,Drummond:2007au}. Recall that the total derivative of the dilogarithm can be written as $d \,{\rm Li}_2(x) = -\log(1-x) d \log x$. Applying this we find for the first term in $d I_{i,j}$:
\begin{align}
 \log& \biggl[\frac{(x_{i,j+1}^2x_{i+1,j}^2-x_{ij}^2 x_{i+1,j+1}^2)-(x_{ij}^2-x_{i,j+1}^2)(x_{i,j+1}^2-x_{i+1,j+1}^2)}{x_{i,j+1}^2x_{i+1,j}^2-x_{ij}^2 x_{i+1,j+1}^2}\biggr] \notag \\
\times&\bigl[ d \log (x_{i,j+1}^2x_{i+1,j}^2-x_{ij}^2 x_{i+1,j+1}^2) - d \log (x_{ij}^2-x_{i,j+1}^2) - d \log (x_{i,j+1}^2-x_{i+1,j+1}^2)   \bigr]\,.
\label{dIij1}
\end{align}
The numerator of the argument of the logarithm factorises so we find that the above is equal to
\begin{align}
\bigl[ \log x_{i,j+1}^2 + \log (x_{i+1,j}^2 + x_{i,j+1}^2 - x_{ij}^2 - x_{i+1,j+1}^2) - \log(x_{i,j+1}^2x_{i+1,j}^2-x_{ij}^2 x_{i+1,j+1}^2)\bigr] \notag \\
\times \bigl[d \log (x_{i,j+1}^2x_{i+1,j}^2-x_{ij}^2 x_{i+1,j+1}^2) -d \log (x_{ij}^2-x_{i,j+1}^2) - d \log (x_{i,j+1}^2-x_{i+1,j+1}^2)  \bigr]\,.
\end{align}
Similar factorisation occurs for the other three terms in $d I_{ij}$ and this leads to cancellation among the four terms. In the end only contributions of the form of the first term in the first line above remain and we find the four terms contribute the following to $d I_{ij}$:
\begin{align}
d I_{ij}&= \notag \\
& \log x_{i,j}^2   \bigl[ d \log (x_{ij}^2-x_{i,j+1}^2) + d \log (x_{ij}^2-x_{i+1,j}^2) - d \log (x_{i,j+1}^2x_{i+1,j}^2-x_{ij}^2 x_{i+1,j+1}^2) \bigr] \notag \\
+& \log x_{i+1,j+1}^2   \bigl[ d \log (x_{i,j+1}^2-x_{i+1,j+1}^2) + d \log (x_{i+1,j}^2-x_{i+1,j+1}^2) - d \log (x_{i,j+1}^2x_{i+1,j}^2-x_{ij}^2 x_{i+1,j+1}^2) \bigr] \notag \\
-&\log x_{i,j+1}^2   \bigl[ d \log (x_{ij}^2-x_{i,j+1}^2) + d \log (x_{i,j+1}^2-x_{i+1,j+1}^2) - d \log (x_{i,j+1}^2x_{i+1,j}^2-x_{ij}^2 x_{i+1,j+1}^2) \bigr] \notag \\
-& \log x_{i+1,j}^2   \bigl[ d \log (x_{ij}^2-x_{i+1,j}^2) + d \log (x_{i+1,j}^2-x_{i+1,j+1}^2) - d \log (x_{i,j+1}^2x_{i+1,j}^2-x_{ij}^2 x_{i+1,j+1}^2) \bigr]
\label{dIij}
\end{align}
We may write the sum of non-adjacent pairs $\{i,j\}$ as $\frac{1}{2}\sum_i \sum_{j=i+2}^{i-2}$ (or the equivalent swapping $i$ and $j$).
Note that in (\ref{dIij}), even with $i,j$ separated edges, there appear to be some logarithms which diverge, e.g. $\log x_{i,j+1}^2$ for $j=i-2$. These terms all have vanishing coefficients however.

Now note that in the sum over $i$ and $j$ many terms cancel. For example the second term in the first line cancels the second term in the third line after summation over $j$, except for a boundary term. Similarly the first term in the the first line cancels the second in the fourth line, the first in the second line cancels the first in the fourth and the second in the second cancels the first in the third, again all up to boundary terms. The third terms in each line all combine with the same $d\log$. In total we then have
\begin{align}
df_n &= \sum_{\{i,j\}} d I_{ij} \notag \\
&= \sum_i \log x_{i-1,i+1}^2 d \log x_{i-1,i+1}^2 + \sum_{\{i,j\}} \log \frac{x_{i,j+1}^2 x_{i+1,j}^2}{x_{ij}^2 x_{i+1,j+1}^2} d \log (x_{i,j+1}^2 x_{i+1,j}^2 - x_{ij}^2 x_{i+1,j+1}^2)\,.
\label{dfn}
\end{align}
If we apply the generator of special conformal transformations,
\be
K^\mu = \sum_i (2 x_i^\mu x_i \cdot \partial - x_i^2 \partial_i^\mu)\,,
\ee
we find, again after a similar use of telescoping for the second term,
\be
K^\mu f_n = 2 \sum_{i=1}^n (x_{i-1}^\mu - 2 x_i^\mu + x_{i+1}^\mu ) \log x_{i-1,i+1}^2\,.
\ee 
If we define $F_n$ consistently with (\ref{Fn1loop}) as follows,
\be
\label{Fn}
F_n = {\rm exp} \, \biggl\{ \frac{1}{4} \Gamma_{\rm cusp}(g,N) f_n \biggr\}\,,
\ee
we have the anomalous conformal Ward identity,
\be
\label{CWI}
K^\mu \log F_n = \frac{1}{2} \Gamma_{\rm cusp} \sum_{i=1}^n (x_{i-1}^\mu - 2 x_i^\mu + x_{i+1}^\mu ) \log x_{i-1,i+1}^2\,.
\ee

The remaining factor $R_n$ in (\ref{Wnfactors}) is finite and conformally invariant and since the order $g^2$ terms are all accounted for by $D_i$ and $F_n$, receives non-trivial corrections from order $g^4$ onwards,
\be
R_n = 1 + O(g^4)\,.
\ee
In the Abelian theory, since the full result is obtained by exponentiating the order $g^2$ term, we have $R_n=1$ simply.

\subsection{Details on multiple light-like loops}
\label{App-multi-WL}

If we consider multiple loops, we parametrise each one via $x_r^\mu : [0,1] \rightarrow \mathbb{R}^{1,3}$, with $1 \leq r \leq m$ for $m$ loops. Let us consider the order $g^2$ contribution to the connected part of the correlator for $m=2$. We have
\begin{align}
    \langle \mathcal{L}(C_1) \mathcal{L}(C_2) \rangle^{\rm conn} = -\frac{1}{N^2} \tr(t_{a_1}) \tr(t_{a_2}) \int_0^1 dt_1 \dot{x}^{\mu_1}(t_1)  \int_0^1 dt_2 \dot{x}^{\mu_2}(t_2) G^{1,2}_{\mu_1 \mu_2,a_1 a_2} + O(g^4) \notag \,.
\end{align}
For piecewise light-like contours, all diagrams are finite and can be evaluated for $d=4$,
\begin{align}    
 \langle \mathcal{L}(C_1) \mathcal{L}(C_2) \rangle^{\rm conn} =  g^2 \frac{1-\alpha}{N^2} f_{n_1,n_2} + O(g^4)\,, \qquad f_{n_1,n_2} = \sum_{i,j} I_{ij}\,,
\end{align}
where $I_{ij}$ is given by the same formula as (\ref{Iij}) but where $x_i$ lies on $C_1$ while $x_j$ lies on $C_2$ and the sum runs over all possible pairs $(i,j)$. As is clear from the colour factor, this contribution vanishes for $G=SU(N)$ ($\alpha=1$) and is subleading in $N$ for $G=U(N)$ ($\alpha=0$).  

By inspecting the total derivative of $f_{n_1,n_2}$, it is straightforward to see that the function $G_{n_1,n_2}$ is indeed conformally invariant as expected. Following very similar steps as outlined in equations (\ref{dIij1}) to (\ref{dIij}), we find a form for the total derivative $dG_{n_1,n_2}$ very similar to that for the anomalous conformal part given in eq. (\ref{dfn})
\be
d f_{n_1,n_2} = \sum _{i,j} d I_{ij} =  \sum _{i,j} \log \frac{x_{i,j+1}^2 x_{i+1,j}^2}{x_{ij}^2 x_{i+1,j+1}^2} d \log (x_{i,j+1}^2x_{i+1,j}^2-x_{ij}^2 x_{i+1,j+1}^2) \,.
\label{SumdI}
\ee
This is almost manifestly conformally invariant. The argument of the logarithm is a conformal cross-ratio while the argument of the $d \log$ is conformally covariant. Note that (\ref{SumdI}) implies that if we apply the special conformal generator $K^\mu$ to the sum over all terms we find
\be
K^{\mu} f_{n_1,n_2} = \sum_{i,j} \log \frac{x_{i,j+1}^2 x_{i+1,j}^2}{x_{ij}^2 x_{i+1,j+1}^2} (x_i^\mu + x_{i+1}^\mu + x_j^\mu + x_{j+1}^\mu) = 0
\ee
where the cancellation again happens for each $x^\mu$ by telescoping in the sum. So we indeed find conformal invariance. 

Another way to see conformal invariance is to make use of twistor variables. Let us label them as $\{Z_1,\ldots,Z_{n_1}\}$ for the first loop and $\{\tilde{Z}_1,\ldots,\tilde{Z}_{n_2}\}$ for the second. Then we have
\be
x_{ij}^2 = \frac{\langle i-1\, i\, \tilde{\jmath}-1\, \tilde{\jmath}\rangle}{\langle i-1\, i \rangle \langle \tilde{\jmath}-1\, \tilde{\jmath}\rangle}.
\ee
We then have
\begin{align}
x_{i,j+1}^2x_{i+1,j}^2-x_{ij}^2 x_{i+1,j+1}^2 & = \frac{\langle i-1\, i \,\tilde{\jmath}\,\tilde{\jmath}+1 \rangle \langle i \,i+1\, \tilde{\jmath}-1 \, \tilde{\jmath} \rangle - \langle i-1 \, i \, \tilde{\jmath}-1\, \tilde{\jmath} \rangle \langle i\, i+1 \, \tilde{\jmath}\, \tilde{\jmath}+1\rangle}{\langle i-1\,i\rangle \langle i\, i+1\rangle \langle \tilde{\jmath}-1\, \tilde{\jmath} \rangle \langle \tilde{\jmath} \, \tilde{\jmath}+1 \rangle} \notag \\
&= -\frac{\langle i-1\, i \, i+1 \tilde{\jmath} \rangle \langle i \, \tilde{\jmath}-1 \, \tilde{\jmath} \, \tilde{\jmath}+1\rangle}{\langle i-1\,i\rangle \langle i\, i+1\rangle \langle \tilde{\jmath}-1\, \tilde{\jmath} \rangle \langle \tilde{\jmath} \, \tilde{\jmath}+1 \rangle}\,,
\end{align}
making use of a Pl\"ucker relation.
Hence
\be
d f_{n_1,n_2} = \sum _{i,j} \log \frac{x_{i,j+1}^2 x_{i+1,j}^2}{x_{ij}^2 x_{i+1,j+1}^2} d \log \bigl[ \langle i-1\, i \, i+1 \tilde{\jmath} \rangle \langle i \, \tilde{\jmath}-1 \, \tilde{\jmath} \, \tilde{\jmath}+1\rangle\bigr] \,,
\label{twistordW12}
\ee
with the two-brackets $\langle i-1\,i \rangle$ etc. cancelling by summing over $i$ or $j$. Thus the expression (\ref{twistordW12}) is manifestly conformally invariant, though not manifestly homogeneous.

To write an expression for $f_{n_1,n_2}$ in a manifestly conformally invariant form we can define conformal cross-ratios via
\be
u_{i,j,k,l} = \frac{x_{ij}^2 x_{kl}^2}{x_{il}^2 x_{kj}^2}\,, \qquad v_{ij} = u_{i,j,i+1,j+1}\,.
\ee
The cross-ratios $v_{ij}$ are not all multiplicatively independent due to the relations
\be
\prod_i v_{ij} = 1\, \qquad \prod_j v_{ij} = 1\,.
\label{vrels-app}
\ee
Then we can write
\begin{align}
d f_{n_1,n_2} &= - \sum _{i,j} \bigl[ \log v_{ij} d \log (1 - v_{ij}) + \log v_{ij} d \log ( x_{i,j+1}^2 x_{i+1,j}^2)\bigr] \notag \\
&= \sum_{i,j} d\, {\rm Li}_2(1-v_{ij}) -  \sum_{i,j}\log v_{ij} d \log ( x_{i,j+1}^2 x_{i+1,j}^2)\,,
\label{dGLi}
\end{align}
In the second term of (\ref{dGLi}) we can use these relations to eliminate $v_{n_1,j}$ and $v_{i,n_2}$ in the first slot. For $i\leq n_1 -1$ and $j \leq n_2 -1$ we can also eliminate $x_{ij}^2$ in favour of $v_{ij}$ and $x_{kl}$ with $k>i$ or $l>j$ in the second slot. Doing so, all remaining $x_{kl}$ cancel and we are left with a manifestly conformally invariant expression which can be written as
\begin{align}
d f_{n_1,n_2} &=  \sum_{i,j}^{n_1,n_2} d\, {\rm Li}_2(1-v_{ij}) + \sum_{\substack{i<n_1\\j<n_2}}\log v_{ij}  \Biggl[\sum_{\substack{1\leq k \leq i\\ j \leq l < n_2}} d \log v_{kl} + \sum_{\substack{i \leq k < n_1\\ 1 \leq l \leq j}} d \log v_{kl}\Biggr] \notag \\
&=  \sum_{i,j}^{n_1,n_2} d\, {\rm Li}_2(1-v_{ij}) + \sum_{\substack{k \leq i < n_1\\ j \leq l <n_2}} d \bigl[ \log v_{ij}  \log v_{kl} \bigr]
\end{align}
Note that the second term in this expression is not manifestly invariant under cyclic shifts of either loop. Up to a possible constant term, this shows that
\be
f_{n_1,n_2} =  \sum_{i,j}^{n_1,n_2} \, {\rm Li}_2(1-v_{ij}) +  \sum_{\substack{k \leq i < n_1\\ j \leq l <n_2}}  \log v_{ij}  \log v_{kl} \,.
\label{Gexpr1}
\ee
Regarding $f_{n_1,n_2}$ as a function of the square distances $x_{ij}^2$, we may consider kinematics where $x_{ij}^2 = r^2 + \epsilon \Delta r^2_{ij}$ and take the limit $\epsilon \rightarrow 0$. We see that the finite diagram $I_{ij}$ given in eq. (\ref{Iijsec2}) vanishes, as does the expression (\ref{Gexpr1}) and we conclude that (\ref{Gexpr1}) is correct as it stands with no constant term added.

Finally, we may notice that we can include the terms where $i=n_1$ or $l=n_2$ in the sum at no cost due to the relations (\ref{vrels}) and hence we have
\be
f_{n_1,n_2} =  \sum_{i,j}^{n_1,n_2} \, {\rm Li}_2(1-v_{ij}) +  \sum_{\substack{k \leq i \\ j \leq l }}  \log v_{ij}  \log v_{kl} \,,
\label{Gexpr2-app}
\ee
justifying (\ref{Gexpr2})\,.

Note that under a cyclic shift $i \rightarrow i+1$ on the first loop the second term in (\ref{Gexpr1}) becomes
\be
\text{cyc}_1 : \quad   \sum_{\substack{k \leq i < n_1\\ j \leq l <n_2}}  \log v_{ij}  \log v_{kl} \longrightarrow \sum_{\substack{2\leq k \leq i \leq n_1\\ j \leq l <n_2}}  \log v_{ij}  \log v_{kl} =  \sum_{\substack{k \leq i \\ j \leq l }}  \log v_{ij}  \log v_{kl}\,,
\ee
where the final equality with the second term of (\ref{Gexpr2}) again follows from the relations (\ref{vrels}). A check of cyclic invariance for the second loop follows similarly.

\section{Details on N\(^3\)MHV Integrals}
\label{N3MHVintegrals}

Here we present some explicit details on the computation of some of the integrals which arise for tree-level N\(^3\)MHV Wilson loop correlators. In each instance it is also possible to arrive at the final, R-invariant form by inspection by using the Feynman rules outlined in Section \ref{Feynman}.

\subsection{Example: Two Double Insertions}

For instance, for \(\mathcal{W}_{n_1,n_2}^{(3), \textrm{conn}}\), one of the integrals which contributes to an N\(^3\)MHV diagram with two double insertions on one twistor line is (here we recall the notation introduced in Eq. \ref{colourStrippedIntegral})
\be
I\bigl( \Delta_*^{i_1 j_1}(u,s_1), \Delta_*^{j_1 i_2}(s_2,t_1), \Delta_*^{i_2 j_2}(t_2,v)\bigr).
\ee

Here, we have explicitly that 
\begin{align}
& I\bigl( \Delta_*^{i_1 j_1}(u,s_1), \Delta_*^{j_1 i_2}(s_2,t_1), \Delta_*^{i_2 j_2}(t_2,v)\bigr) \notag \\
& = \int \frac{du}{u} \frac{dv}{v} \frac{ds_1}{s_1} \frac{ds_2}{s_1-s_2}\frac{dt_1}{t_1}\frac{dt_2}{t_1-t_2}  \Delta_*^{i_1 j_1}(u,s_1) \Delta_*^{j_1 i_2}(s_2,t_1) \Delta_*^{i_2 j_2}(t_2,v)  \notag  \\
 &= \int \frac{du}{u} \frac{dv}{v} \frac{ds_1}{s_1} \frac{ds_2}{s_1-s_2}\frac{dt_1}{t_1}\frac{dt_2}{t_1-t_2} \int \frac{D^2a}{a_1a_2a_3} \frac{D^2b}{b_1b_2b_3} \frac{D^2c}{c_1c_2c_3} \notag \\
 & \times \bar{\delta}^{4|4}(a_1Z_* + a_2 u\mathcal{Z}_{i_1-1} + a_2\mathcal{Z}_{i_1} + a_3 s_1 \mathcal{Z}_{j_1-1} + a_3 \mathcal{Z}_{j_1})  \notag \\
 & \times \bar{\delta}^{4|4}(b_1Z_* + b_2 s_2\mathcal{Z}_{j_1-1} + b_2\mathcal{Z}_{j_1} + b_3 t_1 \mathcal{Z}_{i_2-1} + b_3 \mathcal{Z}_{i_2})  \notag \\
  & \times \bar{\delta}^{4|4}(c_1Z_* + c_2 t_2 \mathcal{Z}_{i_2-1} + c_2\mathcal{Z}_{i_2} + c_3 v \mathcal{Z}_{j_2-1} + c_3 \mathcal{Z}_{j_2}).  
\end{align} 

Changing variables to \(y_1=u\), \(y_2=v\), \(y_3=s_1\), \(y_4=s_1-s_2\), \(y_5=t_1\), and \(y_6=t_1-t_2\), we have
\begin{align}
& I\bigl( \Delta_*^{i_1 j_1}(u,s_1), \Delta_*^{j_1 i_2}(s_2,t_1), \Delta_*^{i_2 j_2}(t_2,v)\bigr) \notag \\
 &= \int \prod_{i=1}^6 \frac{dy_i}{y_i} \frac{D^2a}{a_1a_2a_3} \frac{D^2b}{b_1b_2b_3} \frac{D^2c}{c_1c_2c_3} \notag \\
 & \times \bar{\delta}^{4|4}(a_1Z_* + a_2 y_1\mathcal{Z}_{i_1-1} + a_2\mathcal{Z}_{i_1} + a_3 y_3 \mathcal{Z}_{j_1-1} + a_3 \mathcal{Z}_{j_1})  \notag \\
 & \times \bar{\delta}^{4|4}(b_1Z_* + b_2 (y_3-y_4)\mathcal{Z}_{j_1-1} + b_2\mathcal{Z}_{j_1} + b_3 y_5 \mathcal{Z}_{i_2-1} + b_3 \mathcal{Z}_{i_2})  \notag \\
  & \times \bar{\delta}^{4|4}(c_1Z_* + c_2 (y_5-y_6) \mathcal{Z}_{i_2-1} + c_2\mathcal{Z}_{i_2} + c_3 y_2 \mathcal{Z}_{j_2-1} + c_3 \mathcal{Z}_{j_2}). 
\end{align} 
Rescaling \(y_1 \to \frac{y_1}{a_2}\) and \(y_3 \to \frac{y_3}{a_3}\),
\begin{align}
& I\bigl( \Delta_*^{i_1 j_1}(u,s_1), \Delta_*^{j_1 i_2}(s_2,t_1), \Delta_*^{i_2 j_2}(t_2,v)\bigr) \notag \\
 &= \int \prod_{i=1}^6 \frac{dy_i}{y_i} \frac{D^2a}{a_1a_2a_3} \frac{D^2b}{b_1b_2b_3} \frac{D^2c}{c_1c_2c_3} \notag \\
 & \times \bar{\delta}^{4|4}(a_1Z_* + y_1\mathcal{Z}_{i_1-1} + a_2\mathcal{Z}_{i_1} + y_3 \mathcal{Z}_{j_1-1} + a_3 \mathcal{Z}_{j_1})  \notag \\
 & \times \bar{\delta}^{4|4}(b_1Z_* + b_2 (\frac{y_3}{a_3}-y_4)\mathcal{Z}_{j_1-1} + b_2\mathcal{Z}_{j_1} + b_3 y_5 \mathcal{Z}_{i_2-1} + b_3 \mathcal{Z}_{i_2})  \notag \\
  & \times \bar{\delta}^{4|4}(c_1Z_* + c_2 (y_5-y_6) \mathcal{Z}_{i_2-1} + c_2\mathcal{Z}_{i_2} + c_3 y_2 \mathcal{Z}_{j_2-1} + c_3 \mathcal{Z}_{j_2}).  
\end{align} 
The first \(\bar{\delta^{4|4}}\) clearly gives an R-invariant after integrating out the bosonic part, upon which we find
\begin{align}
& I\bigl( \Delta_*^{i_1 j_1}(u,s_1), \Delta_*^{j_1 i_2}(s_2,t_1), \Delta_*^{i_2 j_2}(t_2,v)\bigr) \notag \\
 &= [*,i_1-1,i_1,j_1-1,j_1] \int \frac{dy_2dy_4dy_5dy_6}{y_2y_4y_5y_6} \frac{D^2b}{b_1b_2b_3} \frac{D^2c}{c_1c_2c_3} \notag \\
 & \times \bar{\delta}^{4|4}(b_1Z_* + b_2 (-\frac{\langle *\, i_1-1\, i_1\, j_1 \rangle}{\langle *\, i_1-1\, i_1\, j_1-1 \rangle}-y_4)\mathcal{Z}_{j_1-1} + b_2\mathcal{Z}_{j_1} + b_3 y_5 \mathcal{Z}_{i_2-1} + b_3 \mathcal{Z}_{i_2})  \notag \\
  & \times \bar{\delta}^{4|4}(c_1Z_* + c_2 (y_5-y_6) \mathcal{Z}_{i_2-1} + c_2\mathcal{Z}_{i_2} + c_3 y_2 \mathcal{Z}_{j_2-1} + c_3 \mathcal{Z}_{j_2}). 
\end{align} 
Then, rescaling \(y_4 \to -\frac{y_4}{b_2}\), then \(b_2 \to \langle *\, i_1-1\, i_1\, j_1-1 \rangle b_2 \) as well as \(y_5 \to \frac{y_5}{b_3}\),
\begin{align}
& I\bigl( \Delta_*^{i_1 j_1}(u,s_1), \Delta_*^{j_1 i_2}(s_2,t_1), \Delta_*^{i_2 j_2}(t_2,v)\bigr) \notag \\
 &= [*,i_1-1,i_1,j_1-1,j_1] \int \frac{dy_2dy_4dy_5dy_6}{y_2y_4y_5y_6} \frac{D^2b}{b_1b_2b_3} \frac{D^2c}{c_1c_2c_3} \notag \\
 & \times \bar{\delta}^{4|4}(b_1Z_* + b_2 \widehat{\mathcal{Z}_{j_1,i_1}}+ y_4Z_{j_1-1} + y_5 \mathcal{Z}_{i_2-1} + b_3 \mathcal{Z}_{i_2})  \notag \\
  & \times \bar{\delta}^{4|4}(c_1Z_* + c_2 (\frac{y_5}{b_3}-y_6) \mathcal{Z}_{i_2-1} + c_2\mathcal{Z}_{i_2} + c_3 y_2 \mathcal{Z}_{j_2-1} + c_3 \mathcal{Z}_{j_2}). 
\end{align} 
Integrating out the bosonic part of the first \(\bar{\delta}\), we get another R-invariant, though this time with a shift:
\begin{align}
& I\bigl( \Delta_*^{i_1 j_1}(u,s_1), \Delta_*^{j_1 i_2}(s_2,t_1), \Delta_*^{i_2 j_2}(t_2,v)\bigr) \notag \\
 &= [*,i_1-1,i_1,j_1-1,j_1] [*,\widehat{j_1}_{i_1},j_1-1,i_2-1,i_2] \int \frac{dy_2dy_6}{y_2y_6}  \frac{D^2c}{c_1c_2c_3} \notag \\
  & \times \bar{\delta}^{4|4}(c_1Z_* + c_2 (-\frac{\langle *\, \widehat{j_1}_{i_1}\, j_1-1\, i_2 \rangle}{\langle *\, \widehat{j_1}_{i_1}\, j_1-1\, i_2-1 \rangle}-y_6) \mathcal{Z}_{i_2-1} + c_2\mathcal{Z}_{i_2} + c_3 y_2 \mathcal{Z}_{j_2-1} + c_3 \mathcal{Z}_{j_2}). 
\end{align} 
Rescaling \(y_6 \to -\frac{y_6}{c_2}\), then \(c_2 \to \langle *\, \widehat{j_1}_{i_1}\, j_1-1\, i_2-1 \rangle c_2\) and \(y_2 \to \frac{y_2}{c_3}\), we have
\begin{align}
& I\bigl( \Delta_*^{i_1 j_1}(u,s_1), \Delta_*^{j_1 i_2}(s_2,t_1), \Delta_*^{i_2 j_2}(t_2,v)\bigr) \notag \\
 &= [*,i_1-1,i_1,j_1-1,j_1] [*,\widehat{j_1}_{i_1},j_1-1,i_2-1,i_2] \int \frac{dy_2dy_6}{y_2y_6}  \frac{D^2c}{c_1c_2c_3} \notag \\
  & \times \bar{\delta}^{4|4}(c_1Z_* + c_2\left(\langle *\, \widehat{j_1}_{i_1}\, j_1-1\, i_2-1 \rangle \mathcal{Z}_{i_2} - \langle *\, \widehat{j_1}_{i_1}\, j_1-1\, i_2 \rangle Z_{i_2-1}\right) + y_6\mathcal{Z}_{i_2-1}  + y_2 \mathcal{Z}_{j_2-1} + c_3 \mathcal{Z}_{j_2}). 
\end{align} 
Let us inspect the coefficient of \(c_2\) more closely. Indeed, substituting the formula for the shifted argument it becomes
\be
\langle *\, i_1-1\, i_1\, j_1-1\rangle \widehat{\mathcal{Z}_{i_2}}_{j_1} 
\ee
so that overall we have
\begin{align}
  &I\bigl( \Delta_*^{i_1 j_1}(u,s_1), \Delta_*^{j_1 i_2}(s_2,t_1), \Delta_*^{i_2 j_2}(t_2,v)\bigr) \notag \\
  &= [*,i_1-1,i_1,j_1-1,j_1] [*,i_2-1,i_2,j_1-1,\widehat{j_1}_{i_1}] [*, j_2-1,j_2,i_2-1,\widehat{i_2}_{j_1}]. 
\end{align} 

\subsection{Example: Triple Insertion}
Next, let us consider an example of a diagram with a triple insertion on one twistor line, e.g. 
\be
I\bigl( \Delta_*^{i,j_1}(s_1,u), \Delta_*^{i,j_2}(s_2,v), \Delta_*^{i, j_3}(s_3,w)\bigr).
\ee

Following the same procedure as in the last example,
\begin{align}
& I\bigl( \Delta_*^{i,j_1}(s_1,u), \Delta_*^{i,j_2}(s_2,v), \Delta_*^{i, j_3}(s_3,w)\bigr) \notag \\
& =\int \frac{ds_1}{s_1} \frac{ds_2}{s_1-s_2} \frac{ds_3}{s_2-s_3} \frac{du}{u}\frac{dv}{v}\frac{dw}{w}  \Delta_*^{i,j_1}(s_1,u) \Delta_*^{i,j_2}(s_2,v)\Delta_*^{i, j_3}(s_3,w)  \notag  \\
 &= \int \frac{ds_1}{s_1} \frac{ds_2}{s_1-s_2} \frac{ds_3}{s_2-s_3} \frac{du}{u}\frac{dv}{v}\frac{dw}{w}  \int \frac{D^2a}{a_1a_2a_3} \frac{D^2b}{b_1b_2b_3} \frac{D^2c}{c_1c_2c_3} \notag \\
 & \times \bar{\delta}^{4|4}(a_1Z_* + a_2 s_1 \mathcal{Z}_{i-1} + a_2\mathcal{Z}_{i} + a_3 u \mathcal{Z}_{j_1-1} + a_3 \mathcal{Z}_{j_1})  \notag \\
 & \times \bar{\delta}^{4|4}(b_1Z_* + b_2 s_2\mathcal{Z}_{i-1} + b_2\mathcal{Z}_{i} + b_3 v \mathcal{Z}_{j_2-1} + b_3 \mathcal{Z}_{j_2})  \notag \\
  & \times \bar{\delta}^{4|4}(c_1Z_* + c_2 s_3 \mathcal{Z}_{i-1} + c_2\mathcal{Z}_{i} + c_3 w \mathcal{Z}_{j_3-1} + c_3 \mathcal{Z}_{j_3}). 
\end{align} 

Changing variables to \(y_1=s_1\), \(y_2=s_1-s_2\), \(y_3=s_2-s_3\), \(y_4=u\), \(y_5=v\), and \(y_6 = w\),
\begin{align}
 &I\bigl( \Delta_*^{i,j_1}(s_1,u), \Delta_*^{i,j_2}(s_2,v), \Delta_*^{i, j_3}(s_3,w)\bigr) \notag \\
 &= \int \prod_{i=6}^6 \frac{dy_i}{y_i}  \int \frac{D^2a}{a_1a_2a_3} \frac{D^2b}{b_1b_2b_3} \frac{D^2c}{c_1c_2c_3} \notag \\
 & \times \bar{\delta}^{4|4}(a_1Z_* + a_2 y_1 \mathcal{Z}_{i-1} + a_2\mathcal{Z}_{i} + a_3 y_4 \mathcal{Z}_{j_1-1} + a_3 \mathcal{Z}_{j_1})  \notag \\
 & \times \bar{\delta}^{4|4}(b_1Z_* + b_2 (y_1-y_2)\mathcal{Z}_{i-1} + b_2\mathcal{Z}_{i} + b_3 y_5 \mathcal{Z}_{j_2-1} + b_3 \mathcal{Z}_{j_2})  \notag \\
  & \times \bar{\delta}^{4|4}(c_1Z_* + c_2 (y_1-y_2-y_3) \mathcal{Z}_{i-1} + c_2\mathcal{Z}_{i} + c_3 y_6 \mathcal{Z}_{j_3-1} + c_3 \mathcal{Z}_{j_3}).  
\end{align} 
Rescaling \(y_1 \to \frac{y_1}{a_2}\) and \(y_4 \to \frac{y_4}{a_3}\),
\begin{align}
&I\bigl( \Delta_*^{i,j_1}(s_1,u), \Delta_*^{i,j_2}(s_2,v), \Delta_*^{i, j_3}(s_3,w)\bigr) \notag \\
 &= \int \prod_{i=6}^6 \frac{dy_i}{y_i}  \int \frac{D^2a}{a_1a_2a_3} \frac{D^2b}{b_1b_2b_3} \frac{D^2c}{c_1c_2c_3} \notag \\
 & \times \bar{\delta}^{4|4}(a_1Z_* + y_1 \mathcal{Z}_{i-1} + a_2\mathcal{Z}_{i} + y_4 \mathcal{Z}_{j_1-1} + a_3 \mathcal{Z}_{j_1})  \notag \\
 & \times \bar{\delta}^{4|4}(b_1Z_* + b_2 (\frac{y_1}{a_2}-y_2)\mathcal{Z}_{i-1} + b_2\mathcal{Z}_{i} + b_3 y_5 \mathcal{Z}_{j_2-1} + b_3 \mathcal{Z}_{j_2})  \notag \\
  & \times \bar{\delta}^{4|4}(c_1Z_* + c_2 (\frac{y_1}{a_2}-y_2-y_3) \mathcal{Z}_{i-1} + c_2\mathcal{Z}_{i} + c_3 y_6 \mathcal{Z}_{j_3-1} + c_3 \mathcal{Z}_{j_3}).
\end{align} 
Integrating out the bosonic part of the first \(\bar{\delta}\),
\begin{align}
&I\bigl( \Delta_*^{i,j_1}(s_1,u), \Delta_*^{i,j_2}(s_2,v), \Delta_*^{i, j_3}(s_3,w)\bigr) \notag \\
 &= [*, i-1, i, j_1-1, j_1]\int \frac{dy_2dy_3dy_5dy_6}{y_2y_3y_5y_6}  \frac{D^2b}{b_1b_2b_3} \frac{D^2c}{c_1c_2c_3} \notag \\
 & \times \bar{\delta}^{4|4}(b_1Z_* + b_2 (-\frac{\langle *\, i\, j_1-1\, j_1 \rangle }{ \langle *\, i-1\, j_1-1\, j_1 \rangle }-y_2)\mathcal{Z}_{i-1} + b_2\mathcal{Z}_{i} + b_3 y_5 \mathcal{Z}_{j_2-1} + b_3 \mathcal{Z}_{j_2})  \notag \\
  & \times \bar{\delta}^{4|4}(c_1Z_* + c_2 (-\frac{\langle *\, i\, j_1-1\, j_1 \rangle }{ \langle *\, i-1\, j_1-1\, j_1 \rangle }-y_2-y_3) \mathcal{Z}_{i-1} + c_2\mathcal{Z}_{i} + c_3 y_6 \mathcal{Z}_{j_3-1} + c_3 \mathcal{Z}_{j_3}).  
\end{align} 
Rescaling \(y_2 \to -\frac{y_2}{b_2}\) followed by \(b_2 \to \langle *\, i-1\, j_1-1, j_1 \rangle b_2\) and \(y_5 \to \frac{y_5}{b_3}\),
\begin{align}
&I\bigl( \Delta_*^{i,j_1}(s_1,u), \Delta_*^{i,j_2}(s_2,v), \Delta_*^{i, j_3}(s_3,w)\bigr) \notag \\
 &= [*, i-1, i, j_1-1, j_1]\int  \frac{dy_2dy_3dy_5dy_6}{y_2y_3y_5y_6}  \frac{D^2b}{b_1b_2b_3} \frac{D^2c}{c_1c_2c_3} \notag \\
 & \times \bar{\delta}^{4|4}(b_1Z_* + b_2 \widehat{\mathcal{Z}_i}_{j_1} + y_2\mathcal{Z}_{i-1} + y_5 \mathcal{Z}_{j_2-1} + b_3 \mathcal{Z}_{j_2})  \notag \\
  & \times \bar{\delta}^{4|4}(c_1Z_* + c_2 (-\frac{\langle *\, i\, j_1-1\, j_1 \rangle }{ \langle *\, i-1\, j_1-1\, j_1 \rangle }+\frac{y_2}{\langle *\, i-1\, j_1-1, j_1 \rangle b_2}-y_3) \mathcal{Z}_{i-1} + c_2\mathcal{Z}_{i} + c_3 y_6 \mathcal{Z}_{j_3-1} + c_3 \mathcal{Z}_{j_3}).  
\end{align} 

Integrating out the bosonic part of the first \(\bar{\delta}\) above,
\begin{align}
& I\bigl( \Delta_*^{i,j_1}(s_1,u), \Delta_*^{i,j_2}(s_2,v), \Delta_*^{i, j_3}(s_3,w)\bigr) \notag \\
 &= [*, i-1, i, j_1-1, j_1] [*, \widehat{i}_{j_1}, i-1, j_2-1, j_2]\int  \frac{dy_3dy_6}{y_3y_6}  \frac{D^2c}{c_1c_2c_3} \notag \\
  & \times \bar{\delta}^{4|4}(c_1Z_* + c_2 (-\frac{\langle *\, i\, j_1-1\, j_1 \rangle }{ \langle *\, i-1\, j_1-1\, j_1 \rangle }-\frac{\langle *\, \widehat{i}_{j_1}\, j_2-1, j_2 \rangle}{\langle *\, i-1 \, j_2-1, j_2 \rangle \langle * i-1 j_1-1 j_1 \rangle }-y_3) \mathcal{Z}_{i-1}  \notag \\
 &+ c_2\mathcal{Z}_{i} + c_3 y_6 \mathcal{Z}_{j_3-1} + c_3 \mathcal{Z}_{j_3}). 
\end{align} 

Rescaling \(y_3 \to -\frac{y_3}{c_2}\) followed by \(c_2 \to \langle *\, i-1\, j_1-1\, j_1 \rangle \langle *\, i-1\, j_2-1\, j_2 \rangle c_2\) and \(y_6 \to \frac{y_6}{c_3}\),

\begin{align}
&I\bigl( \Delta_*^{i,j_1}(s_1,u), \Delta_*^{i,j_2}(s_2,v), \Delta_*^{i, j_3}(s_3,w)\bigr) \notag \\
 &= [*, i-1, i, j_1-1, j_1] [*, \widehat{i}_{j_1}, i-1, j_2-1, j_2]\int  \frac{dy_3dy_6}{y_3y_6}  \frac{D^2c}{c_1c_2c_3} \notag \\
  & \times \bar{\delta}^{4|4}(c_1Z_* + \langle *\, i-1\, j_1-1\, j_1 \rangle c_2 \widehat{\mathcal{Z}_{i}}_{j_2} + y_3\mathcal{Z}_{i-1} + c_2\mathcal{Z}_{i} + y_6 \mathcal{Z}_{j_3-1} + c_3 \mathcal{Z}_{j_3}) 
\end{align} 

where the coefficient of \(c_2\) follows from a simple calculation. Integrating out the bosonic part of the final \(\bar{\delta}\), we arrive at
\begin{align}
&I\bigl( \Delta_*^{i,j_1}(s_1,u), \Delta_*^{i,j_2}(s_2,v), \Delta_*^{i, j_3}(s_3,w)\bigr) \notag \\ 
 &= [*, i-1, i, j_1-1, j_1][*, j_2-1, j_2, i-1,  \widehat{i}_{j_1}][*, j_3-1, j_3, i-1, \widehat{i}_{j_2}] .
\end{align}

\end{document}

%% file: diagramDrawer.tex
\usepackage{pgfkeys}
\usepackage{pgffor}
\usepackage{tikz}
\usepackage{ifthen}
\usetikzlibrary{intersections}
\usetikzlibrary{decorations.pathmorphing}
\usetikzlibrary{calc}

\tikzset{%
  set style/.style={#1},
  expand style/.style={
    set style/.expanded={#1}
  },
  expand style once/.style={
    set style/.expand once={#1}
  },
  expand style twice/.style={
    set style/.expand twice={#1}
  }
}

\pgfkeys{
  /ww/.is family,
  /ww/.unknown/.code = {} 
}

\ExplSyntaxOn
\cs_new_eq:NN \ipLoop \int_step_inline:nn
\ExplSyntaxOff

\pgfkeys{
 /ww/defineline/.is family,/ww/defineline/.cd,
  /ww/defineline/.unknown/.code = {},/ww/defineline/linelength/.store in =\linelength, /ww/defineline/linelength/.default =1%
}

\newcommand{\defineline}[2]{%

  \pgfkeys{/ww/defineline/.cd,linelength}
  \pgfkeys{/ww/defineline/.cd,#2} 

  \pgfkeys{/ww/#1/.is family}%
  \pgfkeys{/ww/#1/.unknown/.code = \pgfkeyssetvalue{/ww/#1/\pgfkeyscurrentname}{##1}}%

  \pgfkeys{/ww/#1/.cd, #2}%

  \pgfkeysgetvalue{/ww/#1/pointcount}{\pointcount}%
  \def\pointsrange{(1.4*\linelength)}

  \pgfmathparse{\pointsrange/(\pointcount-1)}
  \def\pointgap{\pgfmathresult}
  
  \ipLoop{\pointcount}{

    \pgfkeysgetvalue{/ww/#1/originx}{\originx}
    \pgfkeysgetvalue{/ww/#1/originy}{\originy}
    \pgfkeysgetvalue{/ww/#1/angle}{\angle}

   \pgfmathsetmacro{\pointx}{\originx +(-\pointsrange/2 + (##1-1)*\pointgap)*cos(\angle)}   
   \pgfmathsetmacro{\pointy}{\originy +(-\pointsrange/2 + (##1-1)*\pointgap)*sin(\angle)}
   \pgfmathsetmacro{\linelengthtwo}{\linelength}

   \edef\keypx{/ww/#1/ipx##1}
   \edef\keypy{/ww/#1/ipy##1}
   \pgfkeyslet{\keypx}{\pointx}
   \pgfkeyslet{\keypy}{\pointy}

   \edef\keylinelength{/ww/#1/ilinelength}
   \pgfkeyslet{\keylinelength}{\linelengthtwo}

  }

}

\tikzset{gluon/.style={decorate, decoration={coil,amplitude=2pt, segment length=4pt},color=purple}}

\pgfkeys{
  /ww/prop/.cd,
  /ww/prop/.is family,
  /ww/prop/fromline/.initial = 1,
  /ww/prop/fromline/.store in = \fromline,
  /ww/prop/toline/.initial = 1,
  /ww/prop/toline/.store in=\toline,
  /ww/prop/fromip/.initial = 1,
  /ww/prop/fromip/.store in=\fromip,
  /ww/prop/toip/.initial = 1,
  /ww/prop/toip/.store in=\toip,
  /ww/prop/stublength/.store in=\stublength,
  /ww/prop/stublength/.default = 0.5,
  /ww/prop/markendpoints/.store in = \markendpoints,
  /ww/prop/markendpoints/.default = false,
  /ww/prop/startpointlabel/.store in=\startpointlabel,
  /ww/prop/startpointlabel/.default = \phantom{.},  
  /ww/prop/endpointlabel/.store in=\endpointlabel,
  /ww/prop/endpointlabel/.default = \phantom{.},  /ww/prop/labeldistone/.store in = \labeldistone,
  /ww/prop/labeldistone/.default = 0.5,
  /ww/prop/labeldisttwo/.store in = \labeldisttwo,
  /ww/prop/labeldisttwo/.default = 0.5,
  /ww/prop/looseval/.store in = \looseval,
  /ww/prop/looseval/.default = 1.4
}%

\pgfkeys{
  /ww/drawline/.is family,/ww/drawline/.cd,
  /ww/drawline/.unknown/.code = {},%
  /ww/drawline/customstyle/.initial = {line width=1},
  /ww/drawline/customstyle/.default = {line width=1},
  /ww/drawline/showtwistors/.store in = \showtwistors,
  /ww/drawline/showtwistors/.default = false,
  /ww/drawline/showtwistorsout/.store in = \showtwistorsout,
  /ww/drawline/showtwistorsout/.default = false,
  /ww/drawline/showwings/.store in = \showwings,
  /ww/drawline/showwings/.default = false,
  /ww/drawline/showvertex/.store in = \showvertex,
  /ww/drawline/showvertex/.default = false,
  /ww/drawline/labeldist/.store in = \labeldist,
  /ww/drawline/labeldist/.default = 0.5,
  /ww/drawline/wingorientation/.store in = \wingorientation,
  /ww/drawline/wingorientation/.default = inner,
  /ww/drawline/tolabel/.store in = \tolabel,
  /ww/drawline/tolabel/.default = ,
  /ww/drawline/fromlabel/.store in = \fromlabel,
  /ww/drawline/fromlabel/.default = ,
  /ww/drawline/linelength/.store in = \linelength,
}

\newcommand{\drawline}[2][]{%
  \pgfkeys{/ww/drawline/.cd,tolabel,fromlabel,customstyle,showtwistors,showtwistorsout,showwings,wingorientation,showvertex,labeldist}
  \pgfkeys{/ww/drawline/.cd,#1}
  
  \pgfkeysgetvalue{/ww/#2/originx}{\ox}%
  \pgfkeysgetvalue{/ww/#2/originy}{\oy}%
  \pgfkeysgetvalue{/ww/#2/angle}{\ang}%
  \pgfkeysgetvalue{/ww/#2/ilinelength}{\linelength}%

  \pgfmathsetmacro{\oxf}{\ox}%
  \pgfmathsetmacro{\oyf}{\oy}%
  \pgfmathsetmacro{\angf}{\ang}%
  \pgfmathsetmacro{\linelengthf}{\linelength}

  \pgfkeysgetvalue{/ww/drawline/customstyle}{\customstyle}
  
  \draw[expand style = \customstyle, name path=#2]
    ($(\oxf,\oyf) - (\angf:\linelengthf+0.3)$) 
    -- 
    ($(\oxf,\oyf) + (\angf:\linelengthf+0.3)$);

    \ifthenelse{\equal{\showtwistors}{true}}
           {
           
           \ifthenelse{\equal{\showtwistorsout}{false}}{\path ($(\oxf,\oyf) + (\angf:\linelengthf+0.0) + (\angf+90:\labeldist)$) node {\fromlabel};
    \path ($(\oxf,\oyf) - (\angf:\linelengthf+0.0) + (\angf+90:\labeldist)$) node {\tolabel}}{\path ($(\oxf,\oyf) + (\angf:\linelengthf+0.3) + (\angf+90:\labeldist)$) node {\fromlabel};
    \path ($(\oxf,\oyf) - (\angf:\linelengthf+0.3) + (\angf+90:\labeldist)$) node {\tolabel}};

    \ifthenelse{\equal{\showvertex}{true}}{
    \draw[fill=black] ($(\oxf,\oyf) - (\angf:\linelengthf)$) circle (0.05);
    \draw[fill=black] ($(\oxf,\oyf) + (\angf:\linelengthf)$) circle (0.05);
    }
    }{}

    \ifthenelse{\equal{\showwings}{true}}
    {
      \ifthenelse{\equal{\wingorientation}{inner}}
      {
        \draw[expand style = \customstyle] ($(\oxf,\oyf) - (\angf:\linelengthf) - (\angf-30:-0.2) $) -- ($(\oxf,\oyf) - (\angf:\linelengthf) - (\angf-30:0.5)$);
        \draw[expand style = \customstyle] ($(\oxf,\oyf) + (\angf:\linelengthf) - (\angf-150:-0.2)$) -- ($(\oxf,\oyf) + (\angf:\linelengthf) - (\angf-150:0.5)$);
      }{
        \draw[expand style = \customstyle] ($(\oxf,\oyf) - (\angf:\linelengthf) - (\angf+30:-0.2) $) -- ($(\oxf,\oyf) - (\angf:\linelengthf) - (\angf+30:0.5)$);
        \draw[expand style = \customstyle] ($(\oxf,\oyf) + (\angf:\linelengthf) - (\angf+150:-0.2)$) -- ($(\oxf,\oyf) + (\angf:\linelengthf) - (\angf+150:0.5)$);
        }
    }{}

}%

\newcommand{\drawpropagator}[1]{

  \pgfkeys{/ww/prop/.cd,markendpoints,startpointlabel,endpointlabel,labeldistone,labeldisttwo,looseval}
  \pgfkeys{/ww/prop/.cd,#1}

  \pgfkeysgetvalue{/ww/\fromline/ipx\fromip}{\ix}%
  \pgfkeysgetvalue{/ww/\fromline/ipy\fromip}{\iy}%
  \pgfkeysgetvalue{/ww/\toline/ipx\toip}{\fx}%
  \pgfkeysgetvalue{/ww/\toline/ipy\toip}{\fy}%

  \pgfkeysgetvalue{/ww/\fromline/angle}{\angi}%
  \pgfkeysgetvalue{/ww/\fromline/pointcount}{\pointcount}%
  \pgfkeysgetvalue{/ww/\toline/angle}{\angt}%

  \pgfmathsetmacro{\angout}{\angi - 90}
  \pgfmathsetmacro{\angin}{\angt - 90}

  \draw[/tikz/gluon] (\ix,\iy) to[out=\angout, in = \angin,looseness=\looseval] (\fx,\fy);

  \ifthenelse{\equal{\markendpoints}{true}}
  {
    \path ($(\fx,\fy) + (\angin:-\labeldistone) $) node {\endpointlabel};
    \path ($(\ix,\iy) + (\angout:-\labeldisttwo) $) node {\startpointlabel};
  }

}%

\newcommand{\drawpropagatortwo}[1]{

  \pgfkeys{/ww/prop/.cd,markendpoints,startpointlabel,endpointlabel,labeldistone,labeldisttwo,looseval}
  \pgfkeys{/ww/prop/.cd,#1}

  \pgfkeysgetvalue{/ww/\fromline/ipx\fromip}{\ix}%
  \pgfkeysgetvalue{/ww/\fromline/ipy\fromip}{\iy}%
  \pgfkeysgetvalue{/ww/\toline/ipx\toip}{\fx}%
  \pgfkeysgetvalue{/ww/\toline/ipy\toip}{\fy}%

  \pgfkeysgetvalue{/ww/\fromline/angle}{\angi}%
  \pgfkeysgetvalue{/ww/\fromline/pointcount}{\pointcount}%
  \pgfkeysgetvalue{/ww/\toline/angle}{\angt}%

  \pgfmathsetmacro{\angout}{\angi -90}
  \pgfmathsetmacro{\angin}{\angt +90}

  \draw[/tikz/gluon] (\ix,\iy) to[out=\angout, in = \angin,looseness=\looseval] (\fx,\fy);

  \ifthenelse{\equal{\markendpoints}{true}}
  {
    \path ($(\fx,\fy) + (\angin:-\labeldistone) $) node {\endpointlabel};
    \path ($(\ix,\iy) + (\angout:-\labeldisttwo) $) node {\startpointlabel};
  }

}%

\newcommand{\drawpropagatorstub}[1]{
  \pgfkeys{/ww/prop/.cd,markendpoints,startpointlabel,labeldistone,stublength}
  \pgfkeys{/ww/prop/.cd,#1}
  
  \pgfkeysgetvalue{/ww/\fromline/ipx\fromip}{\ix}%
  \pgfkeysgetvalue{/ww/\fromline/ipy\fromip}{\iy}%

  \pgfkeysgetvalue{/ww/\fromline/angle}{\angi}%
  \pgfkeysgetvalue{/ww/\fromline/pointcount}{\pointcount}%

  \pgfmathsetmacro{\angout}{\angi - 90}

  \draw[/tikz/gluon] (\ix,\iy) -- ++ (\angout:\stublength);

  \ifthenelse{\equal{\markendpoints}{true}}
  {
    \path ($(\ix,\iy) + (\angout:-\labeldistone) $) node {\startpointlabel};
  }
  
}%

\newcommand{\drawpropellipsis}[1]{
  \pgfkeys{/ww/prop/.cd,#1}
  \pgfkeysgetvalue{/ww/\fromline/ipx\fromip}{\ix}%
  \pgfkeysgetvalue{/ww/\fromline/ipy\fromip}{\iy}%

  \pgfkeysgetvalue{/ww/\fromline/angle}{\angi}%
  \pgfkeysgetvalue{/ww/\fromline/pointcount}{\pointcount}%

  \pgfmathsetmacro{\angout}{\angi - 90}

  \node[rotate=\angout-90] at ({\ix + 0.3*cos(\angout)},{\iy + 0.3*sin(\angout)}) {$\hdots$};
}%

\newcommand{\generalMHVrule}{
       \begin{tikzpicture}[baseline={([yshift=-.5ex]current bounding box.center)}]

  \draw[line width = 1] (0,0) -- ++(120:2cm);
  \draw[line width = 1] (0,0) -- ++(140:2cm);
  \draw[line width = 1] (0,0) -- ++(-120:2cm);
  \draw[line width = 1] (0,0) -- ++(3,0);
  \draw[line width = 1] (3,0) -- ++(60:2cm);
  \draw[line width = 1] (3,0) -- ++(300:2cm);
  \draw[line width = 1] (3,0) -- ++(320:2cm);

  \filldraw[black] (1.5,1) circle (1.5pt) node[anchor=west]{$x_i$};
  \filldraw[black] (4.25,1) circle (1.5pt) node[anchor=west]{$x_j$};
  \filldraw[black] (1.5,-1) circle (1.5pt) node[anchor=west]{$x_k$};
  \filldraw[black] (-1.25,-1) circle (1.5pt) node[anchor=east]{$x_l$};
  
  \filldraw[black] (-1.25,1.5) circle (0.75pt);
  \filldraw[black] (-1.25+0.1,1.5+0.1) circle (0.75pt);
  \filldraw[black] (-1.25-0.1,1.5-0.1) circle (0.75pt);

  \filldraw[black] (4.25,-1.5) circle (0.75pt);
  \filldraw[black] (4.25+0.1,-1.5+0.1) circle (0.75pt);
  \filldraw[black] (4.25-0.1,-1.5-0.1) circle (0.75pt);
           
       \end{tikzpicture}
   }

\newcommand{\PentSquareOneloopDiagram}{
       \begin{tikzpicture}[baseline={([yshift=-.5ex]current bounding box.center)}]

  \clip (-3.25,-3.75) rectangle (10.75, 4);
  \defineline{lin1}{originx=0.809, originy=2.490, angle=-36, pointcount=3}
  \defineline{lin2}{originx=1+0.309, originy=0.951, angle=72, pointcount=3}
  \defineline{lin3}{originx=0, originy=0, angle=0, pointcount=3}
  \defineline{lin4}{originx=-1-0.309, originy=0.951, angle=-72, pointcount=3}
  \defineline{lin5}{originx=-0.809, originy=2.490, angle=216, pointcount=3}
  
  \drawline[wingorientation=outer,showtwistors=true,tolabel=$\hspace{-6mm}Z_{1}$]{lin1}
  \drawline[wingorientation=outer,showtwistors=true,fromlabel=$\hspace{18.5mm}Z_{2}$,tolabel=$\hspace{22mm}Z_{3}$]{lin2}
  \drawline[showtwistors=true,wingorientation=outer]{lin3}
  \drawline[wingorientation=outer,showtwistors=true,fromlabel=$\hspace{-23mm}Z_{4}$,tolabel=$\hspace{-19mm}Z_{5}$]{lin4}
  \drawline[wingorientation=outer,showtwistors=true]{lin5}

  \defineline{linA}{originx=4, originy=-3, angle=180, pointcount=4,linelength=2}
  \drawline[showtwistors=true,showvertex=true,wingorientation=outer,tolabel=$Z_B$,fromlabel=$Z_A$]{linA}

  \defineline{lin8}{originx=8, originy=0, angle=0, pointcount=3}
  \defineline{lin9}{originx=7, originy=1, angle=270, pointcount=3}
  \defineline{lin10}{originx=8, originy=2, angle=180, pointcount=3}
  \defineline{lin7}{originx=9, originy=1, angle=270, pointcount=3}
  \drawline[showtwistors=true,wingorientation=outer]{lin8}
  \drawline[showtwistors=true,wingorientation=outer,fromlabel=$\hspace{-23mm}Z_{9}$,tolabel=$\hspace{-23mm}Z_{6}$]{lin9}
  \drawline[showtwistors=true,wingorientation=outer]{lin10}
  \drawline[showtwistors=true,wingorientation=outer,tolabel=$\hspace{2mm}Z_{7}$,fromlabel=$\hspace{2mm}Z_{8}$]{lin7}

   \drawpropagator{fromline=lin3, toline=lin5, fromip=1, toip=2,looseval=7.9}
  \drawpropagator{fromline=lin3, toline=lin9, fromip=3, toip=2,looseval=1}
  \drawpropagator{fromline=lin3, toline=linA, fromip=2, toip=3}
  \drawpropagator{fromline=linA, toline=lin8, fromip=2, toip=2}
           
       \end{tikzpicture}
   }

\newcommand{\GeneralPropRule}{
        \begin{tikzpicture}[baseline={([yshift=-.5ex]current bounding box.center)}]
  \defineline{lin1}{originx=-4, originy=-1.75, angle=180, pointcount=5,linelength=1.4}
  \defineline{lin2}{originx=0, originy=-1.75, angle=180, pointcount=6,linelength=1.4}
  \defineline{lin3}{originx=0, originy=1.75, angle=0, pointcount=6,linelength=1.4}
  \defineline{lin4}{originx=4, originy=1.75, angle=0, pointcount=5,linelength=1.4}

  \drawline[showtwistors=true, fromlabel=$Z_{l}$,tolabel=$Z_{l-1}$,showvertex=true]{lin1}
  \drawline[showtwistors=true,fromlabel=$Z_{k}$,tolabel=$Z_{k-1}$,showvertex=true]{lin2}
  \drawline[showtwistors=true,fromlabel=$Z_{i}$,tolabel=$Z_{i-1}$,showvertex=true]{lin3}
  \drawline[showtwistors=true,fromlabel=$Z_{j}$,tolabel=$Z_{j-1}$,showvertex=true]{lin4}

  \drawpropagator{fromline=lin1, toline=lin2, fromip=3, toip=4}
  \drawpropagator{fromline=lin2, toline=lin3, fromip=3, toip=3}
  \drawpropagator{fromline=lin3, toline=lin4, fromip=4, toip=3}

  \drawpropagatorstub{fromline=lin1, fromip=1}
  \drawpropellipsis{fromline=lin1, fromip=2}
  \drawpropellipsis{fromline=lin1, fromip=4}
  \drawpropagatorstub{fromline=lin1, fromip=5}

  \drawpropagatorstub{fromline=lin2, fromip=1}
  \drawpropellipsis{fromline=lin2, fromip=2}
  \drawpropellipsis{fromline=lin2, fromip=5}
  \drawpropagatorstub{fromline=lin2, fromip=6}

  \drawpropagatorstub{fromline=lin3, fromip=1}
  \drawpropellipsis{fromline=lin3, fromip=2}
  \drawpropellipsis{fromline=lin3, fromip=5}
  \drawpropagatorstub{fromline=lin3, fromip=6}

  \drawpropagatorstub{fromline=lin4, fromip=1}
  \drawpropellipsis{fromline=lin4, fromip=2}
  \drawpropellipsis{fromline=lin4, fromip=4}
  \drawpropagatorstub{fromline=lin4, fromip=5}
  
\end{tikzpicture}
}

\newcommand{\nvertexdiag}{\begin{tikzpicture}[scale=0.9][baseline={([yshift=-.5ex]current bounding box.center)}]
  \defineline{lin1}{originx=0, originy=0.5, angle=0, pointcount=9,linelength=2}
  \defineline{lin2}{originx=-4, originy=-0.5, angle=60, pointcount=5,linelength=1}
  \defineline{lin3}{originx=-4, originy=-4, angle=120, pointcount=5,linelength=1}
  \defineline{lin4}{originx=4, originy=-4, angle=240, pointcount=5,linelength=1}
  \defineline{lin5}{originx=4, originy=-0.5, angle=300, pointcount=5,linelength=1}

  \drawline[showtwistors=true, showtwistorsout=true,labeldist=0.6,showvertex=true,fromlabel=$Z_{B}$,tolabel=$Z_{A}$]{lin1}
  \drawline[showtwistors=true,showtwistorsout=true,labeldist=0.6,fromlabel=$Z_{i_n}$,tolabel=$Z_{i_n-1}$,showwings=true,wingorientation=outer]{lin2}
  \drawline[showtwistors=true,showtwistorsout=true,labeldist=0.6,fromlabel=$Z_{i_{n-1}}$,tolabel=$Z_{i_{n-1}-1}$,showwings=true,wingorientation=outer]{lin3}
  \drawline[showtwistors=true,showtwistorsout=true,labeldist=0.6,fromlabel=$Z_{i_2}$,tolabel=$Z_{i_2-1}$,showwings=true,wingorientation=outer]{lin4}
  \drawline[showtwistors=true,showtwistorsout=true,labeldist=0.6,fromlabel=$Z_{i_1}$,tolabel=$Z_{i_1-1}$,showwings=true,wingorientation=outer]{lin5}

  \drawpropagator{fromline=lin1, toline=lin2, fromip=1, toip=3,markendpoints=true,labeldisttwo=0.3,startpointlabel=$u_{n}$,endpointlabel=$t_{n,\tilde{k}_n}$,looseval=1.8}
  \drawpropagator{fromline=lin1, toline=lin3, fromip=3, toip=3,markendpoints=true,labeldisttwo=0.3,startpointlabel=$u_{n-1}$,endpointlabel=$t_{n-1,\tilde{k}_{n-1}}$,labeldistone=0.65,looseval=1.4}
  \drawpropellipsis{fromline=lin1, fromip=5}
  \drawpropagator{fromline=lin1, toline=lin4, fromip=7, toip=3,markendpoints=true,labeldisttwo=0.3,startpointlabel=$u_{2}$,endpointlabel=$t_{2,\tilde{k}_2}$,looseval=1.4}
  \drawpropagator{fromline=lin1, toline=lin5, fromip=9, toip=3,markendpoints=true,labeldisttwo=0.3,startpointlabel=$u_{1}$,endpointlabel=$t_{1,\tilde{k}_1}$,looseval=1.8}

  \drawpropagatorstub{fromline=lin2, fromip=1,markendpoints=true,startpointlabel=$t_{n,k_n}$}
  \drawpropellipsis{fromline=lin2, fromip=2}
  \drawpropellipsis{fromline=lin2, fromip=4}
  \drawpropagatorstub{fromline=lin2, fromip=5,markendpoints=true,startpointlabel=$t_{n,1}$}

  \drawpropagatorstub{fromline=lin3, fromip=1, markendpoints=true,labeldistone=0.6,startpointlabel=$t_{n-1,1}$}
  \drawpropellipsis{fromline=lin3, fromip=2}
  \drawpropellipsis{fromline=lin3, fromip=4}
  \drawpropagatorstub{fromline=lin3, fromip=5,markendpoints=true,labeldistone=0.6,startpointlabel=$t_{n-1,k_{n-1}}\hspace{3mm}$}

  \drawpropagatorstub{fromline=lin4, fromip=1,markendpoints=true,startpointlabel=$t_{2,k_2}$}
  \drawpropellipsis{fromline=lin4, fromip=2}
  \drawpropellipsis{fromline=lin4, fromip=4}
  \drawpropagatorstub{fromline=lin4, fromip=5,markendpoints=true,startpointlabel=$t_{2,1}$}

  \drawpropagatorstub{fromline=lin5, fromip=1,,markendpoints=true,startpointlabel=$t_{1,k_1}$}
  \drawpropellipsis{fromline=lin5, fromip=2}
  \drawpropellipsis{fromline=lin5, fromip=4}
  \drawpropagatorstub{fromline=lin5, fromip=5,markendpoints=true,startpointlabel=$t_{1,1}$}

\end{tikzpicture}
      }

\newcommand{\nvertexdiagwbox}{\begin{tikzpicture}[scale=0.9][baseline={([yshift=-.5ex]current bounding box.center)}]
  \defineline{lin1}{originx=0, originy=0.5, angle=0, pointcount=9,linelength=2}
  \defineline{lin2}{originx=-4, originy=-0.5, angle=60, pointcount=5,linelength=1}
  \defineline{lin3}{originx=-4, originy=-4, angle=120, pointcount=5,linelength=1}
  \defineline{lin4}{originx=4, originy=-4, angle=240, pointcount=5,linelength=1}
  \defineline{lin5}{originx=4, originy=-0.5, angle=300, pointcount=5,linelength=1}

  \drawline[showtwistors=true, showtwistorsout=true,labeldist=0.6,showvertex=true,fromlabel=$Z_{B}$,tolabel=$Z_{A}$]{lin1}
  \drawline[showtwistors=true,showtwistorsout=true,labeldist=0.6,fromlabel=$Z_{i_n}$,tolabel=$Z_{i_n-1}$,showwings=true,wingorientation=outer]{lin2}
  \drawline[showtwistors=true,showtwistorsout=true,labeldist=0.6,fromlabel=$Z_{i_{n-1}}$,tolabel=$Z_{i_{n-1}-1}$,showwings=true,wingorientation=outer]{lin3}
  \drawline[showtwistors=true,showtwistorsout=true,labeldist=0.6,fromlabel=$Z_{i_2}$,tolabel=$Z_{i_2-1}$,showwings=true,wingorientation=outer]{lin4}
  \drawline[showtwistors=true,showtwistorsout=true,labeldist=0.6,fromlabel=$Z_{i_1}$,tolabel=$Z_{i_1-1}$,showwings=true,wingorientation=outer]{lin5}

  \drawpropagator{fromline=lin1, toline=lin2, fromip=1, toip=3,markendpoints=true,labeldisttwo=0.3,startpointlabel=$u_{n}$,endpointlabel=$t_{n,\tilde{k}_n}$,looseval=2.8}
  \drawpropagator{fromline=lin1, toline=lin3, fromip=3, toip=3,markendpoints=true,labeldisttwo=0.3,startpointlabel=$u_{n-1}$,endpointlabel=$t_{n-1,\tilde{k}_{n-1}}$,labeldistone=0.65,looseval=1.8}
  \drawpropellipsis{fromline=lin1, fromip=5}
  \drawpropagator{fromline=lin1, toline=lin4, fromip=7, toip=3,markendpoints=true,labeldisttwo=0.3,startpointlabel=$u_{2}$,endpointlabel=$t_{2,\tilde{k}_2}$,looseval=1.8}
  \drawpropagator{fromline=lin1, toline=lin5, fromip=9, toip=3,markendpoints=true,labeldisttwo=0.3,startpointlabel=$u_{1}$,endpointlabel=$t_{1,\tilde{k}_1}$,looseval=2.8}

  \drawpropagatorstub{fromline=lin2, fromip=1,markendpoints=true,startpointlabel=$t_{n,k_n}$}
  \drawpropellipsis{fromline=lin2, fromip=2}
  \drawpropellipsis{fromline=lin2, fromip=4}
  \drawpropagatorstub{fromline=lin2, fromip=5,markendpoints=true,startpointlabel=$t_{n,1}$}

  \drawpropagatorstub{fromline=lin3, fromip=1, markendpoints=true,labeldistone=0.6,startpointlabel=$t_{n-1,1}$}
  \drawpropellipsis{fromline=lin3, fromip=2}
  \drawpropellipsis{fromline=lin3, fromip=4}
  \drawpropagatorstub{fromline=lin3, fromip=5,markendpoints=true,labeldistone=0.6,startpointlabel=$t_{n-1,k_{n-1}}\hspace{3mm}$}

  \drawpropagatorstub{fromline=lin4, fromip=1,markendpoints=true,startpointlabel=$t_{2,k_2}$}
  \drawpropellipsis{fromline=lin4, fromip=2}
  \drawpropellipsis{fromline=lin4, fromip=4}
  \drawpropagatorstub{fromline=lin4, fromip=5,markendpoints=true,startpointlabel=$t_{2,1}$}

  \drawpropagatorstub{fromline=lin5, fromip=1,,markendpoints=true,startpointlabel=$t_{1,k_1}$}
  \drawpropellipsis{fromline=lin5, fromip=2}
  \drawpropellipsis{fromline=lin5, fromip=4}
  \drawpropagatorstub{fromline=lin5, fromip=5,markendpoints=true,startpointlabel=$t_{1,1}$}

   \draw[fill=lightgray, draw=black] (-2,-1.25) rectangle (2,-0.1);
   \node[draw=lightgray] at (0,-0.675) {permutation $\sigma$};
\end{tikzpicture}
      }
